\documentclass{article}

\usepackage[main,final]{neurips_2026}

\usepackage[utf8]{inputenc} % allow utf-8 input
\usepackage[T1]{fontenc}    % use 8-bit T1 fonts
\usepackage{hyperref}       % hyperlinks
\usepackage{url}            % simple URL typesetting
\usepackage{booktabs}       % professional-quality tables
\usepackage{amsfonts}       % blackboard math symbols
\usepackage{nicefrac}       % compact symbols for 1/2, etc.
\usepackage{microtype}      % microtypography
\usepackage{xcolor}         % colors

\usepackage{graphicx}
\usepackage{subcaption}

\newcommand{\algcomment}[1]{\hspace{1.5em}{\small (#1)}}

\usepackage{algorithm}
\usepackage{algorithmic}

\usepackage{amsmath}
\usepackage{amssymb}
\usepackage{mathtools}
\usepackage{amsthm}

\usepackage{makecell}
\usepackage{booktabs}

\usepackage{placeins}   % gives you \FloatBarrier
\usepackage{float}      % gives you the [H] specifier

\theoremstyle{plain}
\newtheorem{theorem}{Theorem}[section]

\newtheorem{corollary}[theorem]{Corollary}
\theoremstyle{definition}

\theoremstyle{remark}

\usepackage{chemformula}
\usepackage{tabularx}
\usepackage{multirow}

\title{CupOFMoCA: Coupled Objective-Guided Discrete Flows for Molecular Conjugate Assembly}

\author{%
  Ruoxi Rosie Zhang,$^{1,\ast}$
  Ziang Li,$^{2,\ast}$
  Jiatao Gu,$^{1}$
  Pranam Chatterjee$^{1,3,\dagger}$ \\[0.4em]
  $^{1}$Department of Computer and Information Science,
  University of Pennsylvania \\
  $^{2}$College of Arts and Sciences, University of Pennsylvania \\
  $^{3}$Department of Bioengineering, University of Pennsylvania \\[0.4em]
  $^{\ast}$Equal contribution \\
  $^{\dagger}$Corresponding Author:
  \href{mailto:pranam@upenn.edu}
       {\texttt{pranam@upenn.edu}} \\
  Code: \url{https://huggingface.co/ChatterjeeLab/CupOFMoCA}
}

\begin{document}

\maketitle

\begin{abstract}
Molecular conjugates, including PROTACs and peptide-drug conjugates (PDCs), derive their function from the joint behavior of multiple coupled components, yet most generative approaches design these components independently and combine them only after generation. Such staged pipelines ignore cross-component dependencies and often produce conjugates that are chemically invalid, fall outside empirical conjugate distributions, or lose function upon assembly. In this work, we introduce $\textbf{C}$o$\textbf{up}$led $\textbf{O}$bjective-Guided Discrete $\textbf{F}$lows for $\textbf{Mo}$lecular $\textbf{C}$onjugate $\textbf{A}$ssembly $(\textbf{CupOFMoCA})$, a discrete generative method that enforces cross-component chemical constraints during conjugate generation. CupOFMoCA restricts guided updates to chemically feasible conjugate states and reweights local transitions using molecular design objectives. Across three PDC systems, coupled generation achieved $100\%$ conjugation validity among valid peptides and increased valid yield over matched post-hoc filtering, with comparable predicted binding scores across the full valid sets. In three-seed evaluations of three PROTAC systems, CupOFMoCA achieved the highest assembly validity on two systems and generated more assembly-valid molecules per second than the evaluated baselines. We further evaluated the generated PROTACs with post-hoc DegradeMaster scoring and explicit-solvent molecular dynamics, supporting their predicted activity and conformational properties. Together, these results establish CupOFMoCA as an efficient generator of molecular conjugates, supporting therapeutic development via coordinated control of assembly and molecular properties. 
\end{abstract}

\section{Introduction}

Molecular conjugates play a central role in modern therapeutics and chemical biology by integrating complementary functional components into a single molecular entity \cite{cooper2021peptides, hua2022beyond}. Conjugation strategies improve specificity, modulate pharmacokinetics, and enable mechanisms unavailable to individual components alone \citep{marei2022potential, eras2022chemical, drago2021unlocking}. Representative examples include peptide-drug conjugates (PDCs), which combine sequence-based targeting with small-molecule potency \citep{rizvi2024peptide, heh2023peptide, dean2024peptide}, and proteolysis-targeting chimeras (PROTACs), which induce proximity between a target protein and the ubiquitin-proteasome machinery \citep{vikal2025protacs, wang2024new}. In these systems, function emerges from the coupled behavior of the assembled conjugate rather than from its components in isolation, making conjugate design an inherently joint problem.

Structure-based drug discovery using three-dimensional protein-ligand complexes is widely used \cite{wang2018structure, albanese2025computational}, but depends on high-quality structural data and is less effective for flexible or disordered systems \citep{isert2023structure, wang2018structure}. Generative modeling has therefore emerged as a complementary approach for molecular design. Autoregressive, masked, and diffusion-based models \citep{madani2020progen, ferruz2022protgpt2, pmlr-v235-wang24ct, alamdari2023evodiff, tang2025peptunenovogenerationtherapeutic, tang2025tr2d2treesearchguided, lee2025genmol} and discrete flow matching methods \citep{campbell2024generative, stark2024dirichlet, davis2024fisherfm, tang2025gumbelfm, chen2025multiobjectiveguideddiscreteflowmatching, chen2025areurediannealedrectifiedupdates} enable efficient generation within individual molecular modalities. However, these approaches typically generate components independently and assemble them post hoc, traversing chemically implausible intermediate states and failing to enforce compatibility at the system level. This limitation is particularly pronounced for conjugates, where validity and function depend on the assembled structure. In PROTACs, for example, linker properties directly control ternary complex formation, and small perturbations can abolish degradation activity despite preserved binding \citep{bekes2022protac, troup2020current, cao2022chemistries}.

To incorporate these dependencies during design, we seek to generate conjugates under joint chemical constraints and molecular objectives. Towards this goal, we introduce \textbf{C}o\textbf{up}led \textbf{O}bjective-Guided Discrete \textbf{F}lows for \textbf{Mo}lecular \textbf{C}onjugate \textbf{A}ssembly (\textbf{CupOFMoCA}), a discrete generative framework that samples over assembled states under explicit conjugation constraints and target-specific objectives. By operating on discrete flows, CupOFMoCA restricts generation to a chemically feasible conjugate manifold, enabling anticipatory assembly in which intermediate states remain valid and design decisions account for downstream constraints.

Our contributions are threefold:
\begin{enumerate}
    \item \textbf{Coupled discrete flows for molecular conjugate assembly.}  
    We generate conjugates through discrete flows over assembled states and enforce cross-component chemical compatibility during guided sampling.

    \item \textbf{Objective-guided sampling under hard validity constraints.}  
    We introduce a sampling scheme that incorporates functional objectives while enforcing strict chemical and cross-component validity constraints.

    \item \textbf{A unified framework across conjugate modalities.}  
    We evaluate CupOFMoCA across three PDC and three PROTAC systems. Coupled PDC generation increased valid yield with comparable predicted binding scores across the full valid sets. For PROTACs, we improved assembly validity through fixed-anchor sampling and linker-specific fine-tuning, then evaluated the generated molecules with post-hoc degradation prediction and molecular dynamics.
\end{enumerate}

\section{Related Works}

\textbf{Discrete Flow Matching.}
Discrete flow matching has emerged as a principled approach for modeling and sampling from complex discrete spaces. One line of work focuses on jump-process models, which learn time-dependent transition rates for token-level stochastic updates and induce continuous-time Markov chains with prescribed marginals \citep{gat2024discrete, campbell2024generative}. In contrast, continuous-time simplex methods diffuse discrete data through continuous embeddings over the probability simplex, enabling training via flow-matching or optimal transport objectives defined on relaxed categorical representations \citep{tang2025gumbelfm, davis2024fisherfm, stark2024dirichlet}. More recently, rectified formulations such as Rectified Discrete Flows (ReDi) \citep{yoo2025redi} simplify probability paths and improve sampling efficiency, while AReUReDi \citep{chen2025areurediannealedrectifiedupdates} extends this paradigm to multi-objective settings with Pareto-guided sampling. Using multi-objective discrete flow matching (MOG-DFM) \citep{chen2025multiobjectiveguideddiscreteflowmatching} as a base framework, moPPIt establishes motif-specific peptide design \citep{chen2026moppit}, while SOAPIA generates target-selective binders with off-target avoidance \citep{vincoff2026soapia}. To address viral variation, PROPHET uses MOG-DFM to optimize predicted binding across phylogenetically plausible variants \citep{liang2026prophet}. Finally, for editing existing sequences, pCoMole combines preference guidance with hard terminal feasibility \citep{chen2026pcomole}. Here, we extend control to cross-component attachment constraints during conjugate sampling.

\textbf{Constrained Generation for Biomolecule Design.}
Generative models for biomolecule design have largely focused on optimizing continuous objectives through reward-weighted sampling, classifier guidance, or guided diffusion \citep{nisonoff2024unlocking, gruver2023proteindesignguideddiscrete}. However, many design requirements are more naturally expressed as feasibility constraints rather than scalar objectives. Examples include representation validity, where generated graphs or SMILES strings must parse into chemically valid molecules \citep{jin2018junction, tang2025peptunenovogenerationtherapeutic, chen2025areurediannealedrectifiedupdates}, and developability thresholds such as thermostability requirements in antibody design \citep{zeng2024antibody}. Constraint-aware methods incorporate differentiable constraints into language-model sampling \citep{kumar2022gradient} and discrete molecular diffusion \citep{cardei2025constrained}, while projected diffusion explicitly enforces feasibility \citep{christopher2024constrained}. We extend constraint-aware generation to cross-component attachment chemistry in assembled molecular conjugates.

\textbf{Generative Design of Molecular Conjugates and Linkers.}
Generative approaches for molecular conjugates have typically focused on individual components. In PROTAC design, SMILES-based methods such as SyntaLinker \citep{yang2020syntalinker}, PROTAC-INVENT \citep{li2023protacinvent}, and PROTAC-TS \citep{murakami2025protacts} treat linker generation as a sequence completion or reinforcement learning problem, while 3D approaches such as DeLinker \citep{imrie2020delinker}, DiffLinker \citep{igashov2024difflinker}, DiffPROTACs \citep{li2024diffprotacs}, and LinkerNet \citep{guan2023linkernet} generate linkers conditioned on fragment geometry. These methods support fragment-conditioned linker design, including ternary-complex docking in PROTAC-INVENT and permeability optimization in PROTAC-TS. CupOFMoCA combines fixed-anchor preservation with explicit assembly checks and objective-guided updates. Similarly, peptide-drug conjugate design pipelines generate peptides independently and attach payloads post hoc, leading to frequent failures at the conjugation stage.

CupOFMoCA integrates discrete generative transport, molecular objectives, and hard cross-component validity checks within a common sampling procedure for PDCs and PROTACs.

% \begin{figure}[H]
%     \centering
%     \includegraphics[width=0.8\textwidth]{figures/CupOFMoCA_algorithm.png}
%     \caption{\textbf{CupOFMoCA Training Overview.} \textbf{(A)} Conceptual flow showing the transformation of a simple, unassembled prior distribution ($\mu_0$) into a distribution of valid, assembled molecular conjugates ($\mu_1$) over time $t=0$ to $T$, guided by validity constraints and objectives $S(z)$. \textbf{(B)} Schematic of the flow-map consistency training.}
%     \label{fig:CupOFMoCA_algorithm}
% \end{figure}

\begin{figure*}[h]
  \centering
  \includegraphics[width=\textwidth]{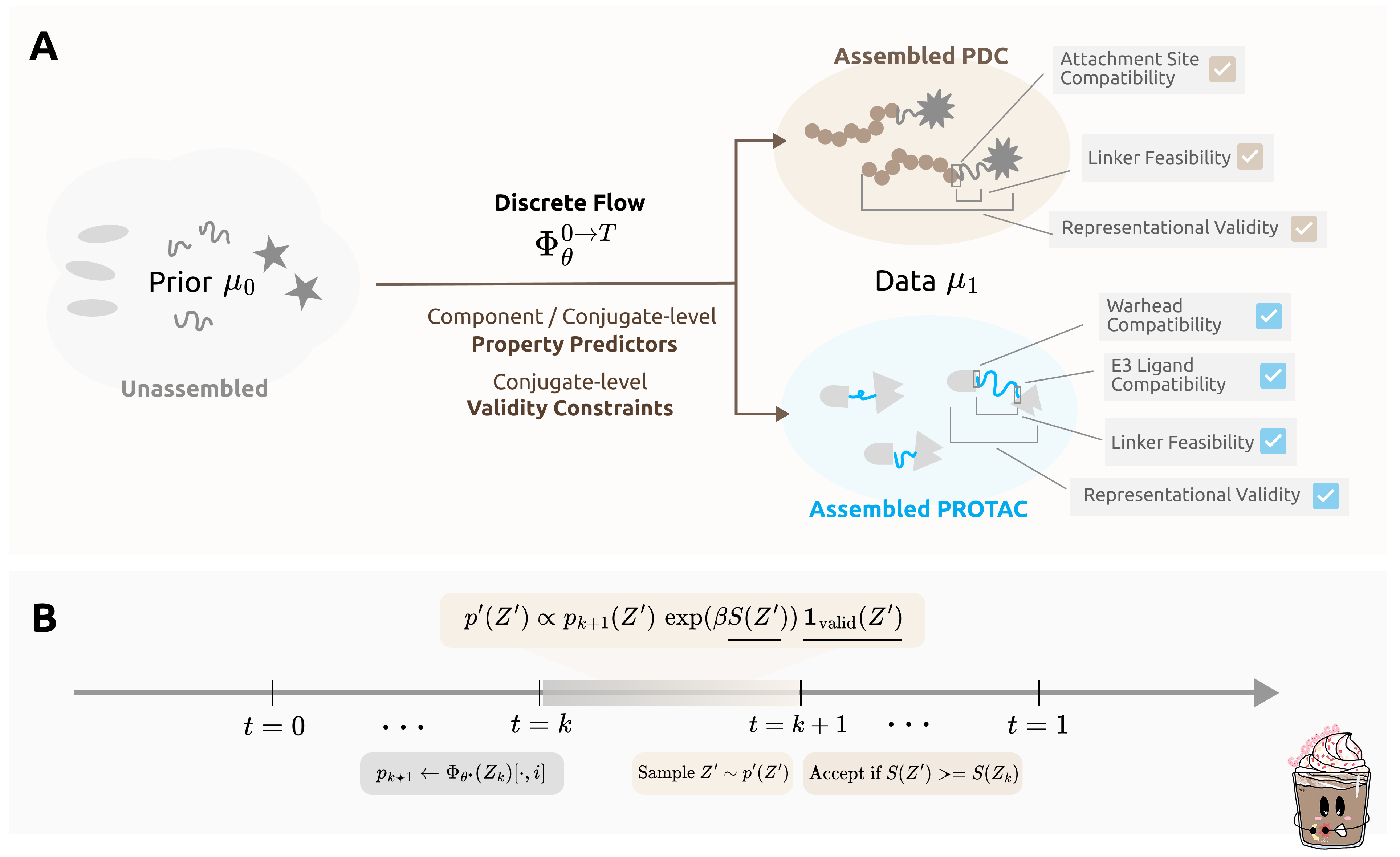}
    \caption{\textbf{Overview of CupOFMoCA for molecular conjugate generation.}
    \textbf{(A)} A pre-trained discrete flow model is guided from unassembled components to valid assembled conjugates, including peptide-drug conjugates (PDCs) and PROTACs, under objective guidance and hard validity constraints (attachment compatibility, linker feasibility, and representational validity). \textbf{(B)} Objective-guided sampling with hard validity constraints. We reweight candidate probabilities with objective scores and exclude invalid updates with a validity mask. The illustrated PDC acceptance rule retains updates that preserve or increase the score. For PROTACs, we accept valid tilted proposals and return the highest-scoring valid state encountered.}
    \label{fig:wide}
\end{figure*}

\section{CupOFMoCA}
\label{sec:problem}

CupOFMoCA generates discrete molecular conjugates by updating free components under constraints imposed by the assembled state. The framework is representation-agnostic, as each component of a conjugate may be expressed as a SMILES token sequence or as a labeled molecular graph, and different components within a single conjugate may use different representations. We instantiate CupOFMoCA in two settings: peptide-drug conjugates (PDCs), where peptides are represented as peptide SMILES strings and the model assembles peptide, linker, and drug components; and the PROTAC-specific instantiation, where states are molecular graphs and the model edits a linker subgraph between fixed warhead and E3-ligand subgraphs (Figure~\ref{fig:wide}A).

\subsection{Conjugate State Space}
\label{subsec:data_space}
Let $\mathcal{X}^{(c)}$ be the state space of component $c$, either a SMILES token space $\mathcal{V}^{L}$ or a graph space $\mathcal{G}_{N}=\mathcal{A}^{N}\times\mathcal{B}^{N\times N}$ with empty tokens enabling insertion/deletion. A conjugate is
\begin{equation}
    z=\bigl(x^{(1)},\dots,x^{(M)}\bigr)\in\mathcal{Z}:=\textstyle\prod_{c}\mathcal{X}^{(c)},
\end{equation}
with components partitioned into \emph{fixed} ($\mathcal{C}_{\mathrm{fix}}$, immutable inputs) and \emph{free} ($\mathcal{C}_{\mathrm{free}}$, edited by the model). PDCs use $z=(x^{\mathrm{pep}},x^{\mathrm{link}},x^{\mathrm{drug}})$ with $\mathcal{C}_{\mathrm{fix}}=\{\mathrm{link},\mathrm{drug}\}$; PROTACs use $z=(x^{\mathrm{wh}},x^{\mathrm{link}},x^{\mathrm{e3}})$ with $\mathcal{C}_{\mathrm{fix}}=\{\mathrm{wh},\mathrm{e3}\}$. Cross-component bonds at fixed-free junctions are assigned to the adjacent free component.

We assume an unknown data distribution $\mu_{1}$ over $\mathcal{Z}$ (PDCdb decompositions for PDCs; PROTAC-DB decompositions for PROTACs \citep{weng2021protacdb}), and a prior $\mu_{0}$ in which fixed components are set to the target system and free components are drawn from the uninformative DFM prior (uniform random tokens for SMILES, empty subgraph for graphs). The flow learns transport $Z_{0}\sim\mu_{0}\Rightarrow Z_{T}\sim\mu_{1}$ that preserves fixed components throughout.

\subsection{Fixed Anchors and Edit Neighborhoods}
\label{subsec:anchors}
For each fixed-free junction we specify an \emph{anchor atom} $a^{(c)}$ in $c\in\mathcal{C}_{\mathrm{fix}}$ at which conjugating chemistry occurs; the atom type at $a^{(c)}$ is immutable. In PROTACs on EGFR/VHL, $a^{\mathrm{wh}}$ is the Gefitinib-derived \textsf{-COOH} and $a^{\mathrm{e3}}$ the VH032 \textsf{-NH}$_{2}$, with assembly via amide closure at both anchors. For PDCs, the anchor lives on the fixed linker-drug template (e.g., the Aoa carboxyl in Aoa-Dau, the Cys carboxyl in SSCpt, or the terminal thiol in DSPA-SN38), and the model must generate a peptide exposing a compatible reactive residue (Lys-$\varepsilon$ amine or Cys thiol) at the junction.
The edit neighborhood
\begin{equation}
\begin{split}
    \mathcal{N}(z):=\bigl\{z'\in\mathcal{Z}\,:\,&d_{\mathrm{free}}(z,z')=1,\\
    &x'^{(c)}=x^{(c)}\ \forall\,c\in\mathcal{C}_{\mathrm{fix}}\bigr\},
\end{split}
\end{equation}
admits a single token (SMILES) or atom/bond (graph) substitution on a free component, including bonds incident to anchors (which are themselves immutable).
\subsection{Discrete Flows and Transport}
\label{subsec:flow_maps}

We use a \emph{discrete flow} to transport samples from the prior to the data distribution through a sequence of deterministic maps. Let
\(0 = t_0 < t_1 < \dots < t_K = T\) denote a discrete time grid. We
parameterize a family of transport maps
\begin{equation}
    \Phi_{\theta}^{t_k \to t_{k+1}} : \mathcal{Z} \to \mathcal{Z},
    \qquad k = 0,\dots,K-1,
\end{equation}
where each map advances a state from time $t_k$ to $t_{k+1}$ while
preserving the fixed components and anchor atoms. Starting from
$z_0 \sim \mu_0$, the discrete flow evolves by composition,
\begin{equation}
    z_{k+1} \;=\; \Phi_{\theta}^{t_k \to t_{k+1}}(z_k),
\end{equation}
so that the state at time $t_k$ is given by
$z_k = \Phi_{\theta}^{t_{k-1} \to t_k} \circ \cdots \circ
\Phi_{\theta}^{t_0 \to t_1}(z_0)$. The induced distribution at $t_k$ is
the pushforward of the prior,
\begin{equation}
    \mu_{t_k}^{\theta}
    \;:=\;
    \bigl(
    \Phi_{\theta}^{t_{k-1} \to t_k} \circ \cdots \circ
    \Phi_{\theta}^{t_0 \to t_1}
    \bigr)_{\#} \mu_0,
\end{equation}
and the terminal distribution $\mu_T^{\theta} = \mu_{t_K}^{\theta}$
defines the model distribution over assembled conjugates. Each map
$\Phi_{\theta}^{t_k \to t_{k+1}}$ approximates the finite-time transport
induced by an underlying continuous probability flow on
$\mathcal{Z}$, enabling few-step generation by composing learned
transports.

\subsection{Hard Constraints and Soft Objectives}
\label{subsec:objectives}
We encode validity through a binary indicator $\mathbf{1}_{\mathrm{valid}}(z)\in\{0,1\}$ that gates sampling: candidates with $\mathbf{1}_{\mathrm{valid}}(z)=0$ are never accepted. The indicator factorizes into three task-agnostic buckets, each instantiated per conjugation system (Appendix~\ref{sec:conjugation_validity_constraints}):

\textbf{Attachment site compatibility.} Conjugation-specific SMARTS checks confirm that the prescribed bond chemistry is realized at every fixed-free junction and that valency at each junction atom is consistent with the reaction (e.g., secondary/tertiary amide N for Lys-$\varepsilon$ amide coupling in Aoa-Dau and SSCpt; \texttt{SX2}-\texttt{SX2} disulfide with no residual free thiol in DSPA-SN38; the analogous warhead/E3 amide patterns in CupOFMoCA). Failure indicators (e.g., unreacted carboxylic acid or free \ch{-SH}) trigger rejection.

\textbf{Linker feasibility.} Conjugation-agnostic structural filters reject infeasible linker regions: strained or oversized ring closures (sp$^2$-bearing 3-membered rings, 4-membered rings with $>2$ sp$^2$ centers, $>12$-atom rings) and excessively long aliphatic runs ($\ge16$ consecutive \ch{CH2}). For CupOFMoCA these subsume the curated liability filter (hydrazides, azo, $N$-oxides, $N$-$O$ singles, carbamates).

\textbf{Representational validity.} Each assembled conjugate must (i) pass standard chemical sanitization, (ii) survive a SMILES parse-serialize round-trip without alteration, (iii) be free of valence and aromaticity violations, and (iv) admit a valid 2D coordinate embedding.

We guide valid proposals with the aggregate score $S(z)=\sum_j w_j S_j(z)$, where $S_j$ is objective $j$ and $w_j$ is its weight. For PDCs, $S$ is the predicted peptide-target binding score used for guidance and evaluation. For PROTACs, we combine synthetic accessibility, chemotype closeness, and connectivity using weights $(4,2,4)$ selected through ablation (Appendix~\ref{sec:weight_ablation}). We also apply a separate PROTAC-likeness classifier as a binary validity check.

\subsection{Endpoint Couplings}
\label{subsec:couplings}
We use a coupling $\gamma\in\Pi(\mu_{0},\mu_{1})$ that pairs each prior sample with a data conjugate sharing the same fixed components and anchors. Intermediate states $z_t$ are constructed by bounded admissible edits on free components from $z_0$ toward $z_1$, yielding training triples $(z_0,z_t,z_1)$.

\subsection{Generative Sampling}
\label{subsec:generation}

Following \citet{chen2025areurediannealedrectifiedupdates}, generation proceeds in two phases. An unguided phase first denoises samples from the pretrained flow (Model in Appendix \ref{subsec:pepredi_base}, \ref{sec:basemodel}), $\Phi_{\theta^*}$ until a valid starting state is reached. During guided sampling, we exponentially reweight each per-site distribution with objective scores and exclude invalid candidates with $\mathbf{1}_{\mathrm{valid}}$ (Figure~\ref{fig:wide}B). We accept PDC updates that preserve or increase $S$. For PROTACs, we accept valid tilted proposals and retain the highest-scoring valid state, following the assembly-validity comparison in Table~\ref{tab:acceptance_ablation}. Implementation details are given in Appendix \ref{sec:algorithm} for PDC and Appendix \ref{sec:sampling_appendix} for PROTAC.

\section{Peptide-Drug Conjugate Design with CupOFMoCA}
\label{sec:results}

For evaluation of CupOFMoCA on PDC design, we focus on three key questions: (1) whether coupled generation guarantees conjugate validity while avoiding functional penalties that could be associated with post-hoc staged conjugation; (2) whether discrete flows with exponential guidance enable high-quality generation; and (3) whether the model exhibits competitive design against therapeutic baselines under \textit{in silico} structural validation.

\subsection{Conjugation and Objective Guidance}

% Current standard practices for PDC design rely on staged pipelines: a peptide is optimized for a target in isolation and subsequently chemically conjugated to a linker-payload. We hypothesize that this decoupling forces trajectories through invalid intermediates and ignores steric or electrostatic clashes introduced by the payload. To test this, we compared CupOFMoCA against a strong staged baseline on two therapeutically-relevant targets: ERBB2 and GNRHR. Specifically, ERBB2 (also known as HER2) is a well-established oncogenic driver in breast and gastric cancers \citep{harari2000molecular} and a validated target for PDCs \cite{armstrong2025peptide}, while GNRHR is a clinically important endocrine receptor targeted in reproductive disorders and hormone-driven cancers \citep{maggi2016gnrh}.

% For the baseline, we generated high-affinity peptide SMILES strings using a pre-trained discrete flow matching model \citep{chen2025areurediannealedrectifiedupdates} and attached fixed linker-payload post-generation. For CupOFMoCA, during generation process, the model produces peptides that can be attached to given linker-payload SMILES under hard chemical validity constraints including attachment site compatibility, linker feasibility, and representational validity. We evaluated both methods using an independent oracle predictor for peptide-target binding affinity. And we compared our results with pharmaceutical reference PDCs with the same linker, payload, and conjugation method from PDCdb \citep{sun2025pdcdb}.

PDC design conventionally proceeds in a \emph{staged} pipeline: a peptide is first
optimized for the target in isolation, then conjugated post hoc to a fixed
linker-payload. We hypothesize that this decoupling drives sampling
trajectories through chemically infeasible intermediates and overlooks
cross-component constraints, including attachment-site compatibility, linker
feasibility, and representational validity, that can only be enforced when
peptide, linker, and drug evolve jointly. To probe this hypothesis, we
benchmarked CupOFMoCA against a broad set of baselines on three therapeutically
relevant targets, each paired with a clinically motivated linker-payload drawn
from PDCdb \citep{sun2025pdcdb}: \textbf{ERBB2 + SSCpt} (Disulfide-Camptothecin),
a validated oncology target in breast and gastric cancers
\citep{harari2000molecular, armstrong2025peptide}; \textbf{GNRHR + Aoa-Dau}
(Aminooxyacetic acid-Daunorubicin), an endocrine receptor implicated in
reproductive disorders and hormone-driven cancers \citep{maggi2016gnrh}; and
\textbf{NRP1 + DSPA-SN-38} (3-Disulfanylpropanoic Acid-SN-38), a
tumor-vasculature co-receptor \citep{niland2019neuropilins} that recognizes C-terminal C-end-rule motifs \citep{teesalu2009c}. For each target, we enforced conjugation chemistry with the corresponding builder under the hard validity constraints in Appendix~\ref{sec:conjugation_validity_constraints}, and we scored peptide-target binding with the affinity predictor used for guidance and report its output in pKd units.

We compared three classes of methods. (i) \emph{Standalone peptide generators}:
moPPIt \citep{chen2026moppit} and BoltzGen \citep{stark2025boltzgen} (AA tokenized) and PepTune \citep{tang2025peptunenovogenerationtherapeutic} (SMILES tokenized), each
producing target-binding peptides that are conjugated post hoc. (ii)
\emph{AReUReDi} \citep{chen2025areurediannealedrectifiedupdates} in both staged
(peptide-only) and guided (conjugation-aware acceptance) modes. (iii)
\emph{CupOFMoCA} in the same staged and guided modes. We evaluated AReUReDi and
CupOFMoCA under both AA/UAA tokenization with 10-residue peptides
(Table~\ref{tab:top20-aa}) and SMILES tokenization with 80- and 100-token
peptides (Tables~\ref{tab:top20-smiles80},~\ref{tab:top20-smiles100}).

For AA/UAA generation, we used the pretrained PepReDi peptide prior. PDCdb contains 393 unique canonical peptides within its pretraining range of 6-49 residues, and their zero-shot negative log likelihood was comparable to that of held-out pretraining sequences (Table~\ref{tab:pdc_domain_gap}). We therefore retained the broad peptide prior for PDC generation (Appendix~\ref{subsec:pdc_runtime_domain}).

\begin{table}[!ht]
\centering
\caption{Top-20 \textit{de novo} PDCs across three receptor targets generated by AA-based methods (peptide length: 10 AAs). Each conjugate comprises a peptide linked to a target-specific cytotoxic payload via the corresponding linker (see Section~\ref{sec:conjugation_validity_constraints}). Conjugation Validity: fraction of valid peptides yielding chemically valid conjugates. Binding affinity: predicted peptide-receptor binding score (pKd). Values are mean (std). \textbf{Bold}: best per column; \underline{underline}: second-best.}
\label{tab:top20-aa}
\footnotesize
\setlength{\tabcolsep}{4pt}
\begin{tabular}{lcccccc}
\toprule
 & \multicolumn{2}{c}{\textbf{ERBB2}} & \multicolumn{2}{c}{\textbf{GNRHR}} & \multicolumn{2}{c}{\textbf{NRP1}} \\
\cmidrule(lr){2-3} \cmidrule(lr){4-5} \cmidrule(lr){6-7}
\textbf{Method} & Conj.\ Val. & Binding & Conj.\ Val. & Binding & Conj.\ Val. & Binding \\
\midrule
moPPIt & 7.1\% & 7.59 (0.07) & 8.1\% & 6.78 (0.46) & 65.0\% & 7.94 (0.06) \\
\addlinespace[2pt]
\multicolumn{7}{l}{\textit{BoltzGen}} \\
\quad Exp & \underline{51.0\%} & 7.81 (0.05) & 16.0\% & 6.63 (1.05) & - & - \\
\quad AF  & 21.2\% & \underline{7.82 (0.11)} & 3.0\% & 7.44 (0.45) & 13.7\% & 7.05 (1.81) \\
\addlinespace[2pt]
\multicolumn{7}{l}{\textit{AReUReDi}} \\
\quad Staged (AA)  & 17.0\% & 7.65 (0.10) & \underline{17.6\%} & 6.58 (0.42) & 19.0\% & 8.23 (0.38) \\
\quad Guided (AA)  & \textbf{100.0\%} & 7.74 (0.09) & \textbf{100.0\%} & 7.15 (0.60) & 98.0\% & \underline{8.81 (0.35)} \\
\quad Staged (UAA) & 3.1\%  & 7.68 (0.08) & 5.1\% & 7.68 (0.33) & 37.4\% & 8.07 (0.37) \\
\quad Guided (UAA) & \textbf{100.0\%} & \textbf{7.83 (0.05)} & \textbf{100.0\%} & \underline{7.93 (0.13)} & \textbf{100.0\%} & 8.34 (0.34) \\
\addlinespace[2pt]
\multicolumn{7}{l}{\textit{CupOFMoCA}} \\
\quad Staged (AA)  & 15.0\% & 7.56 (0.19) & 15.0\% & 6.30 (0.35) & 25.0\% & 8.25 (0.31) \\
\quad Guided (AA)  & \textbf{100.0\%} & 7.75 (0.12) & \textbf{100.0\%} & 7.07 (0.60) & \textbf{100.0\%} & \textbf{8.93 (0.39)} \\
\quad Staged (UAA) & 5.1\%  & 7.77 (0.05) & 4.0\% & 6.98 (0.84) & 19.4\% & 8.02 (0.40) \\
\quad Guided (UAA) & \textbf{100.0\%} & \underline{7.82 (0.03)} & \textbf{100.0\%} & \textbf{8.12 (0.20)} & \textbf{100.0\%} & 8.20 (0.33) \\
\bottomrule
\end{tabular}
\end{table}

\begin{table}[!ht]
\centering
\caption{Top-20 \textit{de novo} PDCs across three receptor targets generated by SMILES-based methods (peptide length: 80 SMILES tokens). Each conjugate comprises a peptide linked to a target-specific cytotoxic payload via the corresponding linker.}
\label{tab:top20-smiles80}
\footnotesize
\setlength{\tabcolsep}{4pt}
\begin{tabular}{lcccccc}
\toprule
 & \multicolumn{2}{c}{\textbf{ERBB2}} & \multicolumn{2}{c}{\textbf{GNRHR}} & \multicolumn{2}{c}{\textbf{NRP1}} \\
\cmidrule(lr){2-3} \cmidrule(lr){4-5} \cmidrule(lr){6-7}
\textbf{Method} & Conj.\ Val. & Binding & Conj.\ Val. & Binding & Conj.\ Val. & Binding \\
\midrule
PepTune & 20.8\% & 6.75 (0.27) & 5.6\% & 4.55 & 13.7\% & 7.05 (1.81) \\
\addlinespace[2pt]
\multicolumn{7}{l}{\textit{AReUReDi}} \\
\quad Staged & \underline{30.1\%} & 7.73 (0.11) & \underline{26.8\%} & 7.69 (0.75) & 23.0\% & 8.62 (0.50) \\
\quad Guided & \textbf{100.0\%} & \textbf{8.02 (0.19)} & \textbf{100.0\%} & \textbf{8.56 (0.35)} & \textbf{100.0\%} & \underline{9.16 (0.36)} \\
\addlinespace[2pt]
\multicolumn{7}{l}{\textit{CupOFMoCA}} \\
\quad Staged & 22.7\% & 7.73 (0.14) & 18.3\% & 8.09 (1.17) & \underline{23.7\%} & 8.76 (0.31) \\
\quad Guided & \textbf{100.0\%} & \underline{7.97 (0.09)} & \textbf{100.0\%} & \underline{8.39 (0.36)} & \textbf{100.0\%} & \textbf{9.36 (0.46)} \\
\bottomrule
\end{tabular}
\end{table}

\paragraph{Coupled generation increased conjugation-valid yield.}
Under identical endpoint checks, staged generation achieved $3.0$-$65.0\%$ conjugation validity for AA/UAA methods and $5.6$-$30.1\%$ for SMILES-80 methods. Guided AReUReDi and CupOFMoCA reached $98$-$100\%$ across these settings, increasing the fraction of valid peptides that formed valid conjugates (Tables~\ref{tab:top20-aa} and~\ref{tab:top20-smiles80}; Figure~\ref{fig:validity-binding}). In the SMILES-80 runs, incompatible attachment chemistry accounted for most staged-generation assembly failures (Figure~\ref{fig:validity-stack}). To assess computational efficiency, we also compared valid-conjugate yield under matched compute. CupOFMoCA-Guided achieved the highest yield per GPU-hour in 8 of 12 target-representation settings (Table~\ref{tab:pdc_runtime_wins}).

\paragraph{Coupled generation preserved predicted binding.}
To assess whether the increased valid yield preserved the binding objective, we evaluated all valid conjugates. Across the six SMILES settings, CupOFMoCA increased conjugation validity by $2.7$-$5.5$-fold over its staged counterpart, with full-set mean binding scores within $0.54$ pKd units of the corresponding staged means (Tables~\ref{tab:full-smiles80} and~\ref{tab:full-smiles100}). The top-20 comparisons and full AA/UAA results further supported retention of predicted binding under conjugation constraints (Tables~\ref{tab:top20-aa},~\ref{tab:top20-smiles80}, and~\ref{tab:full-aa}). To distinguish the contributions of gating and guidance, we compared each component separately. Validity gating raised conjugation validity to $100\%$, and affinity guidance increased mean predicted binding by approximately $0.3$-$1.1$ pKd units within the feasible region (Table~\ref{tab:pdc_gate_affinity_ablation}).

\paragraph{Tokenization regimes.} To examine how representation affected design, we tested AA, UAA, and SMILES tokenizations. AA covers the 20 canonical residues, UAA adds non-canonical monomers, and SMILES operates at the atom level. Predicted binding increased with expressivity for both AReUReDi and CupOFMoCA, with the highest pKd in the SMILES setting on every target. CupOFMoCA-Guided reached $100\%$ conjugation validity across these regimes and matched AReUReDi-Guided overall, with higher predicted binding on NRP1 in the AA and SMILES settings (Tables~\ref{tab:top20-aa},~\ref{tab:top20-smiles80}). We observed the same pattern with 100-token SMILES (Table~\ref{tab:top20-smiles100}).

\paragraph{Molecular properties and linker diversity.} Generated PDCs spanned a physicochemical regime similar to PDCdb pharmaceutical references on molecular weight, LogP, TPSA, and heavy-atom count, with target-specific distributions reported in Appendix~\ref{sec:results_appendix} (Figure~\ref{fig:property-distributions}; per-target breakdowns in Tables~\ref{tab:full-aa-erbb2}-\ref{tab:full-smiles100-nrp1}). To assess flexibility beyond fixed linkers, we sampled linker types from PDCdb frequencies. The resulting PDCs matched the binding and validity of the fixed-linker setting (Table~\ref{tab:variable-vs-fixed-linker}), with the expected linker-type distribution (Table~\ref{tab:linker-diversity}). Across all settings, we obtained high uniqueness and diversity with low training-set similarity (Table~\ref{tab:generation-diversity}).

\section{PROTAC Design with CupOFMoCA}

\subsection{Constraints and Objective Guidance}
\label{subsec:objectives_protacs}

PROTAC assembly imposes hard validity constraints coupling the linker with the fixed warhead and E3 ligand, encoded via a binary classifier $\mathbf{1}_{\mathrm{valid}}(z) \in \{0,1\}$ that returns $1$ iff $z$ satisfies all of:
\begin{enumerate}
    \item[(a)] \textbf{Representational validity.} The molecule passes RDKit sanitization and the SMILES round-trip $\mathrm{MolFromSmiles}(\mathrm{MolToSmiles}(z))$ recovers a valid molecule; states with malformed valences or aromaticity are rejected.
    \item[(b)] \textbf{Liability filter.} No SMARTS match against hydrazides (\texttt{[NX3][NX3]}), azo groups (\texttt{[\#7]=[\#7]}), $N$-oxides (\texttt{[N+][O-]}), $N$-$O$ single bonds (\texttt{[\#7]-[OX2]}), or carbamates (\texttt{[NX3]C(=[OX1])[OX2H1]}), which decompose, exhibit toxicity, or fail amide-coupled assembly.
    \item[(c)] \textbf{Reactive termini count.} At least two carbon-bonded carboxylic acids (\texttt{[\#6][CX3](=[OX1])[OX2H1]}) or aldehydes (\texttt{[\#6][CX3H1](=[OX1])}); fewer renders warhead/E3 amide-coupling impossible.
\end{enumerate}
Candidates with $\mathbf{1}_{\mathrm{valid}}(z) = 0$ receive zero probability under the sampling distribution.

For more details on the validity constraints, see Appendix \ref{sec:validity_appendix}. Beyond the hard gate, we use a continuous aggregate objective
\begin{equation}
    S(z) \;=\; \sum_{j=1}^{3} w_{j}\, S_{j}(z) \;\in\; \mathbb{R},
\end{equation}
combining three soft scorers with default weights $(w_1, w_2, w_3) = (4.0,\, 2.0,\, 4.0)$, selected through the weight sweeps in Appendix~\ref{sec:weight_ablation}; all PROTAC results in Section~\ref{subsec:results_protac} use them:

$S_1$ (\textbf{Synthetic accessibility}): a linear ramp on RDKit's synthetic accessibility (SA) score, $1.0$ at $\mathrm{SA}=1$ and $0.0$ at $\mathrm{SA}=6$, with $\mathrm{SA}>6$ mapped to zero.

$S_2$ (\textbf{Chemotype closeness}): a trapezoidal windowed score from PROTAC-DB statistics averaging ether count, $\log P$, linker MW, and rotatable-bond count, each mapped into $[0,1]$ via a window centered on the empirical PROTAC-DB range (assembled mass $[771, 1108]$~Da minus fixed warhead/E3 offsets).

$S_3$ (\textbf{Connectivity}): returns $1.0$ for a single connected component, otherwise $\,\frac{|\text{largest fragment}|}{|\text{total atoms}| \cdot |\text{fragments}|}\,$, halved if no path exists through $x^{\mathrm{link}}$ between $a^{\mathrm{wh}}$ and $a^{\mathrm{e3}}$. We evaluate $S_3$ on partially assembled states with $\mathbf{1}_{\mathrm{valid}}(z)=0$. The elevated $w_3$ gives connected, anchor-bridging linkers greater weight early in transport.

With this aggregate score, we favor connected linkers with lower SA scores and PROTAC-DB-like properties. We evaluated the objective weights and the joint contribution of hard constraints and soft guidance in Appendix~\ref{sec:weight_ablation} and Table~\ref{tab:soft_hard_ablation}.

\FloatBarrier
\subsection{PROTAC Linker Generation}
\label{subsec:results_protac}

We benchmarked CupOFMoCA against LinkerNet \citep{guan2023linkernet}, DiffPROTACs \citep{li2024diffprotacs}, and DiffLinker \citep{igashov2024difflinker} on three PROTAC systems (EGFR-VHL,
BRD9-VHL, CDK6-CRBN). For CupOFMoCA, we generated $100$ candidates per
system with each of three random seeds; baseline results use one batch of
$100$ candidates per system. We evaluated every candidate with the same
attachment, valence, liability, and assembly checks, without method-specific
repair (Appendix~\ref{sec:validity_appendix}). We report standard and assembly validity with post-hoc predicted degradation (Table~\ref{tab:validity_degradmaster}), and diversity and training-set similarity (Table~\ref{tab:diversity_memorization}).

\begin{table*}[!ht]
\centering
% CAMERA-READY DATA CHECK: the OpenReview rebuttal's multi-seed CI table lists
% assembled n=283/281/265, whereas the ICLR revision and the reported
% 85.0/82.3/83.0% assembly-valid rates imply n=255/247/249 out of 300. This
% transfer follows the internally consistent ICLR counts; confirm before final.
\caption{Assembly validity and predicted degradation across PROTAC systems. Standard validity is the fraction of RDKit-parseable molecules; assembly validity additionally requires canonical attachment and an analyzer pass. With the selected objective weights (Appendix~\ref{sec:weight_ablation}), we generated $100$ candidates for each of three seeds ($42$, $123$, $456$), obtaining $255$, $247$, and $249$ assembly-valid molecules for EGFR-VHL, BRD9-VHL, and CDK6-CRBN, respectively. Validity is mean $\pm$ std across seeds; each baseline used one batch of $100$. DegradeMaster scores are mean (std) over assembly-valid molecules, and threshold columns report the fraction scoring $\ge 0.5$. \textbf{Bold}: best per column; \underline{underline}: second-best. Dashes denote no valid assemblies.}
\label{tab:validity_degradmaster}
\footnotesize
\setlength{\tabcolsep}{4pt}
\renewcommand{\arraystretch}{0.95}
{
\begin{tabular}{lcccccc}
\toprule
 & \multicolumn{2}{c}{\textbf{EGFR-VHL}} & \multicolumn{2}{c}{\textbf{BRD9-VHL}} & \multicolumn{2}{c}{\textbf{CDK6-CRBN}} \\
\cmidrule(lr){2-3} \cmidrule(lr){4-5} \cmidrule(lr){6-7}
\textbf{Method} & Std.\ Val. & Asm.\ Val. & Std.\ Val. & Asm.\ Val. & Std.\ Val. & Asm.\ Val. \\
\midrule
LinkerNet     & \textbf{98.0}  & 46.0           & \textbf{100.0} & \underline{16.0} & \textbf{100.0} & \textbf{96.0} \\
DiffPROTACs   & \textbf{98.0}  & \underline{80.0} & \textbf{100.0} & 0.0            & 86.0           & 74.0 \\
DiffLinker    & 86.0  & 0.0            & 74.0           & 0.0            & 92.0           & 0.0 \\
\textbf{Ours} & \underline{95.3 $\pm$ 2.1} & \textbf{85.0 $\pm$ 3.6} & \underline{94.3 $\pm$ 2.1} & \textbf{82.3 $\pm$ 4.0} & \underline{94.7 $\pm$ 1.5} & \underline{83.0 $\pm$ 4.2} \\
\midrule[\heavyrulewidth]
 & \multicolumn{6}{c}{\textit{DegradeMaster predicted degradation probability}} \\
\midrule
 & \multicolumn{2}{c}{\textbf{EGFR-VHL}} & \multicolumn{2}{c}{\textbf{BRD9-VHL}} & \multicolumn{2}{c}{\textbf{CDK6-CRBN}} \\
\cmidrule(lr){2-3} \cmidrule(lr){4-5} \cmidrule(lr){6-7}
\textbf{Method} & mean & $\ge 0.5$ & mean & $\ge 0.5$ & mean & $\ge 0.5$ \\
\midrule
LinkerNet     & 0.278 (0.051) & 0.0\%            & \underline{0.172 (0.053)} & 0.0\%           & \textbf{0.925 (0.055)} & \textbf{100.0\%} \\
DiffPROTACs   & \underline{0.306 (0.236)} & \underline{20.0\%} & -                 & -             & \underline{0.882 (0.185)} & \underline{96.7\%} \\
\textbf{Ours} & \textbf{0.509 (0.275)}    & \textbf{52.3\%}    & \textbf{0.260 (0.235)} & \textbf{18.9\%} & 0.770 (0.282)             & 80.4\% \\

\bottomrule
\end{tabular}
}
\end{table*}

\paragraph{Standard and assembly validity.}
Table~\ref{tab:validity_degradmaster} distinguishes \emph{standard validity}
(RDKit-parseable SMILES) from \emph{assembly validity} (canonical attachment
and analyzer pass). Under the common endpoint analyzer, DiffLinker produced
parseable SMILES at $74$-$92\%$ but no valid assemblies. LinkerNet reached
$46.0\%$, $16.0\%$, and $96.0\%$ assembly validity on EGFR-VHL, BRD9-VHL,
and CDK6-CRBN, while DiffPROTACs reached $80.0\%$, $0.0\%$, and $74.0\%$.
With fixed anchors (Section~\ref{subsec:anchors}) and validity-gated sampling,
CupOFMoCA reached $85.0 \pm 3.6\%$, $82.3 \pm 4.0\%$, and
$83.0 \pm 4.2\%$, ranking first on EGFR-VHL and BRD9-VHL and second on
CDK6-CRBN. Under matched hardware, CupOFMoCA generated $0.247$
assembly-valid PROTACs per second, compared with $0.181$ for DiffPROTACs,
$0.072$ for LinkerNet, and $0.000$ for DiffLinker
(Appendix~\ref{app:compute_PROTACs}).

\paragraph{Predicted degradation and physical evaluation.}
We evaluated assembly-valid PROTACs with DegradeMaster
\citep{liu2025degrademaster}, which we applied after generation for in-domain activity prediction. DegradeMaster and the fine-tuned generator both draw on PROTAC databases. CupOFMoCA reached mean predicted probabilities
of $0.509$, $0.260$, and $0.770$ on EGFR-VHL, BRD9-VHL, and CDK6-CRBN;
$52.3\%$, $18.9\%$, and $80.4\%$ of its assembly-valid generations scored $\ge 0.5$ (Table~\ref{tab:validity_degradmaster}). CupOFMoCA had the highest
predicted probability on EGFR-VHL and BRD9-VHL, whereas LinkerNet performed
best on CDK6-CRBN.

To complement the degradation predictions with a physical assessment, we evaluated four generated EGFR-VHL candidates by explicit-solvent molecular dynamics using MS39 as a cell-active positive control and a
PEG4-linked congener as a negative control. The generated candidates occupied
a conformational regime comparable to MS39, and C1 and C2 had lower 3D polar
surface area and lower ternary-compatible conformational penalties. We detail the simulation protocol and compound definitions in Appendix~\ref{subsec:md} and Table~\ref{tab:md}. The comparison with both controls supported favorable linker-dependent conformational properties.

\begin{table*}[h!]
\centering
\caption{Diversity and training-set similarity across PROTAC systems.
Div: mean pairwise Tanimoto within the generated batch (lower = more diverse).
Sim: mean max-Tanimoto to PROTAC-DB (lower = less similar to training data).
``Sim $>\!0.8$'' is the fraction exceeding the similarity threshold.
Our rows pool three seeds ($255$ / $247$ / $249$ assembly-valid molecules);
baselines are single batches of $N=100$.
\textbf{Bold}: best per column; \underline{underline}: second-best. Dashes
indicate methods that produce $0\%$ valid assemblies.}
\label{tab:diversity_memorization}
\footnotesize
\setlength{\tabcolsep}{4pt}
\begin{tabular}{l ccc ccc ccc}
\toprule
 & \multicolumn{3}{c}{\textbf{EGFR-VHL}} & \multicolumn{3}{c}{\textbf{BRD9-VHL}} & \multicolumn{3}{c}{\textbf{CDK6-CRBN}} \\
\cmidrule(lr){2-4} \cmidrule(lr){5-7} \cmidrule(lr){8-10}
\textbf{Method} & Div & Sim & $>\!0.8$ & Div & Sim & $>\!0.8$ & Div & Sim & $>\!0.8$ \\
\midrule
LinkerNet     & 0.895 & 0.816 & 78.3\%           & 0.829 & 0.825 & 75.0\%             & 0.833 & 0.776 & 45.8\% \\
DiffPROTACs   & \textbf{0.763} & \underline{0.779} & \underline{35.0\%} & - & - & -           & \underline{0.735} & \underline{0.694} & \underline{2.7\%} \\
\textbf{Ours} & \underline{0.772} & \textbf{0.630} & \textbf{0.0\%}     & \textbf{0.746} & \textbf{0.777} & \textbf{26.0\%} & \textbf{0.716} & \textbf{0.545} & \textbf{0.0\%} \\
\bottomrule
\end{tabular}
\end{table*}

\paragraph{Diversity and training-set similarity.}
Table~\ref{tab:diversity_memorization} reports mean pairwise Tanimoto
similarity within each generated batch and maximum Tanimoto similarity to
PROTAC-DB. CupOFMoCA had the lowest mean maximum training-set similarity on
all three systems. None of its EGFR-VHL or CDK6-CRBN generations exceeded
the $0.80$ similarity threshold, while $26.0\%$ of its BRD9-VHL generations
did. PROTAC-DB contains $37$ reference compounds for BRD9-VHL, $200$ for
EGFR-VHL, and $117$ for CDK6-CRBN. CupOFMoCA had the highest diversity on BRD9-VHL and CDK6-CRBN and the second-highest diversity on EGFR-VHL. On CDK6-CRBN, LinkerNet combined a mean DegradeMaster score of $0.925$ with $45.8\%$ of generations above the
similarity threshold; CupOFMoCA reached $0.770$, with no generations above
the threshold. The activity predictions and similarity analysis therefore characterized complementary properties of the generated candidates.

\paragraph{Takeaways.}
Across the three evaluated systems, fixed anchors (Section~\ref{subsec:anchors}), hard validity
constraints, and soft guidance (Section~\ref{subsec:objectives_protacs}) supported high assembly-valid
yield under a common endpoint analyzer, with the strongest performance from their combination (Table~\ref{tab:soft_hard_ablation}). PROTAC-specific fine-tuning of the graph flow on PROTAC-DB linkers
raised assembly validity by $4.7$-$6.3$ percentage points and lowered mean SA
from approximately $3.6$ to $2.8$ (Table~\ref{tab:defog_finetune}), and CupOFMoCA achieved the
highest assembly validity on two systems and the second-highest value on one.
The generated PROTACs contained diverse linkers between the fixed warhead and E3 ligand (Figure~\ref{fig:top3-protacs}), with target-specific molecular properties summarized in Table~\ref{tab:molecularprop}.

\FloatBarrier
\section{Discussion}
In this work, we present \textbf{CupOFMoCA}, a method for objective-guided discrete generation under coupled chemical constraints. During guided sampling, we enforce chemical compatibility between components so that intermediate states correspond to valid partial conjugates. Applied to peptide-drug conjugates (PDCs), this formulation produced fully assembled molecules, increased conjugation-valid yield relative to staged baselines, and maintained predicted binding scores near the corresponding staged levels. Applied to PROTACs, the same fixed-anchor and
hard-validity machinery reduced the gap between standard and assembly validity, achieving the highest assembly validity on two of the three evaluated systems and the second-highest value on the third.
Beyond the sampling objectives, post-hoc in-domain DegradeMaster scoring and explicit-solvent molecular dynamics provided complementary computational evaluations of completed candidates. CupOFMoCA also produced linkers with the lowest mean maximum similarity to training compounds on all three systems.

While we used molecular dynamics to characterize linker conformations and ternary-geometry compatibility, we have yet to establish the biological activity of our generated candidates experimentally. To connect these computational properties to cellular function, we are currently performing full ternary-complex simulations and wet lab experiments to measure stability, target engagement, and degradation. Beyond these evaluations, we will next extend CupOFMoCA to antibody-drug conjugates and multicomponent degraders, incorporate pharmacokinetic and linker-property objectives, and enable adaptive selection of linker-payload combinations. With these extensions, we envision that CupOFMoCA will enable the next-generation of molecular conjugates, optimized for coordinated control over component compatibility and therapeutic function.

\section*{Author Contributions}
R.R.Z. formulated and trained CupOFMoCA and conducted all PDC experiments. Z.L. formulated and conducted all PROTAC experiments. J.G. supervised the research and reviewed the manuscript. P.C. conceived, formulated, designed, directed, and supervised the study.

\section*{Funding Statement.} The work was funded by a grant to the lab of P.C. from the Futures Fund at the University of Pennsylvania.

\section*{Competing Interests.} P.C. is a co-founder of Gameto, Inc., UbiquiTx, Inc., AtomBioworks, Inc., and Kosha Sciences, Inc. and advises companies involved in therapeutics development. P.C.’s interests are reviewed and managed by the University of Pennsylvania in accordance with their conflict-of-interest policies. The remaining authors have no conflicts of interest to declare. 

\bibliographystyle{plainnat} 
\bibliography{references}

%%%%%%%%%%%%%%%%%%%%%%%%%%%%%%%%%%%%%%%%%%%%%%%%%%%%%%%%%%%%

\newpage
\appendix

% - Appendix Title & Setup -
\begin{center}
    {\LARGE \bfseries Appendix}
\end{center}
\vspace{1em}

% Reset figure naming to A.1, A.2, etc.
\counterwithin{figure}{section}
\renewcommand{\thefigure}{\thesection\arabic{figure}}
\setcounter{figure}{0}

% Reset table numbering to A1, A2, etc.
\counterwithin{table}{section}
\renewcommand{\thetable}{\thesection\arabic{table}}
\setcounter{table}{0}

\section{PDC Additional Results}
\label{sec:results_appendix}

\subsection{Validity-Gate and Affinity-Guidance Ablation}
\label{subsec:pdc_gate_affinity_ablation}

We isolated the roles of the learned flow, conjugation-validity gate, and
affinity guidance by independently enabling the latter two components. The
\emph{unconditional} condition uses neither; \emph{staged} uses affinity
guidance without conjugation gating and applies the common endpoint analyzer
post hoc; \emph{gate-only} enforces conjugation validity without affinity
guidance; and \emph{guided} uses both. All four conditions share the same
SMILES flow checkpoint and are evaluated on their full valid sets.

\begin{table}[!ht]
\centering
\caption{PDC validity-gate and affinity-guidance ablation. Each cell reports
conjugation validity (\%) / mean predicted pKd over the full valid set. The pKd
predictor also supplies the affinity-guidance score and therefore evaluates the
optimized objective.}
\label{tab:pdc_gate_affinity_ablation}
\small
\setlength{\tabcolsep}{4pt}
\begin{tabular}{llccc}
\toprule
\textbf{Length} & \textbf{Configuration} & \textbf{ERBB2} & \textbf{GNRHR} & \textbf{NRP1} \\
\midrule
\multirow{4}{*}{80}
 & Unconditional & 22.8 / 6.12 & 22.8 / 4.48 & 6.1 / 5.20 \\
 & Staged       & 22.7 / 7.70 & 18.3 / 8.09 & 23.7 / 8.68 \\
 & Gate-only    & 100.0 / 7.33 & 100.0 / 6.59 & 100.0 / 7.88 \\
 & Guided       & \textbf{100.0 / 7.69} & \textbf{100.0 / 7.55} & \textbf{100.0 / 8.58} \\
\midrule
\multirow{4}{*}{100}
 & Unconditional & 24.5 / 6.22 & 24.5 / 4.93 & 5.0 / 5.47 \\
 & Staged       & 36.7 / 7.92 & 34.7 / 8.15 & 28.1 / 8.65 \\
 & Gate-only    & 100.0 / 7.52 & 100.0 / 7.01 & 100.0 / 8.01 \\
 & Guided       & \textbf{100.0 / 7.86} & \textbf{100.0 / 8.07} & \textbf{100.0 / 8.67} \\
\bottomrule
\end{tabular}
\end{table}

Across both token budgets and all targets, gating increased conjugation
validity from $5.0$-$36.7\%$ to $100\%$. Comparing gate-only with guided
generation showed that affinity guidance contributed approximately
$0.3$-$1.1$ predicted pKd after restricting sampling to the feasible region.
The staged condition attained similar endpoint pKd among the small fraction of
samples that assembled. Guided generation increased conjugation validity by $2.7$-$5.5$-fold relative to staged generation, with mean predicted binding scores within $0.54$ pKd units. These results separate the
contributions: the hard gate removed assembly failures, and affinity guidance
recovered high predicted binding within the valid manifold.

\subsection{PDC Prior Domain Gap and Runtime-Normalized Yield}
\label{subsec:pdc_runtime_domain}

\paragraph{Choice of peptide prior.}
We retained the broad pretrained PepReDi prior for PDCs. PDCdb contains only
393 unique canonical peptides within the 6-49 residue pretraining range and
195 unique peptide SMILES, making PDC-specific fine-tuning susceptible to
memorization. Only 5 of the 393 canonical PDCdb peptides occur in the
pretraining corpus. We therefore measured zero-shot negative log likelihood
(NLL) under the base PepReDi model and used 5,000 sequence-level bootstrap
samples for confidence intervals (Table~\ref{tab:pdc_domain_gap}).

\begin{table}[!ht]
\centering
\caption{Zero-shot NLL under base PepReDi. SEM is computed across sequences;
confidence intervals for the NLL difference use 5,000 sequence-level bootstrap
samples.}
\label{tab:pdc_domain_gap}
\small
\begin{tabular}{lrrrc}
\toprule
\textbf{Group} & $n$ & \textbf{Mean NLL (SEM)} & $\Delta$\textbf{NLL} & \textbf{95\% CI} \\
\midrule
PDCdb, all & 393 & 1.5052 (0.0090) & $-0.0172$ & [$-0.0360$, 0.0017] \\
PDCdb, CPP-like & 149 & 1.4630 (0.0138) & $-0.0594$ & [$-0.0874$, $-0.0313$] \\
PDCdb, non-CPP & 244 & 1.5349 (0.0118) & 0.0124 & [$-0.0115$, 0.0364] \\
Pretraining validation & 1,414 & 1.5224 (0.0039) & \multicolumn{2}{c}{reference} \\
\bottomrule
\end{tabular}
\end{table}

Neither the complete PDCdb set nor its non-CPP subset differed significantly
from held-out pretraining data. The lower NLL of the CPP-like subset may reflect
its lower sequence complexity and did not indicate a PDC-wide domain shift.
These measurements supported retaining the broader peptide prior without
fine-tuning on the small, heterogeneous PDCdb set.

\paragraph{Matched sampling comparison.}
For PDCs, AReUReDi and CupOFMoCA use identical base-flow checkpoints, so their
comparison isolates sampling. AReUReDi uses annealed guidance, Tchebycheff
scalarization, locally balanced proposals, and Metropolis-Hastings acceptance;
CupOFMoCA directly reweights each local proposal with the functional score and
validity mask. Across the 12 target-representation blocks formed by SMILES-80,
SMILES-100, AA, and UAA generation on ERBB2, GNRHR, and NRP1,
CupOFMoCA-Guided achieved the best runtime-normalized valid yield in 8 blocks;
AReUReDi-Guided achieved the best yield in none
(Table~\ref{tab:pdc_runtime_wins}). CupOFMoCA reduced runtime by 30-60\% in
multiple settings and preserved endpoint affinity.

For the runtime-normalized comparison, each representation was evaluated
using the same implementation employed to generate its reported samples.
In particular, AA- and UAA-tokenized CupOFMoCA-Guided runs used the
$\kappa=8$ cascaded validity evaluation described in Appendix~G.2.2,
whereas SMILES-tokenized runs evaluated the full Top-$K$ candidate set.

\begin{table}[!ht]
\centering
\caption{Number of wins across 12 target-representation blocks. Valid
yield/GPU-hour counts chemically valid conjugates produced per unit compute.}
\label{tab:pdc_runtime_wins}
\small
\begin{tabular}{lcc}
\toprule
\textbf{Method} & \textbf{Best mean affinity} & \textbf{Best valid yield/GPU-hour} \\
\midrule
AReUReDi-Staged & 3 & 1 \\
CupOFMoCA-Staged & \textbf{8} & 3 \\
AReUReDi-Guided & 1 & 0 \\
CupOFMoCA-Guided & 0 & \textbf{8} \\
\bottomrule
\end{tabular}
\end{table}

\subsection{Conjugation Validity and Binding Affinity}

\begin{figure}[!ht]
  \centering
  \includegraphics[width=\linewidth]{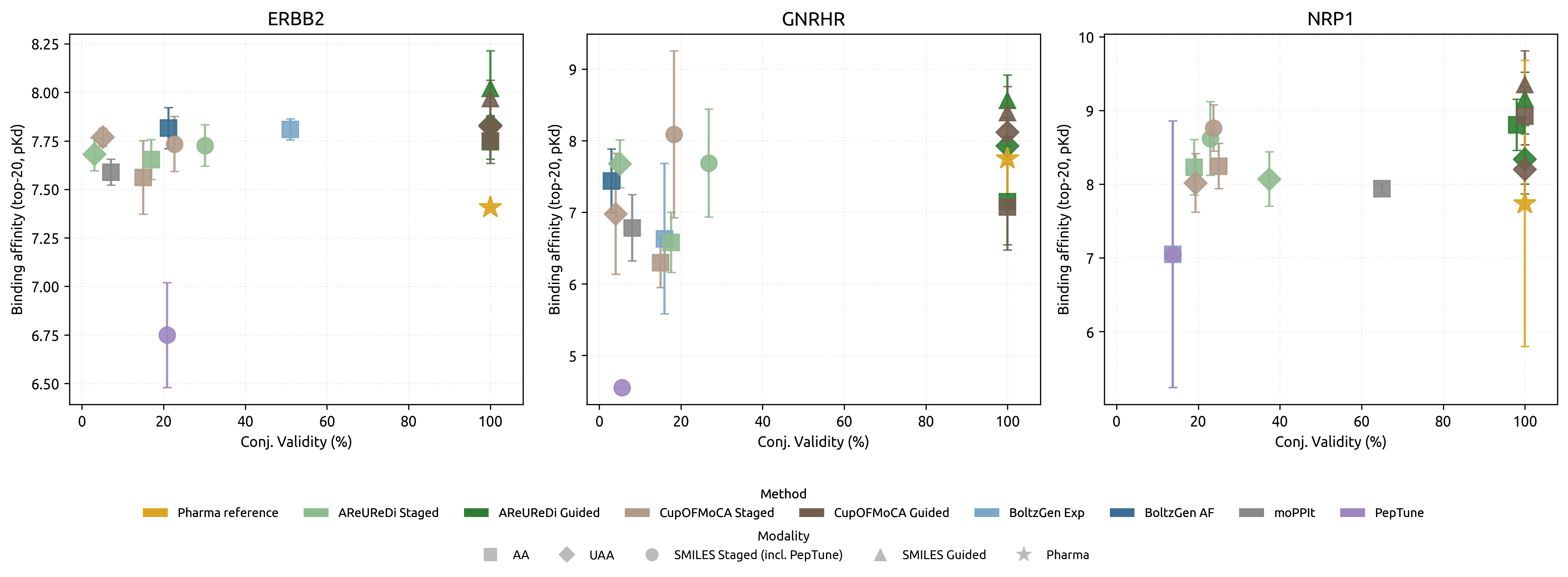}
  \caption{\textbf{Joint validity-binding tradeoff for de novo PDC generation across three receptor targets.}
  Each point shows one method's conjugate validity (x-axis, fraction of valid peptides that yield a valid conjugate) against the mean predicted binding affinity of its top-20 PDCs (y-axis, pKd, error bars: $\pm 1$ std).
  Method family is encoded by colour (CupOFMoCA Staged: tan; CupOFMoCA Guided: dark taupe; AReUReDi Staged/Guided: light/dark green; baselines: blue/grey/lavender; pharmaceutical reference: gold).
  Marker shape encodes the input/output representation (square: amino-acid sequence model; diamond: unnatural-AA model; circle: SMILES staged generation, including PepTune; triangle: SMILES guided/coupled generation; star: pharma reference).
  Guided methods achieved high conjugation validity across all three targets, with representation-dependent predicted binding scores. SMILES-guided samples are triangles; guided AA and UAA samples are squares and diamonds, respectively.
  Pharma validity is set to 100\% by definition (well-formed reference compounds).}
  \label{fig:validity-binding}
\end{figure}

\begin{figure}[!ht]
  \centering
  \includegraphics[width=\linewidth]{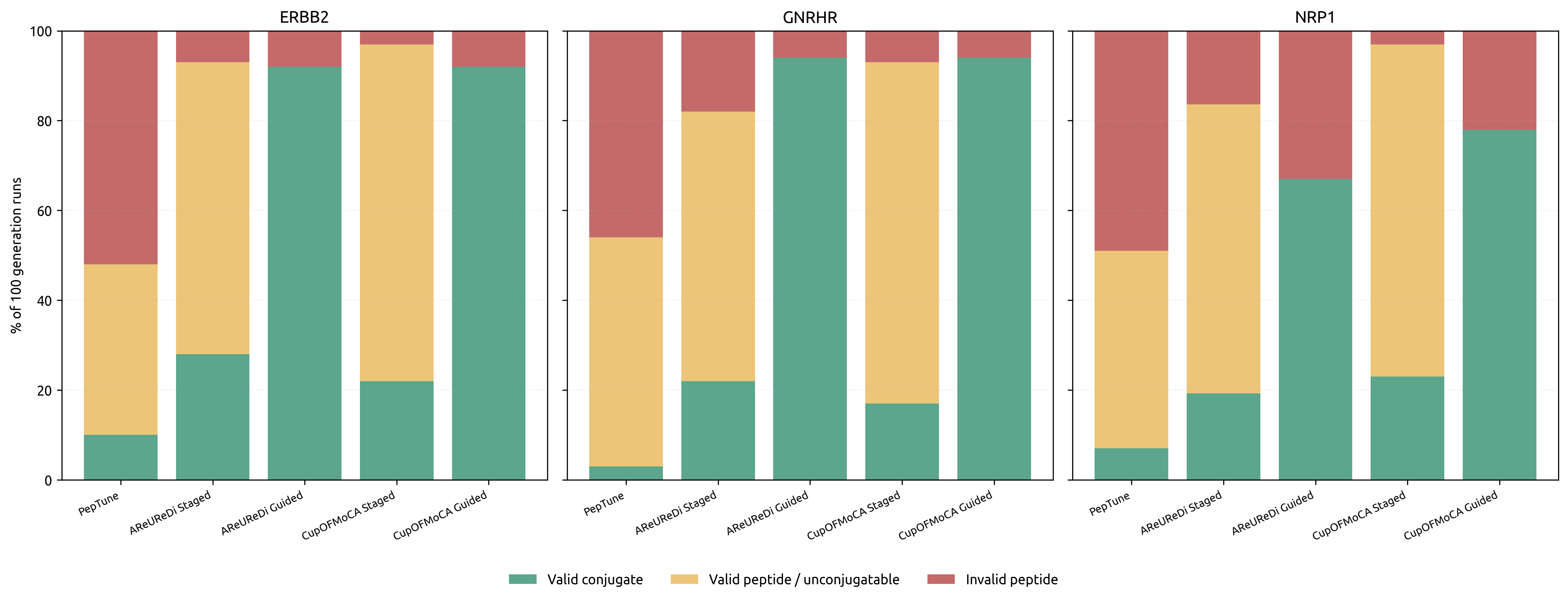}
  \caption{\textbf{Decomposition of 100 generation runs by outcome class for SMILES-based methods at peptide length 80.}
  For each method $\times$ target, the bar splits the 100 runs into three categories: invalid peptide (red, generation failed at the SMILES level), valid peptide that fails post-hoc conjugation (orange, lacks the chemical features required for linker attachment), and valid PDC (green, full success).
  Coupled methods (AReUReDi Guided, CupOFMoCA Guided) eliminate the orange band almost entirely on ERBB2 and GNRHR by generating conjugation-compatible peptides directly, recovering 92-94\% valid PDCs vs.\ 17-28\% for staged baselines and 3-10\% for PepTune.
  On NRP1, coupled methods yielded 67-78\% valid PDCs across all generation attempts, compared with 7-23\% for staged methods.}
  \label{fig:validity-stack}
\end{figure}

\begin{table}[!ht]
\centering
\caption{Top-20 \textit{de novo} PDCs across three receptor targets generated by SMILES-based methods (peptide length: 100 SMILES tokens). Each conjugate comprises a peptide linked to a target-specific cytotoxic payload via the corresponding linker. Conjugation Validity: fraction of valid peptides yielding chemically valid conjugates. Binding affinity: predicted peptide-receptor binding score (pKd). Values are mean (std). \textbf{Bold}: best per column; \underline{underline}: second-best.}
\label{tab:top20-smiles100}
\footnotesize
\setlength{\tabcolsep}{4pt}
\begin{tabular}{lcccccc}
\toprule
 & \multicolumn{2}{c}{\textbf{ERBB2}} & \multicolumn{2}{c}{\textbf{GNRHR}} & \multicolumn{2}{c}{\textbf{NRP1}} \\
\cmidrule(lr){2-3} \cmidrule(lr){4-5} \cmidrule(lr){6-7}
\textbf{Method} & Conj.\ Val. & Binding & Conj.\ Val. & Binding & Conj.\ Val. & Binding \\
\midrule
PepTune & 17.9\% & 6.82 (0.50) & 17.2\% & 5.85 (1.25) & 11.3\% & 7.25(1.11) \\
\addlinespace[2pt]
\multicolumn{7}{l}{\textit{AReUReDi}} \\
\quad Staged & \underline{40.9\%} & 7.94 (0.16) & \underline{39.1\%} & \textbf{9.15 (0.35)} & 18.8\% & 8.75 (0.64) \\
\quad Guided & \textbf{100.0\%} & 7.97 (0.20) & \textbf{100.0\%} & 8.86 (0.32) & \textbf{100.0\%} & \textbf{9.23 (0.24)} \\
\addlinespace[2pt]
\multicolumn{7}{l}{\textit{CupOFMoCA}} \\
\quad Staged & 36.7\% & \underline{8.09 (0.17)} & 34.7\% & 8.66 (0.32) & \underline{28.1\%} & 8.85 (0.38) \\
\quad Guided & \textbf{100.0\%} & \textbf{8.24 (0.14)} & \textbf{100.0\%} & \underline{9.01 (0.36)} & \textbf{100.0\%} & \underline{9.11 (0.19)} \\
\bottomrule
\end{tabular}
\end{table}

\begin{table}[!ht]
\centering
\caption{Full-valid-set \textit{de novo} PDCs across three receptor targets generated by AA-based methods (peptide length: 10 AAs). Metrics are computed over all valid conjugates from each method (not filtered to top-20). Each conjugate comprises a peptide linked to a target-specific cytotoxic payload via the corresponding linker. Conjugation Validity: fraction of valid peptides yielding chemically valid conjugates. Binding affinity: predicted peptide-receptor binding score (pKd). Values are mean (std). \textbf{Bold}: best per column; \underline{underline}: second-best.}
\label{tab:full-aa}
\footnotesize
\setlength{\tabcolsep}{4pt}
\begin{tabular}{lcccccc}
\toprule
 & \multicolumn{2}{c}{\textbf{ERBB2}} & \multicolumn{2}{c}{\textbf{GNRHR}} & \multicolumn{2}{c}{\textbf{NRP1}} \\
\cmidrule(lr){2-3} \cmidrule(lr){4-5} \cmidrule(lr){6-7}
\textbf{Method} & Conj.\ Val. & Binding & Conj.\ Val. & Binding & Conj.\ Val. & Binding \\
\midrule
moPPIt & 7.1\% & 7.59 (0.07) & 8.1\% & 6.78 (0.46) & 65.0\% & 7.73 (0.34) \\
\addlinespace[2pt]
\multicolumn{7}{l}{\textit{BoltzGen}} \\
\quad Exp & \underline{51.0\%} & 7.64 (0.22) & 16.0\% & 6.63 (1.05) & - &  - \\
\quad AF  & 21.2\% & \textbf{7.80 (0.13)} & 3.0\% & 7.44 (0.45) & 13.7\% &  7.05 (1.81) \\
\addlinespace[2pt]
\multicolumn{7}{l}{\textit{AReUReDi}} \\
\quad Staged (AA)  & 17.0\% & 7.65 (0.10) & \underline{17.6\%} & 6.58 (0.42) & 19.0\% & 8.23 (0.38) \\
\quad Guided (AA)  & \textbf{100.0\%} & 7.57 (0.16) & \textbf{100.0\%} & 6.48 (0.49) & 98.0\% & \underline{8.24 (0.37)} \\
\quad Staged (UAA) & 3.1\%  & 7.68 (0.08) & 5.1\% & \textbf{7.68 (0.33)} & 37.4\% & 7.92 (0.32) \\
\quad Guided (UAA) & \textbf{100.0\%} & 7.72 (0.08) & \textbf{100.0\%} & 7.39 (0.47) & \textbf{100.0\%} & 7.94 (0.27) \\
\addlinespace[2pt]
\multicolumn{7}{l}{\textit{CupOFMoCA}} \\
\quad Staged (AA)  & 15.0\% & 7.56 (0.19) & 15.0\% & 6.30 (0.35) & 25.0\% & 8.16 (0.32) \\
\quad Guided (AA)  & \textbf{100.0\%} & 7.56 (0.14) & \textbf{100.0\%} & 6.40 (0.50) & \textbf{100.0\%} & \textbf{8.28 (0.42)} \\
\quad Staged (UAA) & 5.1\%  & \underline{7.77 (0.05)} & 4.0\% & 6.98 (0.84) & 19.4\% & 8.02 (0.40) \\
\quad Guided (UAA) & \textbf{100.0\%} & 7.71 (0.09) & \textbf{100.0\%} & \underline{7.57 (0.45)} & \textbf{100.0\%} & 7.95 (0.22) \\
\bottomrule
\end{tabular}
\end{table}

\begin{table}[!ht]
\centering
\caption{Full-valid-set \textit{de novo} PDCs across three receptor targets generated by SMILES-based methods (peptide length: 80 SMILES tokens). Metrics are computed over all valid conjugates from each method (not filtered to top-20). Each conjugate comprises a peptide linked to a target-specific cytotoxic payload via the corresponding linker. Conjugation Validity: fraction of valid peptides yielding chemically valid conjugates. Binding affinity: predicted peptide-receptor binding score (pKd). Values are mean (std). \textbf{Bold}: best per column; \underline{underline}: second-best.}
\label{tab:full-smiles80}
\footnotesize
\setlength{\tabcolsep}{4pt}
\begin{tabular}{lcccccc}
\toprule
 & \multicolumn{2}{c}{\textbf{ERBB2}} & \multicolumn{2}{c}{\textbf{GNRHR}} & \multicolumn{2}{c}{\textbf{NRP1}} \\
\cmidrule(lr){2-3} \cmidrule(lr){4-5} \cmidrule(lr){6-7}
\textbf{Method} & Conj.\ Val. & Binding & Conj.\ Val. & Binding & Conj.\ Val. & Binding \\
\midrule
PepTune & 20.8\% & 6.75 (0.27) & 5.6\% & 4.55 & 13.7\% & 7.05 (1.81) \\
\addlinespace[2pt]
\multicolumn{7}{l}{\textit{AReUReDi}} \\
\quad Staged & \underline{30.1\%} & 7.65 (0.16) & \underline{26.8\%} & 7.54 (0.86) & 23.0\% & \underline{8.62 (0.50)} \\
\quad Guided & \textbf{100.0\%} & 7.68 (0.25) & \textbf{100.0\%} & \underline{7.64 (0.68)} & \textbf{100.0\%} & 8.59 (0.50) \\
\addlinespace[2pt]
\multicolumn{7}{l}{\textit{CupOFMoCA}} \\
\quad Staged & 22.7\% & \textbf{7.70 (0.17)} & 18.3\% & \textbf{8.09 (1.17)} & \underline{23.7\%} & \textbf{8.68 (0.37)} \\
\quad Guided & \textbf{100.0\%} & \underline{7.69 (0.22)} & \textbf{100.0\%} & 7.55 (0.68) & \textbf{100.0\%} & 8.58 (0.64) \\
\bottomrule
\end{tabular}
\end{table}

\begin{table}[!ht]
\centering
\caption{Full-valid-set \textit{de novo} PDCs across three receptor targets generated by SMILES-based methods (peptide length: 100 SMILES tokens). Metrics are computed over all valid conjugates from each method (not filtered to top-20). Each conjugate comprises a peptide linked to a target-specific cytotoxic payload via the corresponding linker. Conjugation Validity: fraction of valid peptides yielding chemically valid conjugates. Binding affinity: predicted peptide-receptor binding score (pKd). Values are mean (std). \textbf{Bold}: best per column; \underline{underline}: second-best.}
\label{tab:full-smiles100}
\footnotesize
\setlength{\tabcolsep}{4pt}
\begin{tabular}{lcccccc}
\toprule
 & \multicolumn{2}{c}{\textbf{ERBB2}} & \multicolumn{2}{c}{\textbf{GNRHR}} & \multicolumn{2}{c}{\textbf{NRP1}} \\
\cmidrule(lr){2-3} \cmidrule(lr){4-5} \cmidrule(lr){6-7}
\textbf{Method} & Conj.\ Val. & Binding & Conj.\ Val. & Binding & Conj.\ Val. & Binding \\
\midrule
PepTune & 17.9\% & 6.82 (0.50) & 17.2\% & 5.85 (1.25) & 11.3\% & 7.25 (1.11) \\
\addlinespace[2pt]
\multicolumn{7}{l}{\textit{AReUReDi}} \\
\quad Staged & \underline{40.9\%} & 7.79 (0.21) & \underline{39.1\%} & \textbf{8.57 (0.93)} & 18.8\% & \textbf{8.75 (0.64)} \\
\quad Guided & \textbf{100.0\%} & 7.69 (0.21) & \textbf{100.0\%} & 7.95 (0.63) & \textbf{100.0\%} & 8.66 (0.66) \\
\addlinespace[2pt]
\multicolumn{7}{l}{\textit{CupOFMoCA}} \\
\quad Staged & 36.7\% & \textbf{7.92 (0.25)} & 34.7\% & \underline{8.15 (0.74)} & \underline{28.1\%} & 8.65 (0.53) \\
\quad Guided & \textbf{100.0\%} & \underline{7.86 (0.27)} & \textbf{100.0\%} & 8.07 (0.74) & \textbf{100.0\%} & \underline{8.67 (0.47)} \\
\bottomrule
\end{tabular}
\end{table}

\FloatBarrier   % force any pending floats from previous subsection to flush
\subsection{Other molecular properties}
\begin{figure}[!ht]
  \centering
  \includegraphics[width=0.95\linewidth]{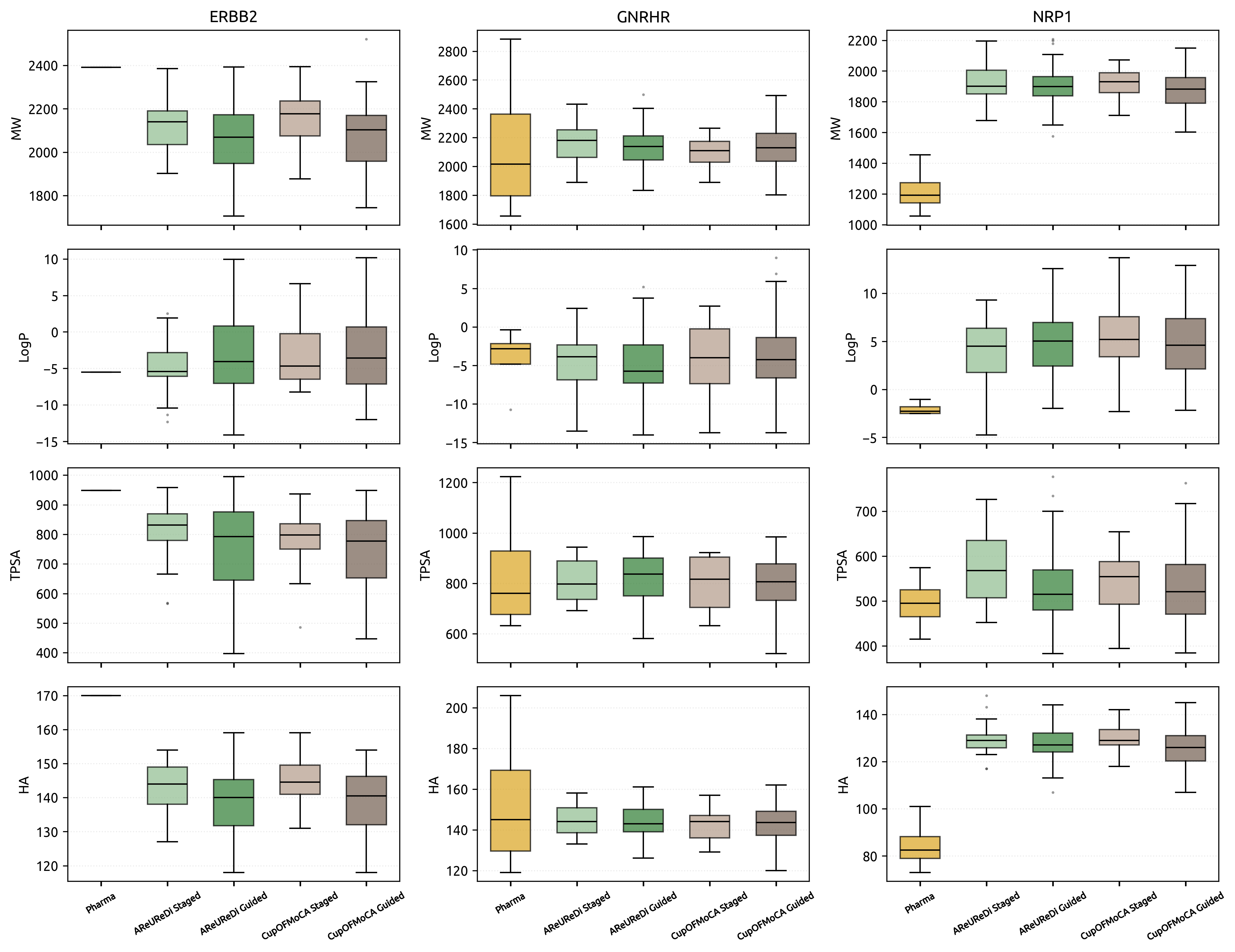}
  \caption{\textbf{Molecular property distributions of generated PDCs vs.\ PDCdb pharmaceutical references at peptide length 80 SMILES tokens.}
  Box plots show the distributions of molecular weight (MW), lipophilicity (LogP), topological polar surface area (TPSA, \r{A}$^2$), and heavy-atom count (HA) over all valid conjugates from each SMILES-based method ($n{=}67{-}94$ per cell for AReUReDi/CupOFMoCA Guided; $n{=}17{-}28$ for staged variants), compared against PDCdb reference compounds (gold, $n{=}1$ for ERBB2; $n{=}4$ for GNRHR; $n{=}4$ for NRP1).
  For ERBB2, the single available reference appears as a degenerate line.
  All methods use the same receptor-specific linker-payload as the references (ERBB2: SSCpt; GNRHR: AoaDau; NRP1: DSPASN38), and we verified that the per-attachment $\Delta\mathrm{MW}$ (conjugate $-$ peptide) is conserved between pharma and generated sets within ${\sim}1$\,Da, confirming identical conjugation chemistry on both sides.
  Differences between gold and generated boxes therefore reflect peptide-length differences rather than chemistry: the GNRHR pharma references are 8-21-residue peptides (mean peptide MW ${\approx}1.56$\,kDa) and overlap the generated distribution; the single ERBB2 reference is an 18-residue peptide (${\approx}1.85$\,kDa) and sits above the generations; the NRP1 references are short 5-9-residue C-end-rule binders (e.g.\ CRGDK; mean peptide MW ${\approx}0.74$\,kDa) and sit below.
  Generated distributions across methods are highly overlapping, indicating that coupled / guided generation does not collapse the chemistry to a narrow region of property space.}
  \label{fig:property-distributions}
\end{figure}

We compared linker-payload mass increments and peptide masses to interpret the target-specific property distributions (Tables~\ref{tab:linker-payload-delta-mw} and~\ref{tab:peptide-length-comparison}; Figure~\ref{fig:property-distributions}).

\begin{table}[!ht]
\centering
\footnotesize
\setlength{\tabcolsep}{5pt}
\caption{Per-attachment mass added by each receptor-specific linker-payload is conserved between PDCdb references and generated PDCs, confirming identical conjugation chemistry. $\Delta\mathrm{MW}$ is computed as conjugate MW $-$ peptide MW. Theoretical values: SSCpt (amide, $-$H$_2$O) ester/carbonate, AoaDau (amide, $-$H$_2$O), DSPASN38 (disulfide, $-2$H).}
\label{tab:linker-payload-delta-mw}
\begin{tabular}{llccc}
\toprule
Target & Linker-payload (reaction) & Theoretical & Pharma & Generated \\
       &                           & $\Delta\mathrm{MW}$ & $\Delta\mathrm{MW}$ ($n$) & $\Delta\mathrm{MW}$ ($n$) \\
\midrule
ERBB2 & SSCpt, ester/carbonate (amide) & 537.10 / 553.10 & 538.8 (1)             & $538.6 \pm 0.64$ (92) \\
GNRHR & AoaDau (amide)                 & 582.18          & $583.5 \pm 0.32$ (4)  & $583.7 \pm 0.48$ (94) \\
NRP1  & DSPASN38, pos10/pos20 (disulfide) & 478.12       & $479.0 \pm 0.10$ (4)  & $479.6 \pm 0.41$ (78) \\
\bottomrule
\end{tabular}
\end{table}
\begin{table}[!ht]
\centering
\footnotesize
\caption{Peptide-length comparison between PDCdb references and generated PDCs. With the linker-payload mass held constant (Table~\ref{tab:linker-payload-delta-mw}), the apparent offsets in Figure~\ref{fig:property-distributions} reduce to peptide-mass differences. Generated peptides are sampled at a fixed budget of 80 SMILES tokens and frequently use non-canonical residues, so a residue count is not directly comparable to the canonical pharma peptides; we report peptide MW (Da) for both sides and \texttt{seq.\ length} only for the pharma references.}
\label{tab:peptide-length-comparison}
\begin{tabular}{lcccl}
\toprule
\multirow{2}{*}{Target} & \multicolumn{2}{c}{Pharma reference} & Generated (80 SMILES tokens) & \multirow{2}{*}{Direction of offset} \\
\cmidrule(lr){2-3} \cmidrule(lr){4-4}
& peptide MW (Da) & seq.\ length & peptide MW (Da) & \\
\midrule
ERBB2 & 1852 ($n{=}1$)    & 18           & 1531 (1205-1986) & generated \emph{lighter} than pharma \\
GNRHR & 1559 (1072-2300) & 13.2 (8-21) & 1544 (1218-1907) & matched (within ${\sim}15$ Da)        \\
NRP1  & 744 (577-974)    & 6.8 (5-10)  & 1396 (1123-1671) & generated \emph{heavier} than pharma  \\
\bottomrule
\end{tabular}
\end{table}

\newpage
\begin{table}[!ht]
\centering
\caption{Full-valid-set molecular properties for AA-based methods on target ERBB2 (peptide length: 10 AAs). All conjugates comprise a peptide linked to a target-specific cytotoxic payload via the corresponding linker). Conjugation Validity: fraction of valid peptides yielding chemically valid conjugates. Binding affinity: predicted peptide-receptor binding score (pKd). MW: molecular weight; HBD/HBA: hydrogen-bond donors/acceptors. Values are mean (std).}
\label{tab:full-aa-erbb2}
\footnotesize
\setlength{\tabcolsep}{3pt}
\resizebox{\textwidth}{!}{%
\begin{tabular}{l c cc cccc cccc}
\toprule
 & & \multicolumn{2}{c}{\textit{BoltzGen}} & \multicolumn{4}{c}{\textit{AReUReDi}} & \multicolumn{4}{c}{\textit{CupOFMoCA}} \\
\cmidrule(lr){3-4} \cmidrule(lr){5-8} \cmidrule(lr){9-12}
\textbf{Property} & \textbf{moPPIt} & Exp & AF & Staged (AA) & Guided (AA) & Staged (UAA) & Guided (UAA) & Staged (AA) & Guided (AA) & Staged (UAA) & Guided (UAA) \\
\midrule
Conj.\ Validity   & 7.1\%           & 51.0\%            & 21.2\%            & 17.0\%           & 100.0\%          & 3.1\%            & 100.0\%          & 15.0\%           & 100.0\%          & 5.1\%            & 100.0\% \\
Binding affinity  & 7.59 (0.07)     & 7.64 (0.22)       & 7.80 (0.13)       & 7.65 (0.10)      & 7.57 (0.16)      & 7.68 (0.08)      & 7.72 (0.08)      & 7.56 (0.19)      & 7.56 (0.14)      & 7.77 (0.05)      & 7.71 (0.09) \\
MW (Da)           & 1797.19 (94.33) & 1862.70 (262.50)  & 1652.95 (339.07)  & 1958.14 (104.89) & 1987.75 (110.63) & 2460.77 (288.16) & 2558.94 (454.66) & 2062.10 (90.96)  & 2037.29 (98.27)  & 2238.06 (380.48) & 2465.91 (404.95) \\
LogP              & 2.23 (1.93)     & $-0.84$ (2.37)    & $-1.03$ (2.89)    & 0.93 (1.94)      & 0.53 (2.54)      & 8.93 (2.53)      & 8.96 (5.73)      & 0.89 (2.23)      & 0.52 (1.99)      & 4.59 (3.27)      & 6.65 (5.25) \\
TPSA (\AA$^2$)    & 556.56 (30.58)  & 663.67 (120.94)   & 571.81 (171.79)   & 648.83 (78.14)   & 671.25 (83.64)   & 668.58 (64.45)   & 723.24 (97.92)   & 708.49 (84.67)   & 698.27 (78.00)   & 670.06 (82.08)   & 724.58 (111.19) \\
HBD               & 17.14 (1.35)    & 19.27 (4.27)      & 16.33 (6.55)      & 19.12 (3.97)     & 20.54 (3.97)     & 21.67 (2.08)     & 21.07 (3.56)     & 23.07 (4.04)     & 22.05 (4.00)     & 20.20 (2.17)     & 22.23 (3.83) \\
HBA               & 26.29 (1.38)    & 28.33 (3.83)      & 25.43 (4.60)      & 27.59 (1.42)     & 27.86 (1.41)     & 32.67 (1.15)     & 34.49 (4.98)     & 28.27 (1.39)     & 27.90 (1.38)     & 32.20 (4.32)     & 33.87 (4.43) \\
Rotatable bonds   & 49.86 (3.34)    & 54.39 (9.50)      & 45.14 (12.97)     & 50.65 (5.20)     & 51.86 (5.25)     & 70.67 (5.03)     & 73.06 (17.43)    & 55.67 (4.70)     & 53.52 (4.94)     & 60.40 (13.16)    & 69.56 (15.28) \\
Heavy atoms       & 124.57 (8.04)   & 130.10 (18.46)    & 115.62 (23.77)    & 138.24 (8.00)    & 140.21 (8.35)    & 170.33 (22.50)   & 178.37 (32.32)   & 145.33 (7.08)    & 144.01 (7.26)    & 155.80 (25.87)   & 172.12 (28.52) \\
\bottomrule
\end{tabular}%
}
\end{table}

\begin{table}[!ht]
\centering
\caption{Full-valid-set molecular properties for AA-based methods on target GNRHR (peptide length: 10 AAs). All conjugates comprise a peptide linked to a target-specific cytotoxic payload via the corresponding linker. Conjugation Validity: fraction of valid peptides yielding chemically valid conjugates. Binding affinity: predicted peptide-receptor binding score (pKd). MW: molecular weight; HBD/HBA: hydrogen-bond donors/acceptors. Values are mean (std).}
\label{tab:full-aa-gnrhr}
\footnotesize
\setlength{\tabcolsep}{3pt}
\resizebox{\textwidth}{!}{%
\begin{tabular}{l c cc cccc cccc}
\toprule
 & & \multicolumn{2}{c}{\textit{BoltzGen}} & \multicolumn{4}{c}{\textit{AReUReDi}} & \multicolumn{4}{c}{\textit{CupOFMoCA}} \\
\cmidrule(lr){3-4} \cmidrule(lr){5-8} \cmidrule(lr){9-12}
\textbf{Property} & \textbf{moPPIt} & Exp & AF & Staged (AA) & Guided (AA) & Staged (UAA) & Guided (UAA) & Staged (AA) & Guided (AA) & Staged (UAA) & Guided (UAA) \\
\midrule
Conj.\ Validity   & 8.1\%            & 16.0\%            & 3.0\%             & 17.6\%           & 100.0\%          & 5.1\%            & 100.0\%          & 15.0\%           & 100.0\%          & 4.0\%            & 100.0\% \\
Binding affinity  & 6.78 (0.46)      & 6.63 (1.05)       & 7.44 (0.45)       & 6.58 (0.42)      & 6.48 (0.49)      & 7.68 (0.33)      & 7.39 (0.47)      & 6.30 (0.35)      & 6.40 (0.50)      & 6.98 (0.84)      & 7.57 (0.45) \\
MW (Da)           & 1807.91 (122.97) & 1914.95 (267.16)  & 1629.48 (178.58)  & 2057.01 (112.57) & 2044.31 (106.20) & 2431.32 (446.81) & 2506.25 (395.02) & 2081.44 (99.03)  & 2074.30 (110.34) & 2852.63 (370.48) & 2650.75 (445.82) \\
LogP              & $-0.08$ (3.66)   & $-3.07$ (3.00)    & 0.36 (1.88)       & $-2.58$ (1.74)   & $-2.51$ (1.40)   & 2.62 (3.27)      & 4.77 (6.72)      & $-1.87$ (1.42)   & $-2.39$ (1.79)   & 8.73 (11.32)     & 7.00 (8.38) \\
TPSA (\AA$^2$)    & 630.78 (69.49)   & 712.12 (138.16)   & 582.29 (78.16)    & 794.03 (91.01)   & 782.39 (73.53)   & 835.99 (164.41)  & 806.31 (98.75)   & 787.20 (74.23)   & 795.54 (86.98)   & 836.49 (137.88)  & 802.97 (104.47) \\
HBD               & 20.62 (2.26)     & 21.62 (4.79)      & 17.00 (3.00)      & 26.56 (4.57)     & 25.80 (3.76)     & 25.00 (4.30)     & 25.44 (3.73)     & 26.07 (4.04)     & 26.75 (4.62)     & 26.25 (4.65)     & 26.01 (4.21) \\
HBA               & 28.25 (1.67)     & 29.62 (4.08)      & 25.00 (2.00)      & 30.61 (1.14)     & 30.29 (1.15)     & 35.80 (6.69)     & 35.72 (4.05)     & 30.27 (0.88)     & 30.24 (1.35)     & 39.25 (1.71)     & 36.58 (3.84) \\
Rotatable bonds   & 46.25 (4.40)     & 49.19 (9.16)      & 41.00 (7.81)      & 52.22 (5.19)     & 51.00 (5.36)     & 64.80 (16.45)    & 67.25 (13.60)    & 52.13 (6.00)     & 51.94 (5.76)     & 77.75 (12.61)    & 72.08 (15.56) \\
Heavy atoms       & 128.00 (9.64)    & 136.12 (19.05)    & 116.33 (12.86)    & 146.56 (8.40)    & 145.83 (7.89)    & 172.20 (31.70)   & 177.40 (27.69)   & 148.73 (7.15)    & 148.08 (8.15)    & 201.00 (25.44)   & 187.29 (31.77) \\
\bottomrule
\end{tabular}%
}
\end{table}
\begin{table}[!ht]
\centering
\caption{Full-valid-set molecular properties for AA-based methods on target NRP1 (peptide length: 10 AAs). All conjugates comprise a peptide linked to SN-38 payload via Disulfanylpropanoic Acid linker. Conjugation Validity: fraction of valid peptides yielding chemically valid conjugates. Binding affinity: predicted peptide-receptor binding score (pKd). MW: molecular weight; HBD/HBA: hydrogen-bond donors/acceptors. Values are mean (std).}
\label{tab:full-aa-nrp1}
\footnotesize
\setlength{\tabcolsep}{3pt}
\resizebox{\textwidth}{!}{%
\begin{tabular}{l cc cccc cccc}
\toprule
 & & & \multicolumn{4}{c}{\textit{AReUReDi}} & \multicolumn{4}{c}{\textit{CupOFMoCA}} \\
\cmidrule(lr){4-7} \cmidrule(lr){8-11}
\textbf{Property} & \textbf{moPPIt} & \textbf{BoltzGen AF} & Staged (AA) & Guided (AA) & Staged (UAA) & Guided (UAA) & Staged (AA) & Guided (AA) & Staged (UAA) & Guided (UAA) \\
\midrule
Conj.\ Validity   & 65.0\%          & 13.7\%             & 19.0\%           & 98.0\%           & 37.4\%           & 100.0\%          & 25.0\%           & 100.0\%          & 19.4\%           & 100.0\% \\
Binding affinity  & 7.73 (0.34)     & 7.05 (1.81)        & 8.23 (0.38)      & 8.24 (0.37)      & 7.92 (0.32)      & 7.94 (0.27)      & 8.16 (0.32)      & 8.28 (0.42)      & 8.02 (0.40)      & 7.95 (0.22) \\
MW (Da)           & 1754.71 (96.21) & 1680.54 (141.90)   & 1838.20 (115.56) & 1842.31 (82.55)  & 2431.86 (271.82) & 2325.85 (252.47) & 1854.46 (94.83)  & 1859.13 (87.62)  & 2269.39 (317.96) & 2244.83 (296.12) \\
LogP              & 4.11 (2.09)     & 2.22 (5.38)        & 2.87 (2.07)      & 3.00 (2.07)      & 10.74 (4.53)     & 8.75 (3.94)      & 3.73 (2.07)      & 3.40 (2.16)      & 10.37 (4.20)     & 8.23 (4.44) \\
TPSA (\AA$^2$)    & 525.45 (55.18)  & 518.17 (130.08)    & 566.37 (70.85)   & 566.61 (55.68)   & 626.38 (64.34)   & 625.54 (78.18)   & 556.44 (65.69)   & 566.99 (72.48)   & 569.27 (61.28)   & 596.79 (61.02) \\
HBD               & 16.12 (2.68)    & 14.71 (5.09)       & 17.05 (3.73)     & 16.53 (2.68)     & 17.35 (1.77)     & 17.40 (2.50)     & 16.40 (3.01)     & 16.58 (3.41)     & 16.37 (1.86)     & 17.00 (2.04) \\
HBA               & 24.94 (1.92)    & 24.86 (4.30)       & 25.42 (1.61)     & 25.50 (1.29)     & 32.62 (3.34)     & 31.85 (3.66)     & 25.04 (1.02)     & 25.24 (1.31)     & 29.74 (3.53)     & 30.82 (3.65) \\
Rotatable bonds   & 44.29 (2.87)    & 41.14 (8.32)       & 46.05 (6.17)     & 46.05 (4.52)     & 68.92 (12.30)    & 63.57 (10.73)    & 46.64 (4.56)     & 46.44 (4.49)     & 63.37 (12.48)    & 61.18 (11.97) \\
Heavy atoms       & 123.40 (7.59)   & 117.43 (9.93)      & 129.37 (8.23)    & 129.64 (6.14)    & 169.35 (19.60)   & 162.31 (17.60)   & 130.64 (7.04)    & 131.02 (6.54)    & 157.95 (22.13)   & 156.61 (20.56) \\
\bottomrule
\end{tabular}%
}
\end{table}
\begin{table}[!ht]
\centering
\caption{Full-valid-set SMILES-based methods, target ERBB2 (peptide length 80 SMILES tokens).}
\label{tab:full-smiles80-erbb2}
\small
\setlength{\tabcolsep}{3pt}
\resizebox{\textwidth}{!}{%
\begin{tabular}{lccccc}
\toprule
\textbf{Property} & \textbf{PepTune} & \textbf{AReUReDi Staged} & \textbf{AReUReDi Guided} & \textbf{CupOFMoCA Staged} & \textbf{CupOFMoCA Guided} \\
\midrule
Conj.\ Validity & 20.8\% & 30.1\% & 100.0\% & 22.7\% & 100.0\% \\
Binding affinity & 6.75 (0.27) & 7.65 (0.16) & 7.68 (0.25) & 7.70 (0.17) & 7.69 (0.22) \\
MW (Da) & 1875.21 (110.72) & 2117.46 (121.51) & 2062.85 (150.76) & 2159.62 (125.87) & 2069.22 (156.24) \\
LogP & $-1.32$ (5.03) & $-4.78$ (3.55) & $-3.25$ (5.32) & $-3.13$ (4.08) & $-3.07$ (5.04) \\
TPSA (\AA$^2$) & 665.82 (97.57) & 807.02 (101.27) & 750.90 (142.53) & 779.71 (92.59) & 750.40 (131.44) \\
HBD & 18.60 (4.03) & 25.64 (3.64) & 25.01 (5.26) & 25.05 (3.36) & 24.35 (4.74) \\
HBA & 28.80 (3.55) & 43.39 (3.85) & 42.76 (5.77) & 44.32 (4.68) & 42.59 (5.70) \\
Rotatable bonds & 53.80 (6.51) & 72.54 (7.94) & 71.77 (8.38) & 75.73 (8.02) & 71.07 (7.44) \\
Heavy atoms & 130.40 (6.82) & 142.96 (7.87) & 138.96 (9.90) & 145.00 (7.66) & 138.72 (9.69) \\
\bottomrule
\end{tabular}%
}
\end{table}

\begin{table}[!ht]
\centering
\caption{Full-valid-set SMILES-based methods, target GNRHR (peptide length 80 SMILES tokens).}
\label{tab:full-smiles80-gnrhr}
\small
\setlength{\tabcolsep}{3pt}
\resizebox{\textwidth}{!}{%
\begin{tabular}{lccccc}
\toprule
\textbf{Property} & \textbf{PepTune} & \textbf{AReUReDi Staged} & \textbf{AReUReDi Guided} & \textbf{CupOFMoCA Staged} & \textbf{CupOFMoCA Guided} \\
\midrule
Conj.\ Validity & 5.6\% & 26.8\% & 100.0\% & 18.3\% & 100.0\% \\
Binding affinity & 4.55 & 7.54 (0.86) & 7.64 (0.68) & 8.09 (1.17) & 7.55 (0.68) \\
MW (Da) & 1857.06 (0.00) & 2161.09 (140.02) & 2136.08 (132.55) & 2100.42 (90.95) & 2128.04 (142.41) \\
LogP & $-4.53$ & $-4.38$ (3.92) & $-5.03$ (4.20) & $-3.92$ (4.57) & $-3.84$ (4.02) \\
TPSA (\AA$^2$) & 777.09 & 814.26 (87.94) & 821.02 (98.05) & 791.03 (105.28) & 795.98 (105.34) \\
HBD & 27.00 & 30.45 (3.73) & 29.63 (3.96) & 27.76 (3.70) & 28.76 (4.21) \\
HBA & 32.00 & 46.41 (4.97) & 45.27 (4.49) & 44.76 (4.32) & 45.69 (4.79) \\
Rotatable bonds & 51.00 & 73.27 (6.51) & 69.04 (7.75) & 70.00 (8.06) & 70.62 (7.48) \\
Heavy atoms & 132.00 & 144.91 (7.62) & 143.71 (8.37) & 142.53 (6.98) & 142.67 (8.80) \\
\bottomrule
\end{tabular}%
}
\end{table}

\begin{table}[!ht]
\centering
\caption{Full-valid-set SMILES-based methods, target NRP1 (peptide length 80 SMILES tokens).}
\label{tab:full-smiles80-nrp1}
\small
\setlength{\tabcolsep}{3pt}
\resizebox{\textwidth}{!}{%
\begin{tabular}{lccccc}
\toprule
\textbf{Property} & \textbf{PepTune} & \textbf{AReUReDi Staged} & \textbf{AReUReDi Guided} & \textbf{CupOFMoCA Staged} & \textbf{CupOFMoCA Guided} \\
\midrule
Conj.\ Validity & 13.7\% & 23.0\% & 100.0\% & 23.7\% & 100.0\% \\
Binding affinity & 7.05 (1.81) & 8.62 (0.50) & 8.59 (0.50) & 8.68 (0.37) & 8.58 (0.64) \\
MW (Da) & 1680.54 (141.90) & 1928.14 (132.97) & 1907.99 (115.10) & 1925.42 (95.30) & 1875.63 (124.12) \\
LogP & 2.22 (5.38) & 3.93 (3.67) & 4.98 (3.32) & 5.09 (3.89) & 4.88 (3.55) \\
TPSA (\AA$^2$) & 518.17 (130.08) & 576.12 (85.53) & 528.84 (72.29) & 543.50 (68.90) & 528.09 (81.09) \\
HBD & 14.71 (5.09) & 19.45 (2.86) & 17.40 (2.96) & 17.61 (4.05) & 17.82 (3.58) \\
HBA & 24.86 (4.30) & 37.40 (5.09) & 36.01 (3.26) & 36.22 (4.32) & 36.21 (3.81) \\
Rotatable bonds & 41.14 (8.32) & 69.85 (8.36) & 67.21 (8.28) & 68.09 (9.02) & 67.00 (8.32) \\
Heavy atoms & 117.43 (9.93) & 129.35 (7.59) & 128.22 (6.81) & 129.78 (6.13) & 125.91 (7.59) \\
\bottomrule
\end{tabular}%
}
\end{table}

\begin{table}[!ht]
\centering
\caption{Full-valid-set SMILES-based methods, target ERBB2 (peptide length 100 SMILES tokens).}
\label{tab:full-smiles100-erbb2}
\small
\setlength{\tabcolsep}{3pt}
\resizebox{\textwidth}{!}{%
\begin{tabular}{lccccc}
\toprule
\textbf{Property} & \textbf{PepTune} & \textbf{AReUReDi Staged} & \textbf{AReUReDi Guided} & \textbf{CupOFMoCA Staged} & \textbf{CupOFMoCA Guided} \\
\midrule
Conj.\ Validity & 17.9\% & 40.9\% & 100.0\% & 36.7\% & 100.0\% \\
Binding affinity & 6.82 (0.50) & 7.79 (0.21) & 7.69 (0.21) & 7.92 (0.25) & 7.86 (0.27) \\
MW (Da) & 2135.35 (135.67) & 2458.65 (158.09) & 2422.54 (176.41) & 2513.95 (147.00) & 2436.57 (171.09) \\
LogP & 0.92 (3.59) & $-3.76$ (4.20) & $-3.98$ (4.88) & $-3.61$ (5.23) & $-3.10$ (5.44) \\
TPSA (\AA$^2$) & 674.01 (54.77) & 905.03 (103.73) & 887.69 (136.12) & 913.99 (110.20) & 875.13 (131.86) \\
HBD & 20.80 (3.56) & 28.56 (4.04) & 29.03 (4.76) & 28.17 (5.85) & 28.34 (5.09) \\
HBA & 32.80 (4.55) & 49.28 (4.67) & 48.36 (5.17) & 50.69 (4.57) & 49.11 (5.44) \\
Rotatable bonds & 59.20 (7.53) & 86.14 (8.64) & 85.29 (8.58) & 88.86 (8.25) & 86.35 (8.95) \\
Heavy atoms & 147.60 (8.14) & 164.97 (8.48) & 163.27 (11.44) & 167.78 (7.57) & 163.54 (10.59) \\
\bottomrule
\end{tabular}%
}
\end{table}

\begin{table}[!ht]
\centering
\caption{Full-valid-set SMILES-based methods, target GNRHR (peptide length 100 SMILES tokens).}
\label{tab:full-smiles100-gnrhr}
\small
\setlength{\tabcolsep}{3pt}
\resizebox{\textwidth}{!}{%
\begin{tabular}{lccccc}
\toprule
\textbf{Property} & \textbf{PepTune} & \textbf{AReUReDi Staged} & \textbf{AReUReDi Guided} & \textbf{CupOFMoCA Staged} & \textbf{CupOFMoCA Guided} \\
\midrule
Conj.\ Validity & 17.2\% & 39.1\% & 100.0\% & 34.7\% & 100.0\% \\
Binding affinity & 5.85 (1.25) & 8.57 (0.93) & 7.95 (0.63) & 8.15 (0.74) & 8.07 (0.74) \\
MW (Da) & 2180.29 (135.67) & 2534.01 (131.60) & 2420.67 (220.22) & 2544.61 (111.73) & 2536.91 (143.44) \\
LogP & $-0.98$ (3.59) & $-5.39$ (3.84) & $-6.13$ (4.31) & $-3.96$ (4.25) & $-4.81$ (4.13) \\
TPSA (\AA$^2$) & 750.86 (54.77) & 981.23 (93.21) & 953.53 (119.62) & 952.33 (119.41) & 962.22 (105.10) \\
HBD & 24.80 (3.56) & 33.76 (4.20) & 32.85 (4.86) & 32.76 (4.34) & 32.96 (4.70) \\
HBA & 34.80 (4.55) & 50.47 (4.07) & 50.31 (5.42) & 50.79 (4.66) & 52.15 (4.75) \\
Rotatable bonds & 57.20 (7.53) & 84.97 (8.64) & 79.82 (10.79) & 87.47 (7.40) & 85.90 (8.97) \\
Heavy atoms & 152.60 (8.14) & 171.24 (8.41) & 163.59 (15.01) & 172.12 (8.03) & 171.00 (9.69) \\
\bottomrule
\end{tabular}%
}
\end{table}

\begin{table}[!ht]
\centering
\caption{Full-valid-set SMILES-based methods, target NRP1 (peptide length 100 SMILES tokens).}
\label{tab:full-smiles100-nrp1}
\small
\setlength{\tabcolsep}{3pt}
\resizebox{\textwidth}{!}{%
\begin{tabular}{lccccc}
\toprule
\textbf{Property} & \textbf{PepTune} & \textbf{AReUReDi Staged} & \textbf{AReUReDi Guided} & \textbf{CupOFMoCA Staged} & \textbf{CupOFMoCA Guided} \\
\midrule
Conj.\ Validity & 11.3\% & 18.8\% & 100.0\% & 28.1\% & 100.0\% \\
Binding affinity & 7.25(1.11) & 8.75 (0.64) & 8.66 (0.66) & 8.65 (0.53) & 8.67 (0.47) \\
MW (Da) & 2001.96(143.32) & 2270.30 (179.81) & 2327.55 (147.16) & 2243.58 (171.81) & 2285.06 (149.25) \\
LogP & 0.04 (2.32) & 4.29 (4.16) & 3.20 (5.86) & 4.36 (4.79) & 2.85 (5.81) \\
683.09 (80.04) & 698.83 (110.29) & 731.19 (147.33) & 684.12 (140.35) & 711.35 (153.47) \\
HBD & 19.83 (1.72) & 22.50 (4.10) & 24.25 (5.10) & 22.26 (4.74) & 23.29 (5.64) \\
HBA & 29.83 (3.60) & 41.69 (4.44) & 44.43 (5.40) & 41.30 (5.43) & 43.68 (5.49) \\
Rotatable bonds &  54.00 (10.30) & 81.69 (7.60) & 85.64 (8.87) & 78.70 (9.24) & 82.60 (9.74) \\
Heavy atoms & 140.50 (9.89) & 154.00 (12.28) & 156.71 (10.30) & 151.30 (10.42) & 153.02 (10.34) \\
\bottomrule
\end{tabular}%
}
\end{table}

\FloatBarrier
\subsection{Variable Linker Generation}

\begin{table}[!ht]
\centering
\caption{Molecular properties of PDCs generated with variable vs.\ fixed linkers (GNRHR-Dau: fixed Aoa; ERBB2-Cpt: fixed SS) for the same receptor-payload pairs. All results are from 100 generation runs at peptide length 80 SMILES tokens. Values shown as mean (std).}
\label{tab:variable-vs-fixed-linker}
\small
\setlength{\tabcolsep}{4pt}
\begin{tabular}{lcccc}
\toprule
\textbf{Property} & \makecell{\textbf{GNRHR-Dau}\\\textbf{Coupled}} & \makecell{\textbf{GNRHR-Dau}\\\textbf{Var.\ Linker}} & \makecell{\textbf{ERBB2-Cpt}\\\textbf{Coupled}} & \makecell{\textbf{ERBB2-Cpt}\\\textbf{Var.\ Linker}} \\
\midrule
Conj. validity   & 100\%          & 100\%          & 100\%          & 100\% \\
Binding affinity & 7.55 (0.68)   & 7.67 (0.96)   & 7.69 (0.22)    & 7.73 (0.23) \\
MW (Da)          & 2128 (142)     & 2136 (178)     & 2069 (156)     & 2090 (149) \\
LogP             & $-3.84$ (4.02) & $-3.58$ (3.92) & $-3.07$ (5.04) & $-2.82$ (4.98) \\
TPSA (\AA$^2$)   & 796.0 (105.3)  & 796.6 (102.1)  & 750.4 (131.4)  & 760.9 (137.8) \\
HBD              & 28.8 (4.2)     & 28.9 (4.2)     & 24.4 (4.7)     & 24.2 (4.6) \\
HBA              & 45.7 (4.8)     & 45.1 (3.8)     & 42.6 (5.7)     & 43.2 (5.6) \\
Rotatable bonds  & 70.6 (7.5)     & 70.5 (9.3)     & 71.1 (7.4)     & 71.7 (7.1) \\
Heavy atoms      & 142.7 (8.8)    & 144.2 (12.0)   & 138.7 (9.7)    & 141.0 (10.0) \\
\bottomrule
\end{tabular}
\end{table}

\begin{table}[!ht]
\centering
\caption{Linker type diversity for the two receptor-payload pairs. \textbf{DB} refers to the PDCdb database distribution of linker types. \textbf{Joint} stands for generation with randomly sampled linker type drawn from the implemented linker types' frequencies in PDCdb.}
\label{tab:linker-diversity}
\small
\setlength{\tabcolsep}{6pt}

\begin{subtable}{\textwidth}
\centering
\caption{GNRHR receptor, Daunorubicin payload.}
\label{tab:linker-diversity-gnrhr}
\begin{tabular}{lcccc}
\toprule
\textbf{Linker} & \textbf{DB Count} & \textbf{DB \%} & \textbf{Joint Count} & \textbf{Joint \%} \\
\midrule
Aoa      & 21 & 77.8\% & 78 & 84.8\% \\
Glutaryl &  2 &  7.4\% &  6 &  6.5\% \\
ValAla   &  2 &  7.4\% &  4 &  4.3\% \\
ValCit   &  2 &  7.4\% &  4 &  4.3\% \\
\bottomrule
\end{tabular}
\end{subtable}

\vspace{1em}

\begin{subtable}{\textwidth}
\centering
\caption{ERBB2 receptor, Camptothecin payload.}
\label{tab:linker-diversity-erbb2}
\begin{tabular}{lcccc}
\toprule
\textbf{Linker} & \textbf{DB Count} & \textbf{DB \%} & \textbf{Joint Count} & \textbf{Joint \%} \\
\midrule
DSEC   & 5 & 18.5\% & 33 & 35.9\% \\
SS     & 7 & 25.9\% & 23 & 25.0\% \\
MalHex & 5 & 18.5\% & 21 & 22.8\% \\
DSPA   & 5 & 18.5\% & 15 & 16.3\% \\
\bottomrule
\end{tabular}
\end{subtable}

\end{table}

% \begin{table}[!ht]
% \centering
% \caption{Computation time vs.\ generation length (SMILES tokens) for receptors ERBB2 and GNRHR. Average time reported as hh:mm:ss and total seconds.}
% \label{tab:computation-time}
% \small
% \setlength{\tabcolsep}{6pt}
% \renewcommand{\arraystretch}{1.1}
% \begin{tabular}{llrr}
% \toprule
% \textbf{Target} & \textbf{Length} & \textbf{Avg Time} & \textbf{Avg Seconds} \\
% \midrule
% \multirow{3}{*}{ERBB2} 
%   &  80 & 00:05:40 &  339.71 \\
%   & 100 & 00:10:59 &  658.85 \\
%   & 200 & 02:07:44 & 7664.42 \\
% \midrule
% \multirow{3}{*}{GNRHR} 
%   &  80 & 00:03:07 &  187.06 \\
%   & 100 & 00:06:45 &  405.46 \\
%   & 200 & 01:09:48 & 4187.99 \\
% \bottomrule
% \end{tabular}
% \end{table}

\begin{table}[!ht]
\centering
\caption{Generation diversity metrics across receptors, peptide lengths, and linker configurations. Validity: fraction of generated peptides forming valid conjugates. Uniqueness: fraction of distinct samples among generated. Diversity: average pairwise Tanimoto distance. SNN: similarity to nearest neighbor in training set (lower = more novel).}
\label{tab:generation-diversity}
\small
\setlength{\tabcolsep}{6pt}
\renewcommand{\arraystretch}{1.1}
\begin{tabular}{llcccccc}
\toprule
\textbf{Receptor} & \textbf{Variant} & \textbf{Length} & \textbf{Validity} & \textbf{Uniqueness} & \textbf{Diversity} & \textbf{SNN} \\
\midrule
\multirow{4}{*}{ERBB2} 
  & \multirow{2}{*}{Coupled}      &  80 & 0.919 & 1.000 & 0.844 & 0.187 \\
  &                                & 100 & 0.949 & 1.000 & 0.835 & 0.194 \\
\cmidrule(l){2-7}
  & \multirow{2}{*}{Var.\ linker} &  80 & 0.919 & 1.000 & 0.842 & 0.188 \\
  &                                & 100 & 0.960 & 1.000 & 0.835 & 0.194 \\
\midrule
\multirow{4}{*}{GNRHR} 
  & \multirow{2}{*}{Coupled}      &  80 & 0.939 & 1.000 & 0.843 & 0.180 \\
  &                                & 100 & 0.960 & 1.000 & 0.832 & 0.187 \\
\cmidrule(l){2-7}
  & \multirow{2}{*}{Var.\ linker} &  80 & 0.919 & 1.000 & 0.845 & 0.179 \\
  &                                & 100 & 0.949 & 1.000 & 0.832 & 0.189 \\
\bottomrule
\end{tabular}
\end{table}

\subsection{Structural Validation}

To evaluate the structural quality of designs generated by CupOFMoCA, we performed structure prediction using Boltz-2, the current state-of-the-art molecule-protein complex predictor \citep{passaro2025boltz}, on receptor-ligand complexes. We compared our generated designs (Coupled) against reference pharmaceutical compounds (Pharma) across four metrics: overall confidence score, interface predicted TM-score (ipTM), complex pLDDT, and interface pLDDT (ipLDDT).

\begin{table}[h]
\centering
\caption{Structure prediction quality for top 10 ERBB2-targeting designs (length 80). Comparison of peptides and PDCs binding to ERBB2 receptor, with metrics measured on the receptor-ligand complex using Boltz-2. Pharma represents reference pharmaceutical designs, while Coupled represents our generation method. conf. = confidence score, ipTM = interface predicted TM-score, pLDDT = complex pLDDT, ipLDDT = complex ipLDDT.}
\label{tab:erbb2_boltz}
\small
\renewcommand{\arraystretch}{1.2}
\begin{tabular}{lcccc}
\hline
Metric & \multicolumn{2}{c}{Peptide} & \multicolumn{2}{c}{\textbf{Conjugate}}\\
 & Pharma &\textbf{Ours} & Pharma&\textbf{Ours}\\
\hline
Conf.& 0.6653  &0.6878& 0.6415 &0.6971\\
ipTM& 0.7217  &0.7571& 0.6281 &0.8043\\
pLDDT& 0.6511  &0.6705& 0.6448 &0.6703\\
ipLDDT& 0.4344  &0.4976& 0.4021 &0.5023\\
\hline
\end{tabular}
\end{table}

\begin{table}[h]
\centering
\caption{Structure prediction quality for top 10 GNRHR-targeting designs (length 80). Comparison of peptides and PDCs binding to GNRHR receptor, with metrics measured on the receptor-ligand complex using Boltz-2. }
\label{tab:gnrhr_boltz}
\small
\renewcommand{\arraystretch}{1.2}
\begin{tabular}{lcccc}
\hline
Metric & \multicolumn{2}{c}{Peptide} & \multicolumn{2}{c}{\textbf{Conjugate}}\\
 & Pharma &\textbf{Ours} & Pharma&\textbf{Ours}\\
\hline
Conf.& 0.8037&0.7764& 0.7111 &0.7251\\
ipTM& 0.9217&0.8826& 0.8581 &0.8695\\
pLDDT& 0.7742&0.7498& 0.6744 &0.6890\\
ipLDDT& 0.6522&0.5827& 0.4639 &0.4961\\
\hline
\end{tabular}
\end{table}

\begin{table}[h]
\centering
\caption{Structure prediction quality for top 10 NRP1-targeting designs (length 80). Comparison of peptides and PDCs binding to NRP1 receptor, with metrics measured on the receptor-ligand complex using Boltz-2. }
\label{tab:nrp1_boltz}
\small
\renewcommand{\arraystretch}{1.2}
\begin{tabular}{lcccc}
\hline
Metric & \multicolumn{2}{c}{Peptide} & \multicolumn{2}{c}{\textbf{Conjugate}}\\
 & Pharma &\textbf{Ours} & Pharma&\textbf{Ours}\\
\hline
Conf.& 0.7314&0.6545& 0.6560 &0.6356\\
ipTM& 0.6673&0.3914& 0.4359 &0.3486\\
pLDDT& 0.7474&0.7203& 0.7111 &0.7074\\
ipLDDT& 0.6563&0.4421& 0.4547 &0.4036\\
\hline
\end{tabular}
\end{table}

As shown in Tables~\ref{tab:erbb2_boltz}, \ref{tab:gnrhr_boltz}, and~\ref{tab:nrp1_boltz}, CupOFMoCA produces structurally valid receptor-binding designs across three diverse targets, with results that vary by target but reveal a consistent advantage on conjugates relative to peptides. For ERBB2, our method outperforms pharmaceutical references across all four metrics for both peptides and PDCs, with particularly substantial improvements on conjugates: a 28\% relative gain in ipTM (0.8043 vs.\ 0.6281) and 25\% in ipLDDT (0.5023 vs.\ 0.4021). For GNRHR, CupOFMoCA-generated conjugates exceeded the pharmaceutical references on every metric. This conjugate-level gain accompanied lower scores for the corresponding peptide designs, supporting the value of evaluating the assembled molecule. For NRP1, pharmaceutical references retain an advantage on both modalities; however, the gap is markedly narrower for conjugates than for peptides (e.g., a confidence gap of 0.020 for PDCs vs.\ 0.077 for peptides), echoing the same pattern seen on the other two targets. Taken together, these results suggest that our coupled generation approach is particularly effective for the more challenging task of designing PDCs, where the model must jointly optimize peptide sequence and conjugation chemistry while maintaining favorable binding geometry; closing the remaining gap on harder targets such as NRP1 is a natural direction for future work.

\newpage
\section{PROTAC Additional Results}

\subsection{Generated Samples}
\begin{figure}[!ht]
    \centering
    \includegraphics[width=\linewidth]{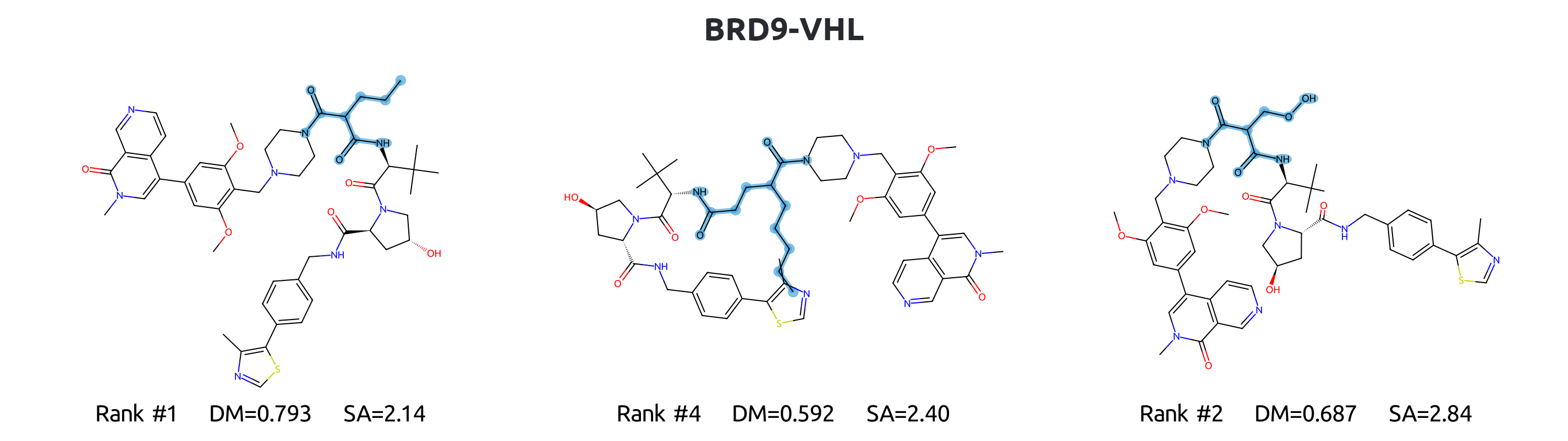}\\[2pt]
    \includegraphics[width=\linewidth]{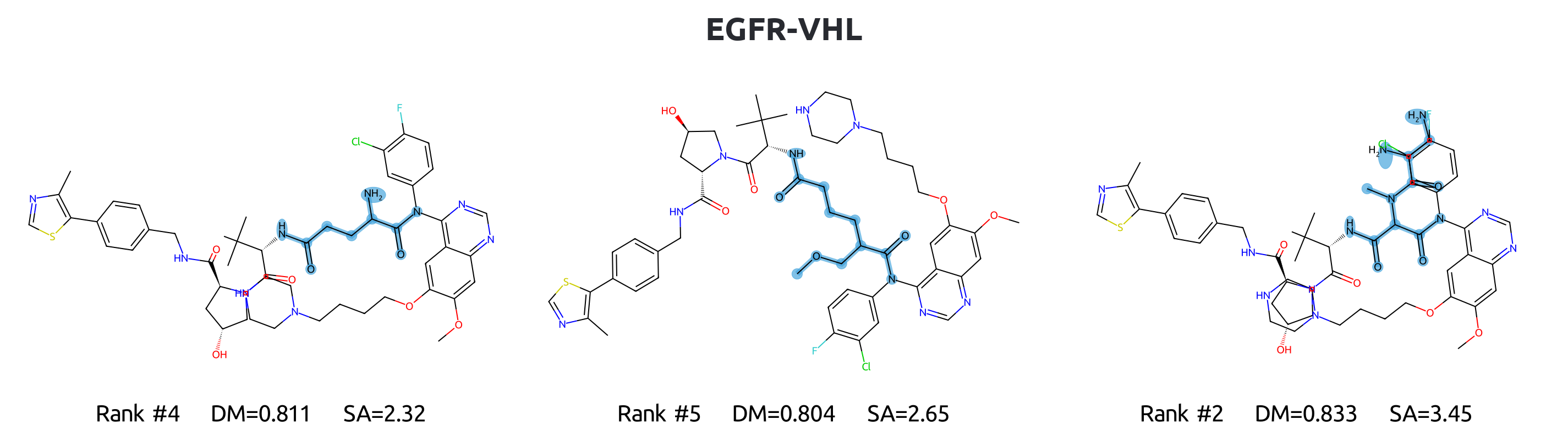}\\[2pt]
    \includegraphics[width=\linewidth]{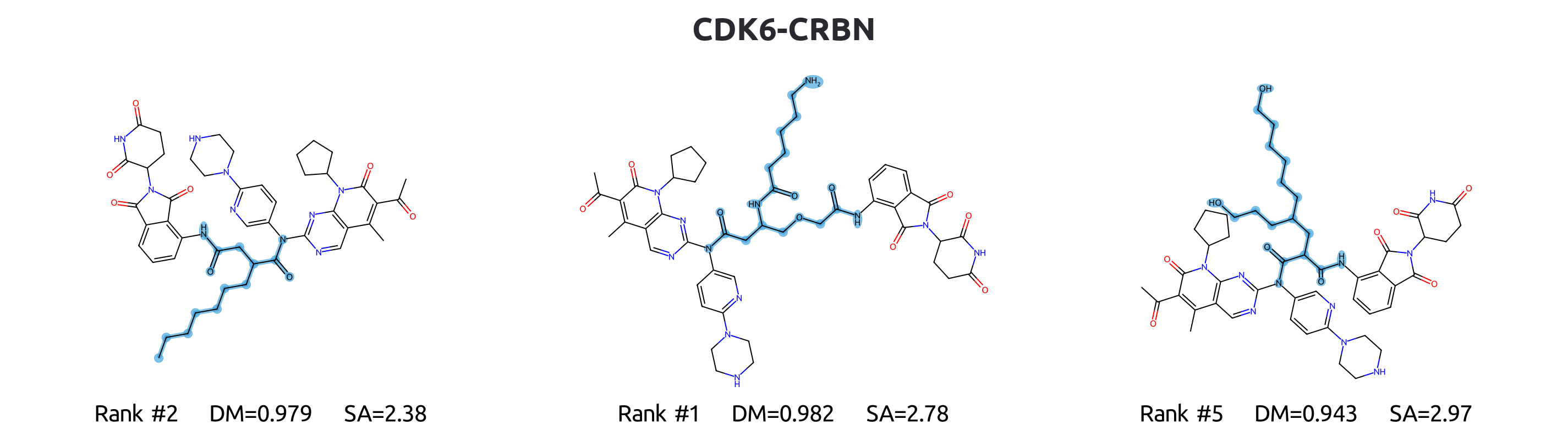}
    \caption{%
        \textbf{Top three PROTACs generated for each target.}
        For each target system (BRD9-VHL, EGFR-VHL, CDK6-CRBN), the
        three PROTACs with the lowest synthetic-accessibility (SA) score
        among the top five by DegradeMaster (DM) probability are shown
        horizontally. Each panel depicts the assembled conjugate with the
        \textit{de novo}-generated linker highlighted in blue; the
        warhead and E3-ligand fragments are the fixed inputs.
        DM denotes the predicted degradation-activity probability
        (higher is better) and SA the synthetic-accessibility score
        (lower is better, range 1-10).%
    }
    \label{fig:top3-protacs}
\end{figure}

We selected three candidates per system by synthetic accessibility from the five highest DegradeMaster-scoring candidates (Figure~\ref{fig:top3-protacs}). Their generated linkers are highlighted between the fixed warhead and E3-ligand fragments.
\newpage

\subsection{Molecular Properties Baselines Comparison}
\begin{table}[h]
\centering
\caption{Molecular properties of valid PROTACs across three target systems
(EGFR-VHL, BRD9-VHL, CDK6-CRBN). SA $\downarrow$, all others $\uparrow$.
SA and QED reported as mean $\pm$ std across molecules. The $\log P$, MW, and
RotB rows report the fraction of valid PROTACs falling within
PROTAC-DB-derived windows ($\log P \in [2.5, 7.4]$,
$\mathrm{MW} \in [771, 1108]$~Da, $\mathrm{RotB} \in [12, 25]$); note that
Table~\ref{tab:defog_finetune} uses a wider MW window ($700$-$1200$~Da) and
is not directly comparable on that row. Our column pools three seeds
($286$ / $283$ / $284$ RDKit-valid linkers) under the optimized objective
weights; baseline columns are single batches of $N=100$. Uniqueness is the
per-seed mean; pooled over all three seeds it is $92.7$ / $92.9$ / $91.9\%$.
Dashes (-) indicate $0\%$ assembly validity on that system.}
\small
\setlength{\tabcolsep}{4pt}
\begin{tabular}{llccc}
\toprule
\textbf{Target} & \textbf{Property} & \textbf{LinkerNet} & \textbf{DiffPROTACs} & \textbf{Ours} \\
\midrule
\multirow{6}{*}{EGFR-VHL}
 & SA score          & \textbf{1.89 $\pm$ 0.46} & 2.95 $\pm$ 0.91   & \underline{2.76 $\pm$ 0.57} \\
 & QED               & 0.487 $\pm$ 0.072        & 0.424 $\pm$ 0.133 & \textbf{0.51 $\pm$ 0.13} \\
 & $\log P$ in range & \textbf{87.0\%}          & 75.0\%            & 70.3\% \\
 & MW in range       & \textbf{100.0\%}         & 80.0\%            & \underline{71.4\%} \\
 & RotB in range     & \textbf{100.0\%}         & 67.5\%            & 84.5\% \\
 & Uniqueness        & 65.2\%                   & \textbf{97.5\%}   & \underline{96.2\%} \\
\midrule
\multirow{6}{*}{BRD9-VHL}
 & SA score          & \textbf{1.94 $\pm$ 0.52} & - & 2.76 $\pm$ 0.57 \\
 & QED               & 0.457 $\pm$ 0.072        & - & \textbf{0.51 $\pm$ 0.12} \\
 & $\log P$ in range & 87.5\%                   & - & \textbf{99.6\%} \\
 & MW in range       & \textbf{100.0\%}         & - & 90.7\% \\
 & RotB in range     & \textbf{100.0\%}         & - & 95.0\% \\
 & Uniqueness        & \textbf{100.0\%}         & - & 96.8\% \\
\midrule
\multirow{6}{*}{CDK6-CRBN}
 & SA score          & \textbf{1.97 $\pm$ 0.44} & \underline{2.58 $\pm$ 0.79} & 2.83 $\pm$ 0.59 \\
 & QED               & 0.484 $\pm$ 0.078        & 0.513 $\pm$ 0.106 & \textbf{0.51 $\pm$ 0.12} \\
 & $\log P$ in range & 87.5\%                   & \textbf{91.9\%}   & 84.2\% \\
 & MW in range       & \textbf{100.0\%}         & \textbf{100.0\%}  & \textbf{100.0\%} \\
 & RotB in range     & 89.6\%                   & \textbf{97.3\%}   & 78.4\% \\
 & Uniqueness        & 77.1\%                   & \textbf{100.0\%}  & 96.5\% \\
\bottomrule
\end{tabular}
\label{tab:molecularprop}
\end{table}

\subsection{Physical Validation by Molecular Dynamics}
\label{subsec:md}

DegradeMaster is a learned predictor trained on overlapping databases. We
therefore evaluated four generated EGFR-VHL candidates, measuring
conformational 3D polar surface area in chloroform (a low-dielectric
membrane-interior mimic, where lower values indicate stronger
permeability-relevant polarity shielding) and the conformational penalty
$-kT\log f_{\mathrm{clash-free}}$ in water, where
$f_{\mathrm{clash-free}}$ is the fraction of conformers compatible with
ternary geometry against EGFR (PDB 4WKQ) and VHL (PDB 6GFY). We included
MS39, an experimentally validated cell-active EGFR-VHL degrader sharing our
warhead and E3 ligand, as a positive control, and a PEG4-linked congener as a
negative control in explicit-solvent molecular dynamics. We ran three
independent $50$\,ns simulations per compound in each solvent following
\citet{poongavanam2022protac}. Method details are in
Appendix~\ref{sec:md_appendix}.

\subsubsection{Molecular Dynamics Protocol}
\label{sec:md_appendix}

We ran all simulations in OpenMM 8.5.2 with OpenFF Sage 2.2.0 and AM1-BCC
charges for PROTACs, Amber ff14SB for proteins, and TIP3P water; chloroform
boxes were built with Packmol. We used a LangevinMiddle integrator with a
$2$\,fs timestep and $1$\,ps$^{-1}$ friction, PME electrostatics with a
$1.0$\,nm cutoff, and NPT equilibration. Analysis used RDKit and MDTraj.
Solvent-accessible 3D PSA is computed with Bondi radii and a $1.4$\,\AA{}
probe (Shrake-Rupley), matching the PyMOL convention; values are therefore
comparable to PyMOL-derived published ranges and read systematically lower
than VEGA-derived ones. We validated the 3D PSA implementation on
grazoprevir, recovering its published range (median $164$\,\AA$^2$, with
$92\%$ of conformers between $144$ and $205$\,\AA$^2$).

All four generated candidates occupied a conformational regime consistent
with the validated degrader. C1 and C2 had lower 3D PSA and lower ternary
penalties than MS39. The negative control had the highest values on both
metrics, supporting the sensitivity of the analysis to linker-dependent
differences. These simulations measured conformational properties and did not
establish cellular degradation. Full target-PROTAC-E3 ternary simulations
remain in progress.

\subsection{Simulated Compounds}

All six compounds share the Gefitinib-derived EGFR warhead and the VH032 VHL
binder and differ only in the linker (Table~\ref{tab:md}). SMILES
for all six are in the code repository.

\begin{table}[!ht]
\centering
\caption{Explicit-solvent MD for generated EGFR-VHL PROTACs against
experimental controls. MS39 is a validated cell-active EGFR-VHL degrader
\citep{cheng2020ms39}; PEG4 is a PEG4-linked congener (negative control).
Ternary penalty in kcal/mol.}
\label{tab:md}
\small
\begin{tabular}{lllcc}
\toprule
& \textbf{Linker} & \textbf{Role} & \textbf{3D PSA (\AA$^2$)} $\downarrow$ & \textbf{Penalty} $\downarrow$ \\
\midrule
\textbf{C1} & $n$-propylsuccinyl    & Generated & \textbf{140} & \textbf{2.2} \\
\textbf{C2} & allyl succinyl        & Generated & \textbf{141} & \textbf{1.9} \\
C3          & gem-dimethylsuccinate & Generated & 178 & \textbf{2.7} \\
C4          & butylmalonate         & Generated & 184 & \textbf{3.3} \\
MS39        & decanedioyl           & Positive  & 178 & 3.7 \\
PEG4        & PEG4                  & Negative  & 193 & 4.1 \\
\bottomrule
\end{tabular}
\end{table}

\FloatBarrier
\section{Algorithm}
\label{sec:algorithm}
\subsection{Preliminaries}
\label{subsec:preliminaries}

We introduce the mathematical objects required to define coupled discrete flows on structured molecular string spaces and to formulate objective-guided transport for molecular conjugate assembly. Throughout, $\mathcal{V}$ denotes a finite vocabulary of molecular string tokens (e.g., SMILES tokens), and $\mathcal{V}^{L}$ denotes the set of molecular strings of length $L$. In this work, peptides are represented directly as peptide SMILES strings, consistent with prior peptide generative frameworks such as PepTune, TR2-D2, and AReUReDi \citep{tang2025peptunenovogenerationtherapeutic, chen2025areurediannealedrectifiedupdates,tang2025tr2d2treesearchguided}. While we present notation for a single molecular string, subsequent sections consider structured states composed of multiple coupled molecular strings corresponding to different components of a conjugate.

\subsection{Molecular Strings as States in a Discrete Space}

A molecular object is represented as a discrete string state
\begin{equation}
    x = (x_{1}, \dots, x_{L}) \in \mathcal{V}^{L},
\end{equation}
where each token $x_i \in \mathcal{V}$ corresponds to a valid SMILES symbol. For peptide components, $x$ encodes a peptide SMILES string rather than a residue-level amino-acid sequence.

Transitions between molecular strings occur through local token-level edit operations. We define the edit neighborhood
\begin{equation}
    \mathcal{N}(x)
    :=
    \left\{
        x' \in \mathcal{V}^{L} \,:\, \text{single token edit}
    \right\},
\end{equation}
where a local edit corresponds to a token substitution. Extensions to insertions or deletions are straightforward and compatible with SMILES-based representations, but are omitted here for clarity.

\subsection{Reference Dynamics}

To define transport in discrete molecular string space, we introduce a simple reference process that captures uninformed local exploration. We consider a continuous-time Markov chain (CTMC) with generator $R_{0}$ defined by
\begin{equation}
    R_{0}(x, x')
    =
    \begin{cases}
        \frac{1}{|\mathcal{N}(x)|} & x' \in \mathcal{N}(x), \\[4pt]
        -\displaystyle\sum_{y \ne x} R_{0}(x, y) & x' = x.
    \end{cases}
\end{equation}
This reference dynamics defines a random walk over molecular strings that performs small, unbiased token edits and serves as a neutral baseline relative to which structured transport is learned.

\subsection{Discrete Flow}
Rather than modeling infinitesimal stochastic dynamics, we focus on learning \emph{discrete flows} that directly transport molecular string states between times. For $0 < t \le T$, a discrete flow is a function
\begin{equation}
    \Phi_{\theta}^{t \to T} : \mathcal{V}^{L} \to \mathcal{V}^{L},
\end{equation}
parameterized by $\theta$, which maps a molecular string at time $t$ to a molecular string at time $T$. These maps are intended to approximate the solution operator of an underlying probability flow, enabling one- or few-step generation without explicit simulation of a continuous-time process.

\subsection{Objective Functions}

To guide transport toward functionally relevant regions of molecular string space, we assume access to one or more objective functions
\begin{equation}
    S : \mathcal{V}^{L} \to \mathbb{R},
\end{equation}
which score molecular strings according to downstream criteria such as peptide-target binding, compatibility with conjugation chemistry, or chemical feasibility. Importantly, these objectives are evaluated on the \emph{current transported molecular state} and do not define a probability density. Instead, they provide directional guidance that biases transport toward desirable configurations.

% \subsection{Objective-Guided Updates}

% To incorporate objectives into transport, the learned flows are trained to prefer transitions that improve objective value while remaining compatible with the reference dynamics and flow-map consistency constraints. Given a current state $x$ at time $s$, candidate successors $x' \in \mathcal{N}(x)$ are evaluated based on changes in $S(x)$, and the flow is optimized to increase the likelihood of transitions that move in favorable directions.

\subsection{Generative Sampling}
After training, sampling proceeds by drawing an initial state $X_{0} \sim \mu_{0}$ from a simple molecular-string prior and applying the learned flow:
\begin{equation}
    X_{t}
    =
    \Phi_{\theta}^{t}(X_{0}),
    \qquad 0 < t \leq T.
\end{equation}
In the multistep regime, we choose a time grid $0 = t_0 < t_1 < \cdots < t_K = T$ and compose maps sequentially. Iterating this procedure from $0$ to $T$ yields samples from the learned objective-guided transport, producing valid molecular strings corresponding to peptides or assembled conjugates in a small number of steps.

\FloatBarrier
\subsection{Main Algorithm}
\begin{algorithm}[H]
\caption{\textbf{Exponential Tilting with Acceptance Criterion}}
\label{alg:ExpTiltSimple_PDC}
\begin{algorithmic}
  \STATE {\bfseries Input:} trained discrete flow $\Phi_{\theta^{\ast}}$, prior $\mu_0$ on $\mathcal{Z}$
  \STATE \hspace{2.7em} time horizon $T$, number of steps $K$
  \STATE \hspace{2.7em} guidance score $S(z)$, strength $\beta$
  \STATE Define time grid $0=t_0<t_1<\cdots<t_K=T$
  \STATE Sample $Z_0 \sim \mu_0$ s.t.\ $\mathbf{1}_{\mathrm{valid}}(Z_0)=1$; \ $Z^\ast \leftarrow Z_0$

  \FOR{$k=0$ {\bfseries to} $K-1$}
    \STATE Select random position $i$
    \STATE $p_{k+1} \leftarrow \Phi_{\theta^\ast}(Z_k)[\cdot,i]$ \algcomment{predicted distribution at $i$}

    \FOR{{\bfseries each} $z' \in \mathrm{TopK}(p_{k+1})$}
      \STATE $p'(z') \propto p_{k+1}(z') \,\exp(\beta S(z'))\, \mathbf{1}_{\mathrm{valid}}(z')$
      \STATE \algcomment{exponential tilting}
    \ENDFOR

    \STATE Sample $z^\dagger \sim p'$
    \STATE $Z' \leftarrow Z_k$ with position $i$ set to $z^\dagger$

    \IF{$S(Z') \ge S(Z_k)$} 
      \STATE \algcomment{acceptance criterion}
      \STATE $Z_{k+1} \leftarrow Z'$
      \IF{$S(Z') > S(Z^\ast)$}
        \STATE $Z^\ast \leftarrow Z'$
      \ENDIF
    \ELSE
      \STATE $Z_{k+1} \leftarrow Z_k$ \algcomment{reject, keep current}
    \ENDIF
  \ENDFOR

  \STATE {\bfseries return} $Z^\ast$ \algcomment{best molecular conjugate found}
\end{algorithmic}
\end{algorithm}

\section{Theoretical Proofs}
\label{sec:proofs}

We provide a minimal theoretical justification for the objective-guided sampling mechanism used in CupOFMoCA. In particular, we show that the exponential tilting rule employed during discrete flow sampling arises as the unique solution to a KL-regularized objective improvement problem under hard validity constraints.

\subsection{Notation and Setup}

We adopt the notation from the main text. Let $\mathcal{Z}$ denote the finite discrete state space of molecular conjugates. A discrete flow step is represented by a Markov kernel $K(\cdot \mid z)$ on $\mathcal{Z}$, where $K(z' \mid z)$ denotes the probability of proposing state $z'$ given current state $z$. Deterministic updates are recovered as the special case where $K(\cdot \mid z)$ is a point mass.

Validity is encoded by an indicator function $\mathbf{1}_{\mathrm{valid}}:\mathcal{Z}\to\{0,1\}$, which specifies whether a candidate state satisfies all chemical and representational constraints. Objective guidance is defined by a score function $S:\mathcal{Z}\to\mathbb{R}$ and a guidance strength $\beta>0$.

\subsection{Objective Guidance as KL-Regularized Exponential Tilting}

We show that exponential tilting is the unique optimal way to incorporate objective information into a discrete flow step while remaining close to a base kernel and enforcing hard validity constraints.

\begin{theorem}[Exponential tilting solves KL-regularized objective improvement]
\label{thm:exp_tilt}
Fix a base kernel $K(\cdot\mid z)$ on $\mathcal{Z}$ and a score function $S:\mathcal{Z}\to\mathbb{R}$. For a fixed input state $z$, consider the optimization problem
\begin{equation}
    \max_{Q}
    \left\{
        \mathbb{E}_{Z'\sim Q}\!\left[S(Z')\right]
        -
        \frac{1}{\beta}\,\mathrm{KL}\!\left(Q \,\Vert\, K(\cdot\mid z)\right)
    \right\}
    \quad
    \text{subject to}
    \quad
    Q(z') = 0 \;\; \text{if} \;\; \mathbf{1}_{\mathrm{valid}}(z') = 0,
    \label{eq:kl_reg_problem}
\end{equation}
where $\beta>0$. Then the unique maximizer is
\begin{equation}
    Q^{\star}_{\beta}(z' \mid z)
    =
    \frac{
        K(z'\mid z)\,\mathbf{1}_{\mathrm{valid}}(z')\,\exp\!\big(\beta S(z')\big)
    }{
        \sum_{y\in\mathcal{Z}}
        K(y\mid z)\,\mathbf{1}_{\mathrm{valid}}(y)\,\exp\!\big(\beta S(y)\big)
    }.
    \label{eq:guided_kernel}
\end{equation}
\end{theorem}

\begin{proof}
Let $\mathcal{Z}_{\mathrm{ok}} := \{z'\in\mathcal{Z} : \mathbf{1}_{\mathrm{valid}}(z') = 1\}$ and restrict $Q$ to be supported on $\mathcal{Z}_{\mathrm{ok}}$. The objective in \eqref{eq:kl_reg_problem} can be written as
\begin{equation}
    \sum_{z'\in\mathcal{Z}_{\mathrm{ok}}} Q(z')\,S(z')
    -
    \frac{1}{\beta}
    \sum_{z'\in\mathcal{Z}_{\mathrm{ok}}}
    Q(z') \log \frac{Q(z')}{K(z'\mid z)}.
\end{equation}
Introduce a Lagrange multiplier $\lambda$ enforcing $\sum_{z'\in\mathcal{Z}_{\mathrm{ok}}} Q(z') = 1$. Differentiating with respect to $Q(z')$ and setting the derivative to zero yields
\begin{equation}
    S(z')
    -
    \frac{1}{\beta}\!\left(\log Q(z') - \log K(z'\mid z) + 1\right)
    -
    \lambda
    = 0.
\end{equation}
Rearranging gives
\begin{equation}
    \log Q(z') = \log K(z'\mid z) + \beta S(z') + c,
\end{equation}
for a constant $c$, implying
\begin{equation}
    Q(z') \propto K(z'\mid z)\exp\!\big(\beta S(z')\big)
    \quad \text{on } \mathcal{Z}_{\mathrm{ok}}.
\end{equation}
Normalizing over $\mathcal{Z}_{\mathrm{ok}}$ yields \eqref{eq:guided_kernel}. The objective is strictly concave in $Q$, so the maximizer is unique.
\end{proof}

\begin{corollary}[Limits of guided kernels]
\label{cor:guided_limits}
Assume $K(\cdot\mid z)$ assigns positive mass to at least one valid state. Then:
(i) as $\beta\to 0$, $Q^{\star}_{\beta}(\cdot\mid z)$ converges to $K(\cdot\mid z)$ restricted and renormalized to valid states;  
(ii) as $\beta\to\infty$, $Q^{\star}_{\beta}(\cdot\mid z)$ concentrates on valid maximizers of $S$ within the support of $K(\cdot\mid z)$.
\end{corollary}

\begin{proof}
Both statements follow directly from standard properties of the log-sum-exp normalization in \eqref{eq:guided_kernel}.
\end{proof}

%%%%%%%%%%%%%%%%%%%%%%%%%%%%%%%%%%%%%%%%%%%%%%%%%%%%%%%%%%%%%%%%%%%%%%%%%%%%%%%
%%%%%%%%%%%%%%%%%%%%%%%%%%%%%%%%%%%%%%%%%%%%%%%%%%%%%%%%%%%%%%%%%%%%%%%%%%%%%%%

% NOTE: Replace "authorname2024" with the actual citation key for the paper you obtained these models from.
% Make sure this citation appears in your main text where you first mention using these models.
% This section describes models you are USING, not models you developed.

\section{PDC Model Architectures and Training Details}

\subsection{PepReDi Base Models}
\label{subsec:pepredi_base}

We adopt PepReDi \citep{chen2025areurediannealedrectifiedupdates} as the discrete-flow base model for the PDC instantiation of CupOFMoCA, and train two variants that differ only in vocabulary and training corpus: \textbf{AA-PepReDi} (canonical $20$ L-amino acids) and \textbf{UAA-PepReDi} (extended to D-amino acids and non-canonical residues). Both share architecture, optimizer, and rectification protocol; we describe the common pieces once and flag deltas where they apply.

\paragraph{Shared architecture and training.}
PepReDi is a discrete-flow model parameterized by a Diffusion Transformer applied in the MDLM regime \citep{peebles2023scalable, sahoo2024simple}. Per-token and per-timestep embeddings are produced by separate heads, summed, and propagated through stacked attention blocks with adaLN-modulated residual streams; a linear head emits per-position categorical logits. Both variants are trained with Adam ($\mathrm{lr}=10^{-4}$, weight decay $10^{-5}$; $2\times10^{-5}$ for rectified runs), a $10$-epoch warm-up, and cosine decay between $10^{-5}$ and the peak. Following ESM-2 \citep{lin2023evolutionary}, batches are formed by length-bucketing with a per-GPU token cap. We monitor conditional total correlation on a held-out subset (20 batches $\times$ 50 samples) and retain the checkpoint with the lowest value. At inference, both variants run $16$ flow steps at temperature $1.0$ to produce the prior $\mu_0$ used downstream by CupOFMoCA.

\paragraph{AA-PepReDi.}
The canonical variant uses the ESM-2-650M tokenizer \citep{lin2023evolutionary}, yielding a vocabulary of $24$ tokens ($20$ L-amino acids plus $4$ special symbols), a model dimension of $512$, $6$ DiT layers, $8$ attention heads, and a maximum sequence length of $100$. Training data combines all peptides from PepNN \citep{abdin2022pepnn} and BioLip2 \citep{zhang2024biolip2} with $6$-$49$-residue sequences from PPIRef \citep{bushuiev2023learning}, partitioned $80{:}10{:}10$ into train/validation/test. Optimization runs for $100$ epochs on a single NVIDIA GPU ($128$~GB) under SLURM.

\paragraph{UAA-PepReDi.}
To support D-amino acids, head-to-tail (N-C) and disulfide (S-S) cyclization, and non-canonical residues required for realistic PDC carriers, we extend the tokenizer to $53$ entries: the $4$ specials and $20$ L-residues from AA-PepReDi, $20$ lowercase D-residue tokens, and $9$ NCAAs available in our wet lab. Because no curated corpus combines all of these features, we synthesize training data with \texttt{p2smi} \citep{feller2025p2smi}, sweeping residue composition over a $5\times 4$ grid of NCAA fractions $\{0.1, 0.2, 0.3, 0.4, 0.5\}$ and D-residue fractions $\{0.1, 0.2, 0.3, 0.4\}$, with L-residues making up the remainder, yielding $20$ composition combinations spanning the conjugate-relevant chemical space.The resulting corpus contains 150K sequences partitioned $80{:}10{:}10$ into train/val/test. Architecture and optimization match AA-PepReDi except for the enlarged vocabulary.

PepReDi operates natively in amino-acid token space, but CupOFMoCA's objectives (binding affinity, conjugation validity) are defined on the assembled SMILES. Generated token sequences are therefore converted to peptide SMILES on the fly via three-letter codes \citep{feller2025p2smi}, applying the appropriate cyclization closure when relevant, and the resulting molecule is passed to the scoring stack described in Section~\ref{subsec:objectives}.

\subsection{SMILES Base Model}

We utilized the pre-trained SMILESReDi model from \citep{chen2025areurediannealedrectifiedupdates} as our peptide SMILES generation model. SMILESReDi is a \emph{rectified discrete flow} (ReDi) model \citep{yoo2025redi} built on top of discrete flow matching (DFM) \cite{gat2024discrete}, using a Diffusion Transformer (DiT) architecture as its backbone. The model operates within a masked discrete generation framework and learns probability paths between source and target distributions rather than simulating a forward diffusion process.

SMILES sequences are tokenized using the PeptideCLM-23M tokenizer \citep{feller2025peptide} and mapped to discrete tokens. Token embeddings and time-step embeddings are projected into continuous representations and processed through stacked DiT transformer blocks with Rotary Positional Embeddings (RoPE) and multi-head self-attention to capture long-range dependencies. The model outputs token-level transition probabilities via a final layer normalization followed by a linear projection.

\subsubsection{Time-Dependent Bond-Aware Probability Path}

A key feature of SMILESReDi is its bond-aware discrete probability path designed for peptide SMILES. Peptides contain a conserved backbone of alternating carbonyl and amide groups connected by peptide bonds, with variable side chains. To preserve this structure, SMILESReDi employs a time-dependent bond-aware probability path that transitions backbone bond tokens more slowly than side-chain tokens.

For each position $j$ with bond indicator $b_j \in \{0,1\}$, the time-$t$ marginal of the probability path is defined as
\begin{equation}
p_t(x_t^{(j)} \mid x_0^{(j)}, x_1^{(j)}) =
[b_j t^\gamma + (1 - b_j)t] \delta_{x_1^{(j)}}
+ [1 - b_j t^\gamma - (1 - b_j)t] \delta_{x_0^{(j)}},
\quad t \in [0,1],\ \gamma > 1,
\end{equation}
ensuring that backbone tokens ($b_j = 1$) transition more slowly along the discrete flow path than non-bond tokens.

\subsubsection{Training Configuration}

The SMILESReDi model was trained and rectified following the configuration reported in \citep{chen2025areurediannealedrectifiedupdates}, and we use the released pre-trained weights without further fine-tuning. The configuration is detailed in Table~\ref{tab:smilesredi_config}.

\begin{table}[h]
\centering
\caption{SMILESReDi training configuration.}
\label{tab:smilesredi_config}
\begin{tabular}{ll}
\toprule
\textbf{Parameter} & \textbf{Value} \\
\midrule
Hardware & 4$\times$A6000 NVIDIA RTX 6000 Ada GPU (48 GB VRAM) \\
Training epochs & 5 \\
Optimizer & Adam \\
Learning rate & $1 \times 10^{-4}$ \\
Learning rate schedule & Cosine decay with 10\% warmup \\
Minimum learning rate & $1 \times 10^{-5}$ \\
Model dimension & 768 \\
Transformer layers & 8 \\
Attention heads & 8 \\
Gradient clipping & 1.0 \\
Noising parameter $\gamma$ & 2.0 \\
\bottomrule
\end{tabular}
\end{table}

\subsection{Binding Affinity Score Model}

We employed the binding affinity prediction model from \citep{chen2025areurediannealedrectifiedupdates} to score protein-peptide binding affinity. This model uses a reciprocal attention transformer architecture that utilizes unpooled token-level representations from the ESM-2 650M protein language model \citep{lin2023evolutionary}. The architecture processes unpooled token embeddings from ESM-2 through convolutional layers followed by cross-attention layers, allowing the model to capture interaction patterns between protein and peptide sequences.

\subsubsection{Model Architecture}

The binding affinity predictor architecture is detailed in Table~\ref{tab:binding_affinity_arch}.

\begin{table}[h]
\centering
\caption{Binding affinity model architecture.}
\label{tab:binding_affinity_arch}
\begin{tabular}{ll}
\toprule
\textbf{Component} & \textbf{Configuration} \\
\midrule
Input & Unpooled token-level embeddings from ESM-2 650M \\
Convolutional layers & 3 initial CNN kernel layers (dimension 384) \\
Cross-attention layers & 4 layers (dimension 2048) \\
Prediction head & Shared prediction head (dimension 1024) \\
\bottomrule
\end{tabular}
\end{table}

\subsubsection{Training Details}

As reported in the original work, the binding affinity data were split into a 0.8/0.2 ratio for training and validation, maintaining similar affinity score distributions across splits. The authors employed OPTUNA \citep{akiba2019optunanextgenerationhyperparameteroptimization} for hyperparameter optimization, tracking validation correlation scores as the objective metric. The training configuration is presented in Table~\ref{tab:binding_affinity_training}.

\begin{table}[h]
\centering
\caption{Binding affinity model training configuration.}
\label{tab:binding_affinity_training}
\begin{tabular}{ll}
\toprule
\textbf{Parameter} & \textbf{Value} \\
\midrule
Training epochs & 50 \\
Learning rate & $3.84 \times 10^{-5}$ \\
Dropout rate & 0.15 \\
CNN layers & 3 initial layers (dimension 384) \\
Cross-attention layers & 4 layers (dimension 2048) \\
Prediction head dimension & 1024 \\
Train/validation split & 0.8/0.2 \\
Validation performance & 0.64 Spearman correlation \\
\bottomrule
\end{tabular}
\end{table}

\section{PDC Conjugation Validity Constraints}
\label{sec:conjugation_validity_constraints}

During the generative workflow, a candidate PDC is first validated by the \texttt{PDCAnalyzer} for general structural feasibility, then subjected to conjugation-specific checks by the builder's validation methods, and finally scored for binding affinity and other therapeutic properties by downstream prediction models.

To ensure chemically valid PDCs from de-novo designed peptide components, we developed a specialized conjugation and validation pipeline using RDKit \citep{landrum2025rdkit}. Both the Aoa-Dau and SSCpt conjugation builders implement multi-layered validation comprising SMARTS-based pattern recognition, structural integrity checks, and valency analysis.

\subsection{Aminooxyacetic acid - Daunorubicin Conjugates (Aoa-Dau)}
\subsubsection{Conjugation Reaction}

Among the 21 PDC entries in PDCdb sharing the same peptide scaffold, Aminooxyacetic acid (Aoa) linker, and Daunorubicin (Dau) payload, four distinct conjugation strategies are represented, including direct conjugation via Lys side chain $\varepsilon$-amine, spacer-mediated conjugation, ranched conjugates and bis-conjugates at two Lys sites ( additional Lys introduced via peptide modification).

We adopted the direct conjugation approach via Lys side chain \ch{NH2}, which is the most prevalent strategy among the database entries. Our implementation follows the architecture of PDC\_00208 (GnRH-III-[8Lys(Dau=Aoa)]) as a reference template. To ensure site-selectivity, conjugation was restricted to the $\varepsilon$-amine of Lys side chains; the $\alpha$-amine at the N-terminus was excluded as a potential reaction site.

\paragraph{Reaction Scheme.}
Conjugation proceeds via amide bond formation between the Lysine $\varepsilon$-amine (side chain \ch{-NH2}) on valid peptide, and the carboxyl group on Aoa (pre-conjugated to Dau via oxime bond):
\begin{equation}
    \underbrace{\text{Peptide-Lys}}_{\text{targeting moiety}}\text{-}\ch{NH2} 
    \;+\; 
    \ch{HOOC}\text{-}\underbrace{\text{Aoa-Dau}}_{\text{linker-payload}}
    \;\xrightarrow{\text{amide coupling}}\;
    \text{Peptide-Lys-}\ch{NH-CO}\text{-Aoa-Dau}
    \;+\;
    \ch{H2O}
\end{equation}

\paragraph{Molecular Architecture.}
The complete PDC structure is represented schematically as:
\begin{equation}
    \boxed{
    \underbrace{\text{Peptide-Lys}}_{\text{De-novo Designed}}
    \text{-}
    \underbrace{\ch{NH-C(=O)}}_{\substack{\text{Amide} \\ \text{(formed)}}}
    \text{-}
    \underbrace{\ch{CH2-O-N=C(CH3)}}_{\substack{\text{Aoa oxime} \\ \text{(pre-formed)}}}
    \text{-}
    \underbrace{\text{Dau}}_{\text{Payload}}
    }
\end{equation}

\subsubsection{Attachment Site Compatibility Checks}

The resulting PDC candidates need to pass all the checks in Table~\ref{tab:aoadau-validation}  for compatibility on the attachment site. For valency requirements, the amide nitrogen at the conjugation site must satisfy one of the following: a secondary amide, where N is bonded to 2 heavy atoms (one carbonyl C) and 1 H, or a tertiary amide, where N is bonded to 3 heavy atoms (one carbonyl C) and 0 H.

\begin{table}[h]
\centering
\scriptsize
\renewcommand{\arraystretch}{1.4}
\begin{tabularx}{\textwidth}{>{\ttfamily\raggedright\arraybackslash}p{4.5cm} >{\ttfamily\raggedright\arraybackslash}X >{\raggedright\arraybackslash}p{3cm}}
\hline
\textnormal{\textbf{Method}} & \textnormal{\textbf{SMARTS Pattern}} & \textbf{Purpose} \\
\hline
has\_aoa\_amide\_conjugation() & 
[NX3][CX3](=[OX1])[CH2][OX2][NX2]=[C] & 
Confirms amide-\ch{CH2}-O-oxime linkage \\

has\_oxime\_bond() & 
[C]-[O]-[N]=[C] & 
Verifies Aoa-Dau linkage intact \\

has\_unreacted\_aoa\_carboxylic\_acid() & 
[CX3](=[OX1])[OX2H1][CH2][OX2][NX2]=[C] & 
Failure indicator: Free \ch{HOOC-CH2-O-N=C} \\

find\_amide\_conjugation\_sites() & 
[NX3:1][CX3:2](=[OX1])[CH2][OX2][NX2]=[C] & 
Locates conjugation sites \\

check\_valency\_at\_conjugation\_site() & 
\textnormal{Structural analysis} & 
Validates N valency \\
\hline
\end{tabularx}
\caption{AoaDau attachment site compatibility checks.}
\label{tab:aoadau-validation}
\end{table}

\subsection{Disulfide-Camptothecin Conjugates (SSCpt)}
\subsubsection{Conjugation Reaction}

 Seven PDCdb entries share the same ERBB2-targeting peptide scaffold, disulfide linker, and camptothecin (CPT) payload. Their conjugation architectures include dimers with internal attachment and monomers with N-terminal or Lys side-chain attachment. Three linker-to-drug chemistries are observed: ester, carbonate, and carbamate.

We adopted the monomer architecture with Lys side chain \ch{NH2} conjugation, following PDC\_00294 (Z7, ester) and PDC\_00295 (Z8, carbonate) as reference templates. To ensure site-selectivity, conjugation was restricted to the $\varepsilon$-amine of Lys side chains; the $\alpha$-amine at the N-terminus was excluded as a potential reaction site.

\paragraph{Reaction Scheme.}
Conjugation proceeds via amide bond formation between the Lysine $\varepsilon$-amine (side chain \ch{-NH2}) on the valid peptide, and the carboxyl group on Cys (pre-conjugated to CPT via disulfide-linked spacer):
\begin{equation}
\begin{aligned}
&\underbrace{\text{Peptide-Lys}}_{\text{targeting moiety}}\text{-}\ch{NH2}
\;+\;
\ch{HOOC}\text{-}\underbrace{\text{Cys-SS-Spacer-CPT}}_{\text{linker-payload}} \\
&\qquad\xrightarrow{\text{amide coupling}}\;
\text{Peptide-Lys-}\ch{NH-CO}\text{-Cys-SS-Spacer-CPT}
\;+\;\ch{H2O}
\end{aligned}
\end{equation}

\paragraph{Molecular Architecture.}
The complete PDC structure is represented schematically as:
\begin{equation}
    \boxed{
    \underbrace{\text{Peptide-Lys}}_{\text{De-novo Designed}}
    \text{-}
    \underbrace{\ch{NH-C(=O)}}_{\substack{\text{Amide} \\ \text{(formed)}}}
    \text{-Cys-}
    \underbrace{\ch{S-S}}_{\substack{\text{Disulfide} \\ \text{(cleavable)}}}
    \text{-Spacer-}
    \underbrace{\text{CPT}}_{\text{Payload}}
    }
\end{equation}

\noindent
Two linker-to-drug chemistries are supported, differing in the bond connecting the spacer to Camptothecin:
\begin{itemize}
    \item Ester (Z7): \ch{Cys-S-S-CH2-CH2-C(=O)-O-CPT}
    \item Carbonate (Z8): \ch{Cys-S-S-CH2-CH2-O-C(=O)-O-CPT}
\end{itemize}

\subsubsection{Attachment Site Compatibility Checks}
The resulting PDC candidates need to pass all the checks in Table~\ref{tab:sscpt-validation}  for compatibility on the attachment site. For valency requirements, the conjugation site must satisfy the following connectivity: Lys nitrogen with 3 bonds (NH connected to amide), carbonyl carbon with 3 bonds (C=O and two single bonds), and Cys $\alpha$-carbon with 4 bonds (tetrahedral geometry).

\begin{table}[h]
\centering
\scriptsize
\renewcommand{\arraystretch}{1.4}
\begin{tabularx}{\textwidth}{>{\ttfamily\raggedright\arraybackslash}p{4.5cm} >{\ttfamily\raggedright\arraybackslash}X >{\raggedright\arraybackslash}p{2cm}}
\hline
\textnormal{\textbf{Method}} & \textnormal{\textbf{SMARTS Pattern}} & \textbf{Purpose} \\
\hline
has\_lys\_cys\_amide\_conjugation() & 
[NX3][CX3](=[OX1])[CH]([CH2][SX2][SX2])[NX3] & 
Confirms Lys-$\varepsilon$-\ch{N-C(=O)}-C$_\alpha$(Cys) pattern \\

has\_disulfide\_bond() & 
[SX2]-[SX2] & 
Verifies S-S linker intact \\

has\_unreacted\_cys\_carboxylic\_acid() & 
[NH2][CH]([CH2][SX2][SX2])[CX3](=[OX1])[OX2H1] & 
Failure indicator: Free Cys-COOH \\

find\_amide\_conjugation\_sites() & 
[NX3][CX3](=[OX1])[CH]([CH2][SX2][SX2])[NX3] & 
Locates conjugation sites \\

check\_valency\_at\_conjugation\_site() & 
\textnormal{Structural analysis} & 
Validates atom connectivity \\
\hline
\end{tabularx}
\caption{SSCpt attachment site compatibility checks.}
\label{tab:sscpt-validation}
\end{table}

\subsection{Disulfanylpropanoic Acid - SN-38 Conjugates (DSPA-SN38)}
\subsubsection{Conjugation Reaction}

Among the 9 PDC entries in PDCdb sharing the NRP1-targeting (RGD-motif) peptide
scaffold family, 3-Disulfanylpropanoic Acid (DSPA, LIN00149) linker, and
7-Ethyl-10-hydroxycamptothecin (SN-38) payload, two distinct linker-to-drug
attachment positions are represented: ester at the C-10 phenolic hydroxyl
(most prevalent, e.g., PDC\_02099) and ester at the C-20 tertiary aliphatic
hydroxyl (e.g., PDC\_02112). All entries adopt a monomer architecture
employing a cleavable disulfide bond as the peptide-to-linker junction.

We adopted the monomer architecture with disulfide conjugation through the
peptide cysteine thiol, following PDC\_02099 (SN38(10)-CRGDK) and PDC\_02112
(SN38(20)-CRGDK) as reference templates. To ensure site-selectivity,
conjugation was restricted to free Cys side-chain thiol (\ch{-SH}) groups;
pre-existing disulfides and other sulfur-containing motifs were excluded as
potential reaction sites. Both C-10 and C-20 ester variants are supported
and pooled by default.

\paragraph{Reaction Scheme.}
Conjugation proceeds via disulfide bond formation between the cysteine thiol
(side chain \ch{-SH}) on the valid peptide, and the free terminal thiol on
DSPA (pre-conjugated to SN-38 via an ester bond):
\begin{equation}
\begin{aligned}
&\underbrace{\text{Peptide-Cys}}_{\text{targeting moiety}}\text{-}\ch{SH}
\;+\;
\ch{HS}\text{-}\underbrace{\text{DSPA-SN38}}_{\text{linker-payload}} \\
&\qquad\xrightarrow{\text{thiol-disulfide exchange}}\;
\text{Peptide-Cys-}\ch{S-S}\text{-DSPA-SN38}
\;+\;\ch{2 H}
\end{aligned}
\end{equation}

\paragraph{Molecular Architecture.}
The complete PDC structure is represented schematically as:
\begin{equation}
    \boxed{
    \underbrace{\text{Peptide-Cys}}_{\text{De-novo Designed}}
    \text{-}
    \underbrace{\ch{S-S}}_{\substack{\text{Disulfide} \\ \text{(formed, cleavable)}}}
    \text{-}
    \underbrace{\ch{CH2-CH2-C(=O)-O}}_{\substack{\text{DSPA ester} \\ \text{(pre-formed)}}}
    \text{-}
    \underbrace{\text{SN-38}}_{\text{Payload}}
    }
\end{equation}

\noindent
Two ester attachment positions on SN-38 are supported, differing in the
hydroxyl that anchors the DSPA propanoate:
\begin{itemize}
    \item C-10 (phenolic OH, default):
        \ch{Cys-S-S-CH2-CH2-C(=O)-O-SN38(10)}
    \item C-20 (tertiary aliphatic OH):
        \ch{Cys-S-S-CH2-CH2-C(=O)-O-SN38(20)}
\end{itemize}

\subsubsection{Attachment Site Compatibility Checks}

The resulting PDC candidates must pass all checks in
Table~\ref{tab:dspasn38-validation} for compatibility on the attachment
site. For valency requirements, both sulfur atoms in the newly formed
disulfide must adopt an \texttt{SX2} configuration (each bonded to exactly
one carbon and one sulfur, with zero residual hydrogens), the carbonyl
carbon of the linker-drug ester must remain \texttt{CX3} (sp$^2$, bonded
to an alkyl carbon, a doubly-bonded oxygen, and the ester oxygen), and no
free thiol (\ch{-SH}) may persist on either the peptide or the linker
after coupling.

\begin{table}[h]
\centering
\scriptsize
\renewcommand{\arraystretch}{1.4}
\begin{tabularx}{\textwidth}{>{\ttfamily\raggedright\arraybackslash}p{4.7cm} >{\ttfamily\raggedright\arraybackslash}X >{\raggedright\arraybackslash}p{4.8cm}}
\hline
\textnormal{\textbf{Method}} & \textnormal{\textbf{SMARTS Pattern}} & \textbf{Purpose} \\
\hline
has\_disulfide\_bond() &
[SX2]-[SX2] &
Confirms Cys-S-S-linker bond formed \\

has\_ester\_bond\_to\_drug() &
[\#6][CX3](=[OX1])[OX2][\#6] &
Verifies DSPA-SN38 ester linkage intact \\

has\_sn38\_core() &
c1ccc2ncc3C(=O)OCc3c2c1 &
Verifies SN-38 camptothecin scaffold preserved \\

has\_free\_thiol() &
\textnormal{Atom-level scan (S, $\geq$1\,H, degree\,1)} &
Failure indicator: unreacted \ch{-SH} remains \\

find\_conjugation\_sites() / find\_peptide\_cysteines() &
\textnormal{Atom iteration (S, $\geq$1\,H, degree\,1)} &
Locates free Cys thiols on peptide \\

check\_attachment\_site\_compatibility() &
\textnormal{Structural analysis} &
Validates disulfide / no-free-thiol / ester jointly \\
\hline
\end{tabularx}
\caption{DSPA-SN38 attachment site compatibility checks.}
\label{tab:dspasn38-validation}
\end{table}

\subsection{PDC Analyzer}
While the builder classes encapsulate linker-payload and conjugation reaction specific validation logic, the \texttt{PDCAnalyzer} remains conjugation-agnostic and can be applied to novel PDC designs under different conjugation reaction and / or linker-payload components

\subsubsection{Hard Constraint Validation}
The analyzer enforces two categories of hard constraints, both of which must be satisfied for a PDC to be considered structurally valid. Violation of any hard constraint results in immediate rejection of the candidate molecule.

\paragraph{Linker Feasibility Constraints.}
Linker regions in PDCs are particularly susceptible to structural infeasibility. The analyzer addresses two primary failure modes:

\textit{Ring-closure constraints} are evaluated by identifying rings that are likely part of the linker rather than the peptide backbone or drug payload. Aromatic rings are excluded from linker classification as they typically belong to the drug core. The analyzer specifically searches for succinimide rings (five-membered rings containing nitrogen and two carbonyl groups), which commonly form from maleimide-based conjugation after thiol addition. For non-aromatic rings, size-based filtering is applied: three-membered rings are checked for sp$^2$ carbons participating in C=C double bonds (which introduce severe angle strain), four-membered rings are rejected if they contain more than two sp$^2$ centers, and rings larger than twelve atoms are flagged as synthetically implausible linker components.

\textit{Long carbon chain detection} identifies excessively long aliphatic chains that would be synthetically challenging and prone to conformational heterogeneity. The analyzer uses a SMARTS pattern matching sixteen consecutive CH$_2$ groups, rejecting any PDC containing such chains. 

\paragraph{Representational Validity Constraints.}
These constraints ensure that the generated SMILES string represents a chemically valid and processable molecule. Four checks are performed sequentially: (1) RDKit sanitization; (2) round-trip SMILES conversion, where the molecule is converted to SMILES and re-parsed to check if reconstruction could be conducted; (3) chemistry problem detection using RDKit's diagnostic tools, specifically checking for \texttt{AtomValenceException} and \texttt{KekulizeException} errors that indicate fundamental chemical inconsistencies; and (4) 2D coordinate generation, which serves as a final structural coherence test. Failure at any stage indicates that the molecule, while potentially representable as a string, does not correspond to a valid chemical structure.

\subsubsection{Physicochemical Properties}

Beyond binary validity assessment, the analyzer can compute eight molecular descriptors that are relevant for drug-likeness and pharmacokinetic profiling. These properties are toggled via boolean flags and include: molecular formula, topological polar surface area (TPSA), molecular complexity (BertzCT index), lipophilicity (XLogP via Wildman-Crippen method), heavy atom count, hydrogen bond donors and acceptors, and rotatable bond count. 

\section{PDC Sampling Implementation}
Table~\ref{tab:notation_pdc} defines the notation, and Tables~\ref{tab:phase1_config} and~\ref{tab:phase2_config} summarize the two sampling phases.

\subsection{Notation}

\begin{table}[!ht]
\centering
\caption{Notation used in the algorithm description.}
\label{tab:notation_pdc}
\begin{tabular}{cl}
\toprule
\textbf{Symbol} & \textbf{Description} \\
\midrule
$\Phi_{\theta^*}$ & Pretrained discrete flow model with parameters $\theta^*$ \\
$\mathcal{Z}$ & Discrete token space (SMILES vocabulary) \\
$\mu_0$ & Prior distribution (uniform over $\mathcal{Z}^L$) \\
$N$ & Number of flow steps for initial generation \\
$K$ & Number of guided optimization steps \\
$S(z)$ & Guidance score function \\
$\beta$ & Guidance strength (temperature parameter) \\
$\mathbf{1}_{\text{valid}}(z)$ & Validity indicator function \\
$Z^*$ & Best sequence found during optimization \\
$L$ & Sequence length \\
\bottomrule
\end{tabular}
\end{table}

\subsection{Two-Phase Generation Framework}

Following \citet{chen2025areurediannealedrectifiedupdates}, our approach separates generation into two phases, each serving a specific purpose. This design decouples \emph{generation} (producing valid molecules) from \emph{optimization} (improving target properties), allowing each phase to be tuned independently.

\subsubsection{Phase 1: Initial Sequence Generation}

The first phase rapidly generates an initial sequence $Z_0$ using unguided discrete flow matching. Starting from random noise, we iteratively denoise through the learned flow to obtain a molecular sequence satisfying the validity constraints. Running $N=16$ steps provides sufficient denoising while remaining computationally efficient, and ensures the initial sequence lies on the data manifold. When the \texttt{require\_valid\_x0} flag is set, we regenerate $Z_0$ until the final sequence in this phase satisfies $\mathbf{1}_{\mathrm{valid}}(Z_0) = 1$, enforcing that the optimization phase starts from a feasible point in the constraint manifold. This was adapted from the sampling process in \citet{chen2025areurediannealedrectifiedupdates}, which was originally used to guarantee SMILES peptide validity.

\begin{table}[!ht]
\centering
\caption{Phase 1 (initial generation) parameters.}
\label{tab:phase1_config}
\begin{tabular}{lcp{6cm}}
\toprule
\textbf{Parameter} & \textbf{Value} & \textbf{Description} \\
\midrule
Flow Steps ($N$) & 16 & Number of denoising steps \\
Temperature & 1.0 & Softmax temperature for predictions \\
\texttt{require\_valid\_x0} & Enabled & Regenerate until $\mathbf{1}_{\text{valid}}(Z_0) = 1$ \\
\texttt{max\_x0\_retries} & 10 & Maximum regeneration attempts \\
\bottomrule
\end{tabular}
\end{table}

\subsubsection{Phase 2: Guided Optimization}

The second phase refines $Z_0$ through $K=128$ steps of guided optimization using exponential tilting with an acceptance criterion, as in Algorithm~\ref{alg:ExpTiltSimple_PDC}. At each step, we select a random position, compute a tilted distribution over candidate tokens, and accept the proposal only if it improves (or maintains) the guidance score. This allows fine-grained control through local modifications, while the acceptance criterion prevents the optimization from degrading sequence quality.

\paragraph{Cascaded validity evaluation for residue-token representations.}
For AA- and UAA-tokenized generation, we use a compute-bounded cascade to avoid performing the expensive assembled-PDC validity check for every candidate proposal. At each guided update, let $\mathcal{C}=\operatorname{TopK}(p_{k+1})$ denote the candidate set, with $|\mathcal{C}|\leq K$ and $K=200$. We first decode each candidate into a peptide SMILES string and apply inexpensive checks for peptide validity and the presence of a compatible conjugation site. Binding affinity is evaluated for all candidates, after which the $\kappa=8$ highest-affinity candidates among those passing the prefilter are retained. Full linker--payload conjugation and assembled-PDC validation are performed
only for these retained candidates. Candidates that fail the prefilter or are not retained are assigned a zero validity indicator in the exponential tilt. SMILES tokenized generation does not use this cascade and applies the full scoring and validity procedure to all candidates in $\mathcal{C}$. 

\paragraph{Log-Space Computation.} For numerical stability, we compute the tilted distribution in log-space:
\begin{align}
\tilde{p}(z') &= \log p_{k}(z') + \beta \cdot S(z') + \log \mathbf{1}_{\mathrm{valid}}(z') \\
\tilde{p}_{\mathrm{stable}}(z') &= \tilde{p}(z') - \max_{z''} \tilde{p}(z'') \\
p'(z') &= \frac{\exp(\tilde{p}_{\mathrm{stable}}(z'))}{\sum_{z''} \exp(\tilde{p}_{\mathrm{stable}}(z''))}
\end{align}

\paragraph{Fallback Behavior.}

If exponential tilting produces invalid probabilities (due to all candidates being invalid or numerical instability), we fall back to sampling from the base distribution $p_k$.

\paragraph{Acceptance Criterion.} After sampling $z^\dagger \sim p'$ and constructing the candidate sequence $Z'$, we apply an acceptance criterion:
\begin{equation}
Z_{k+1} = \begin{cases} 
Z' & \text{if } S(Z') \geq S(Z_k) \\ 
Z_k & \text{otherwise}
\end{cases}
\label{eq:acceptance}
\end{equation}
This ensures the score never decreases, providing monotonic improvement guarantees. Combined with tracking $Z^*$, this guarantees we return the best sequence encountered during optimization.

\begin{table}[!ht]
\centering
\caption{Phase 2 (guided optimization) parameters.}
\label{tab:phase2_config}
\begin{tabular}{lcp{6cm}}
\toprule
\textbf{Parameter} & \textbf{Value} & \textbf{Description} \\
\midrule
Optimization Steps ($K$) & 128 & Number of guided refinement steps \\
Guidance Strength ($\beta$) & 50.0 & Exponential tilting temperature \\
Sequence Length ($L$) & 80, 100 & Generated sequence length (tokens) \\
Top-$p$ & 1.0 & Nucleus sampling threshold \\
Top-$K$ & 200 & Maximum candidates per position \\
Cascade size ($\kappa$)
& 8 (AA/UAA); not used (SMILES)
& Candidates receiving full PDC validation \\
Position Selection & Random & Uniform random position each step \\
\texttt{use\_acceptance} & Enabled & Reject proposals with $S(Z') < S(Z_k)$ \\
\bottomrule
\end{tabular}
\end{table}

\paragraph{Token Filtering.}

Invalid tokens are masked during both phases. The filtered indices $\{0, 1, 2, 3, 4, 585, 586\}$ correspond to special tokens: \texttt{[PAD]}, \texttt{[UNK]}, \texttt{[CLS]}, \texttt{[SEP]}, \texttt{[MASK]}, and reserved indices.

\subsection{Computational Requirements}
\label{app:compute}

Sampling experiments were each conducted on a single A6000 NVIDIA GPU with 48GB of VRAM. For each sample, Phase 1 (16 flow steps) completes in under 1 minute; Phase 2 (128 optimization steps) takes 5-10 minutes depending on the sequence length and candidate size.

\newpage
\section{PROTAC Base Model and Scorer Architectures}

\subsection{DeFoG Base Model}
\label{sec:basemodel}

We use the pre-trained DeFoG discrete flow matching model
\citep{qin2025defog} as our base PROTAC linker generator. DeFoG operates
directly on molecular graphs rather than SMILES strings, representing nodes
(atoms) and edges (bonds) as separate discrete random variables and learning a
continuous-time discrete flow between a structured prior and the data
distribution. The model uses a graph transformer architecture with separate
node and edge update streams. We fine-tuned the released GuacaMol checkpoint of DeFoG
for $100$ epochs on $2{,}070$ PROTAC-DB linker graphs ($258$ validation, $260$
test), adapting the prior to the PROTAC-linker distribution. The linker
subgraph is the only region the flow modifies, with warhead and E3-ligand
subgraphs held fixed throughout sampling (Section~\ref{subsec:flow_maps}).

\begin{table}[h]
\centering
\caption{Effect of PROTAC-specific fine-tuning of the DeFoG base model. Mean
$\pm$ std over three seeds, $100$ generations per seed.}
\label{tab:defog_finetune}
\small
\begin{tabular}{llcc}
\toprule
\textbf{System} & \textbf{Metric} & \textbf{Base} & \textbf{Fine-tuned} \\
\midrule
\multirow{3}{*}{EGFR-VHL}
 & Assembly validity (\%) & 78.7 $\pm$ 5.4 & \textbf{85.0 $\pm$ 3.6} \\
 & SA $\downarrow$        & 3.6 $\pm$ 0.2  & \textbf{2.8 $\pm$ 0.0} \\
 & MW in range (\%)       & 48.7 $\pm$ 2.5 & \textbf{94.7 $\pm$ 1.9} \\
\midrule
\multirow{3}{*}{BRD9-VHL}
 & Assembly validity (\%) & 76.7 $\pm$ 3.3 & \textbf{82.3 $\pm$ 4.0} \\
 & SA $\downarrow$        & 3.6 $\pm$ 0.1  & \textbf{2.8 $\pm$ 0.0} \\
 & MW in range (\%)       & 99.7 $\pm$ 0.5 & 100.0 $\pm$ 0.0 \\
\midrule
\multirow{3}{*}{CDK6-CRBN}
 & Assembly validity (\%) & 78.3 $\pm$ 5.3 & \textbf{83.0 $\pm$ 4.2} \\
 & SA $\downarrow$        & 3.6 $\pm$ 0.0  & \textbf{2.8 $\pm$ 0.0} \\
 & MW in range (\%)       & 100.0 $\pm$ 0.0 & 100.0 $\pm$ 0.0 \\
\bottomrule
\end{tabular}
\end{table}

\subsection{DegradeMaster Activity Predictor}

For evaluation in Table ~\ref{tab:validity_degradmaster} we use
DegradeMaster \citep{liu2025degrademaster}, a neural
classifier trained to predict whether a candidate PROTAC induces
target degradation. DegradeMaster takes the assembled PROTAC
SMILES as input and returns a probability in $[0,1]$. We use the
released pre-trained model and report mean predicted probability
along with the fraction exceeding thresholds $0.3, 0.5, 0.7$.
DegradeMaster is used solely for evaluation in this work; it is
not used as a guidance objective during sampling.

We note that DegradeMaster and our fine-tuned generator both draw on PROTAC
databases, so DegradeMaster constitutes an in-domain evaluation rather than a
fully independent one. The memorization analysis
(Table~\ref{tab:diversity_memorization}) rules out direct reproduction of
training molecules, and the molecular-dynamics analysis
(Section~\ref{subsec:md}) provides an evaluation that does not depend on the
shared database.

\section{PROTAC Validity Constraints}
\label{sec:validity_appendix}

We detail the implementation of the validity classifier
$\mathbf{1}_{\mathrm{valid}}$ and the soft connectivity scorer
$S_3$ from Section~\ref{subsec:objectives_protacs}. All checks are
implemented in RDKit \citep{landrum2025rdkit}.

\subsection{Hard Validity Gate}

A generated PROTAC is accepted as valid only if all three checks
in Table~\ref{tab:hard_validity} pass. Failure on any check
returns $\mathbf{1}_{\mathrm{valid}}(z) = 0$, which masks the
candidate to zero probability under the tilted distribution
during sampling.

\begin{table}[h]
\centering
\scriptsize
\renewcommand{\arraystretch}{1.4}
\begin{tabularx}{\textwidth}{>{\raggedright\arraybackslash}p{4cm} >{\ttfamily\raggedright\arraybackslash}X >{\raggedright\arraybackslash}p{4.5cm}}
\toprule
\textnormal{\textbf{Check}} & \textnormal{\textbf{Pattern / Procedure}} & \textbf{Failure mode} \\
\midrule
Representational validity &
\textnormal{\texttt{Chem.SanitizeMol(mol)} succeeds AND
\texttt{MolFromSmiles(MolToSmiles(mol))} round-trips} &
Bad valences or aromaticity; molecule cannot be persisted \\
\addlinespace
Liability filter &
[NX3][NX3], [\#7]=[\#7], [N+][O-], [\#7]-[OX2], [NX3]C(=[OX1])[OX2H1] &
Hydrazide, azo, $N$-oxide, $N$-$O$ bond, and carbamate patterns associated with instability, toxicity, or failure of amide assembly \\
\addlinespace
Reactive termini count $\ge 2$ &
\textnormal{Count of} [\#6][CX3](=[OX1])[OX2H1] \textnormal{(C-COOH) +}
[\#6][CX3H1](=[OX1]) \textnormal{(C-CHO) $\ge 2$} &
Fewer than two reactive termini; warhead/E3 amide-coupling assembly impossible \\
\bottomrule
\end{tabularx}
\caption{Hard validity gate checks. All three must pass for
$\mathbf{1}_{\mathrm{valid}}(z) = 1$.}
\label{tab:hard_validity}
\end{table}

\subsection{Soft Connectivity Scorer}

The connectivity scorer $S_3$ is computed even on states with
$\mathbf{1}_{\mathrm{valid}}(z) = 0$, providing a gradient toward
single-component linkers during early sampling. Let $|\text{frag}|$
denote the number of connected fragments in the assembled graph,
$|\text{largest}|$ the size of the largest fragment, and
$|\text{total}|$ the total atom count. Then
\begin{equation}
    S_3(z) =
    \begin{cases}
        1.0 & \text{if } |\text{frag}| = 1, \\
        \dfrac{|\text{largest}|}{|\text{total}| \cdot |\text{frag}|}
        & \text{otherwise}.
    \end{cases}
\end{equation}
The score is halved if no path exists through the linker subgraph
between the warhead anchor $a^{\mathrm{wh}}$ and the E3-ligand
anchor $a^{\mathrm{e3}}$.

\subsection{Termini Conventions for Benchmark Systems}

Each PROTAC system in our experiments specifies fixed warhead and
E3-ligand anchor atoms at which amide-coupling assembly occurs.
The conventions used are listed in
Table~\ref{tab:termini_conventions}.

\begin{table}[h]
\centering
\small
\begin{tabular}{lll}
\toprule
\textbf{System} & \textbf{Warhead anchor} & \textbf{E3 anchor} \\
\midrule
EGFR-VHL  & Gefitinib-piperazine NH       & VH032 NH$_2$ (tert-leucine) \\
BRD9-VHL  & BI-7273-piperazine NH         & VH032 NH$_2$ (tert-leucine) \\
CDK6-CRBN & Palbociclib-piperazine NH     & Pomalidomide aryl-NH$_2$ \\
\bottomrule
\end{tabular}
\caption{Fixed termini conventions for benchmark PROTAC systems.}
\label{tab:termini_conventions}
\end{table}

\section{PROTAC Sampling Implementation}
\label{sec:sampling_appendix}

\subsection{Algorithm}

\begin{algorithm}[!tb]
\caption{\textbf{Exponential Tilting with Warmup and Best-So-Far Tracking}}
\label{alg:ExpTiltSimple}
\setlength{\itemsep}{0pt}
\setlength{\parskip}{0pt}
\begin{algorithmic}
  \STATE {\bfseries Input:} trained discrete flow $\Phi_{\theta^{\ast}}$, prior $\mu_0$ on $\mathcal{Z}$
  \STATE \hspace{2.7em} time horizon $T$, total steps $K$, warmup steps $K_{\mathrm{warm}}$
  \STATE \hspace{2.7em} guidance score $S(z)$, strength $\beta$
  \STATE Define time grid $0=t_0<t_1<\cdots<t_K=T$
  \STATE Sample $Z_0 \sim \mu_0$ s.t.\ $\mathbf{1}_{\mathrm{valid}}(Z_0)=1$; \ $Z^\ast \leftarrow Z_0$

  \STATE \algcomment{Warmup: unguided sampling from the pre-trained flow}
  \FOR{$k=0$ {\bfseries to} $K_{\mathrm{warm}}-1$}
    \STATE Select a free linker site $i$
    \STATE $p_{k+1} \leftarrow \Phi_{\theta^\ast}(Z_k)[\cdot, i]$
    \STATE Sample $z^\dagger \sim p_{k+1}$ \algcomment{no tilting, no acceptance}
    \STATE $Z_{k+1} \leftarrow Z_k$ with site $i$ set to $z^\dagger$
    \IF{$\mathbf{1}_{\mathrm{valid}}(Z_{k+1})=1$ {\bfseries and} $S(Z_{k+1}) > S(Z^\ast)$}
      \STATE $Z^\ast \leftarrow Z_{k+1}$ \algcomment{track best-so-far during warmup}
    \ENDIF
  \ENDFOR

  \STATE \algcomment{Guided phase: exponential tilting with hard validity mask}
  \FOR{$k=K_{\mathrm{warm}}$ {\bfseries to} $K-1$}
    \STATE Select a free linker site $i$ \algcomment{atom slot in $\mathcal{I}_{\mathrm{link}}$ or bond entry incident to it}
    \STATE $p_{k+1} \leftarrow \Phi_{\theta^\ast}(Z_k)[\cdot, i]$ \algcomment{predicted distribution at site $i$}

    \FOR{{\bfseries each} $z' \in \mathrm{TopK}(p_{k+1})$}
      \STATE $p'(z') \propto p_{k+1}(z') \,\exp(\beta\, S(z'))\, \mathbf{1}_{\mathrm{valid}}(z')$
    \ENDFOR

    \STATE Sample $z^\dagger \sim p'$
    \STATE $Z' \leftarrow Z_k$ with site $i$ set to $z^\dagger$ \algcomment{atom-type or bond-type edit}

    \STATE $Z_{k+1} \leftarrow Z'$ \algcomment{no acceptance test; see Table~\ref{tab:acceptance_ablation}}
    \IF{$\mathbf{1}_{\mathrm{valid}}(Z_{k+1})=1$ {\bfseries and} $S(Z_{k+1}) > S(Z^\ast)$}
      \STATE $Z^\ast \leftarrow Z_{k+1}$ \algcomment{track best-so-far}
    \ENDIF
  \ENDFOR

  \STATE {\bfseries return} $Z^\ast$ \algcomment{best PROTAC graph found}
\end{algorithmic}
\end{algorithm}

\FloatBarrier
\subsection{Notation}

\begin{table}[!ht]
\centering
\caption{Notation used in the PROTAC sampling algorithm, with parameter values in Table~\ref{tab:sampling_config}.}
\label{tab:notation}
\begin{tabular}{cl}
\toprule
\textbf{Symbol} & \textbf{Description} \\
\midrule
$\Phi_{\theta^*}$ & Pre-trained discrete flow model with parameters $\theta^*$ \\
$\mathcal{Z}$ & Discrete state space (PROTAC molecular graphs) \\
$\mu_0$ & Prior distribution (empty linker, fixed warhead/E3) \\
$K$ & Total number of guided sampling steps \\
$K_{\mathrm{warm}}$ & Number of unguided warmup steps \\
$S(z)$ & Aggregate guidance score (Section~\ref{subsec:objectives_protacs}) \\
$\beta$ & Guidance strength (temperature parameter) \\
$\mathbf{1}_{\mathrm{valid}}(z)$ & Hard validity indicator \\
$Z^{\ast}$ & Best PROTAC graph found during sampling \\
\bottomrule
\end{tabular}
\end{table}

\subsection{Two-Phase Sampling Framework}

Sampling proceeds in two phases (Algorithm~\ref{alg:ExpTiltSimple}).
The first phase performs unguided sampling from the pre-trained
flow for $K_{\mathrm{warm}}$ steps, growing the linker outward
from the anchors during the early-branching regime where the soft
objectives provide limited discriminative signal
(Section~\ref{subsec:warmup}). The second phase applies exponential tilting
with hard validity masking for the remaining $K - K_{\mathrm{warm}}$ steps,
optimizing the aggregate objective $S(z)$ while maintaining chemical validity.

\subsubsection{Phase 1: Warmup}
\label{subsec:warmup}
The warmup phase samples directly from the pre-trained flow:
$Z_{k+1} \sim \Phi_{\theta^*}(Z_k)$ for
$k = 0, \ldots, K_{\mathrm{warm}} - 1$. No tilting and no
acceptance check are applied. The best-so-far state $Z^{\ast}$ is
tracked over warmup steps so that any high-scoring intermediate
configurations are retained.

\subsubsection{Phase 2: Guided Optimization}
The guided phase applies exponential tilting with hard validity
masking. At each step, a free linker
site $i$ (atom slot in $\mathcal{I}_{\mathrm{link}}$ or bond
entry incident to it) is selected, and the model's predicted
distribution $p_{k+1}$ at site $i$ is tilted by the aggregate
score $S$ and masked by validity.

\begin{table}[H]
\centering
\caption{Two-phase sampling parameters.}
\label{tab:sampling_config}
\begin{tabular}{lcp{6cm}}
\toprule
\textbf{Parameter} & \textbf{Value} & \textbf{Description} \\
\midrule
Total steps ($K$)                  & 1500  & Total sampling steps (warmup + guided) \\
Warmup steps ($K_{\mathrm{warm}}$) & 100   & Unguided sampling from pre-trained flow \\
Guided steps                       & 1400  & $K - K_{\mathrm{warm}}$, with tilting + no acceptance \\
Guidance strength ($\beta$)        & 5.0   & Exponential tilting temperature \\
Top-$K$                            & 50    & Candidates per site for tilting \\
Temperature                        & 1.0   & Softmax temperature for predictions \\
Position selection                 & Random & Uniform random free linker site each step \\
\texttt{use\_acceptance}           & Disabled & Every tilted proposal accepted; best-so-far tracked \\
\bottomrule
\end{tabular}
\end{table}

\paragraph{Log-Space Computation.} For numerical stability, we
compute the tilted distribution in log-space:
\begin{align}
\tilde{p}(z') &= \log p_{k}(z') + \beta \cdot S(z') + \log \mathbf{1}_{\mathrm{valid}}(z'), \\
\tilde{p}_{\mathrm{stable}}(z') &= \tilde{p}(z') - \max_{z''} \tilde{p}(z''), \\
p'(z') &= \frac{\exp(\tilde{p}_{\mathrm{stable}}(z'))}{\sum_{z''} \exp(\tilde{p}_{\mathrm{stable}}(z''))}.
\end{align}
Tilting and validity masking are restricted to the TopK candidates
under $p_{k+1}$ to limit per-step scorer evaluation cost.

\paragraph{Fallback Behavior.} If exponential tilting produces
invalid probabilities (all candidates fail validity, or numerical
underflow), we fall back to sampling from the base distribution
$p_{k+1}$ restricted to valid candidates. If no valid candidate
exists in the TopK, the step is rejected and $Z_{k+1} = Z_k$.

\paragraph{Acceptance Criterion (disabled).} We tested the monotonic
acceptance criterion in the PROTAC setting. Requiring
$S(Z') \ge S(Z_k)$ reduced assembly validity by $24$-$26$ percentage
points and slightly increased SA (Table~\ref{tab:acceptance_ablation}).
The final sampler accepts each tilted proposal and returns the
highest-scoring valid state encountered. The PDC sampler
(Appendix~\ref{sec:algorithm}) retains the acceptance criterion.

\begin{table}[h]
\centering
\caption{Monotonic acceptance ablation. Assembly validity (\%), mean $\pm$
std over three seeds, $100$ linkers per seed.}
\label{tab:acceptance_ablation}
\small
\begin{tabular}{lcc}
\toprule
\textbf{System} & \textbf{Acceptance on} & \textbf{Acceptance off} \\
\midrule
EGFR-VHL  & 58.7 $\pm$ 3.4 & \textbf{85.0 $\pm$ 3.6} \\
BRD9-VHL  & 58.3 $\pm$ 4.1 & \textbf{82.3 $\pm$ 4.0} \\
CDK6-CRBN & 58.7 $\pm$ 2.9 & \textbf{83.0 $\pm$ 4.2} \\
\bottomrule
\end{tabular}
\end{table}

\subsection{Objective Weights}

The aggregate objective $S(z) = \sum_{j=1}^{3} w_j S_j(z)$ uses the weights
in Table~\ref{tab:objective_weights}, selected by the ablation in
Appendix~\ref{sec:weight_ablation}. The connectivity weight $w_3 = 4.0$ is
elevated so that the topological gradient toward connected, anchor-bridging
linkers dominates early in sampling, where partial assemblies score similarly
under $S_1$-$S_2$.

\begin{table}[H]
\centering
\caption{Soft objective weights for the aggregate score $S(z)$.}
\label{tab:objective_weights}
\begin{tabular}{llc}
\toprule
\textbf{Scorer} & \textbf{Description} & \textbf{Weight} \\
\midrule
$S_1$ & Synthetic accessibility ramp  & 4.0 \\
$S_2$ & PROTAC-DB chemotype closeness & 2.0 \\
$S_3$ & Soft connectivity gradient    & 4.0 \\
\bottomrule
\end{tabular}
\end{table}

\subsection{Objective Weight Ablation}
\label{sec:weight_ablation}
\label{sec:protaclikeness}

We swept each objective weight independently on EGFR-VHL, generating $100$
linkers across three seeds ($42/43/44$) per setting while holding the remaining
weights at $1$. Tables~\ref{tab:weight_ablation_val},~\ref{tab:weight_ablation_sa}, and~\ref{tab:weight_ablation_dm}
report the completed sweep. Every row shares the weight-$1$ reference
configuration, and the term below the rule in each table is the one excluded
from the final objective. A PROTAC-likeness classifier term is omitted from all
three: it acts as a $0/1$ gate, so its weight cannot change candidate ranking
and every metric is identical to the reference at each weight.

We selected the per-term argmax of this sweep for the final configuration:
connectivity $4$, SA $4$, chemotype $2$, and fingerprint $0$. Combining the
three retained terms reached $85.0 \pm 3.6\%$ assembly validity
(Table~\ref{tab:validity_degradmaster}), above any single-term setting in the
sweep.

\begin{table}[!ht]
\centering
\caption{Objective weight sweep on EGFR-VHL: assembly validity (\%), higher
is better. Seeds $42/43/44$, $100$ linkers per setting, mean $\pm$ std over
seeds. \textbf{Bold}: value used in the final configuration.}
\label{tab:weight_ablation_val}
\small
\setlength{\tabcolsep}{6pt}
\begin{tabular}{lcccc}
\toprule
\textbf{Term} & 0 & 1 & 2 & 4 \\
\midrule
Connectivity            & 36.7 $\pm$ 4.7 & 60.7 $\pm$ 8.5 & 72.0 $\pm$ 5.3 & \textbf{75.3 $\pm$ 5.5} \\
Chemotype closeness      & 51.0 $\pm$ 7.9 & 60.7 $\pm$ 8.5 & \textbf{64.7 $\pm$ 10.0} & 64.3 $\pm$ 8.0 \\
Synthetic accessibility & 62.7 $\pm$ 5.7 & 60.7 $\pm$ 8.5 & 60.7 $\pm$ 9.5 & \textbf{55.0 $\pm$ 8.9} \\
\midrule
Morgan fingerprint      & \textbf{64.3 $\pm$ 8.0} & 60.7 $\pm$ 8.5 & 52.3 $\pm$ 9.1 & 50.7 $\pm$ 5.5 \\
\bottomrule
\end{tabular}
\end{table}

\begin{table}[!ht]
\centering
\caption{Objective weight sweep on EGFR-VHL: SA score, lower is better. Same
runs as Table~\ref{tab:weight_ablation_val}.}
\label{tab:weight_ablation_sa}
\small
\setlength{\tabcolsep}{6pt}
\begin{tabular}{lcccc}
\toprule
\textbf{Term} & 0 & 1 & 2 & 4 \\
\midrule
Connectivity            & 2.94 $\pm$ 0.03 & 3.03 $\pm$ 0.11 & 3.07 $\pm$ 0.20 & \textbf{3.11 $\pm$ 0.07} \\
Chemotype closeness      & 2.96 $\pm$ 0.11 & 3.03 $\pm$ 0.11 & \textbf{3.11 $\pm$ 0.14} & 3.21 $\pm$ 0.16 \\
Synthetic accessibility & 3.36 $\pm$ 0.06 & 3.03 $\pm$ 0.11 & 2.80 $\pm$ 0.12 & \textbf{2.64 $\pm$ 0.08} \\
\midrule
Morgan fingerprint      & \textbf{3.00 $\pm$ 0.07} & 3.03 $\pm$ 0.11 & 3.06 $\pm$ 0.12 & 3.24 $\pm$ 0.10 \\
\bottomrule
\end{tabular}
\end{table}

\begin{table}[!ht]
\centering
\caption{Objective weight sweep on EGFR-VHL: mean DegradeMaster probability
over valid generations, higher is better. Same runs as
Table~\ref{tab:weight_ablation_val}.}
\label{tab:weight_ablation_dm}
\small
\setlength{\tabcolsep}{6pt}
\begin{tabular}{lcccc}
\toprule
\textbf{Term} & 0 & 1 & 2 & 4 \\
\midrule
Connectivity            & 0.518 $\pm$ 0.022 & 0.481 $\pm$ 0.022 & 0.498 $\pm$ 0.029 & \textbf{0.496 $\pm$ 0.031} \\
Chemotype closeness      & 0.522 $\pm$ 0.021 & 0.481 $\pm$ 0.022 & \textbf{0.489 $\pm$ 0.046} & 0.422 $\pm$ 0.046 \\
Synthetic accessibility & 0.469 $\pm$ 0.033 & 0.481 $\pm$ 0.022 & 0.533 $\pm$ 0.026 & \textbf{0.536 $\pm$ 0.028} \\
\midrule
Morgan fingerprint      & \textbf{0.506 $\pm$ 0.053} & 0.481 $\pm$ 0.022 & 0.470 $\pm$ 0.057 & 0.447 $\pm$ 0.012 \\
\bottomrule
\end{tabular}
\end{table}

Removing connectivity guidance reduced assembly validity to $36.7\%$; weight
$4$ reached $75.3\%$. Chemotype closeness produced a smaller gain that
saturated at weight $2$. Increasing the synthetic-accessibility weight from
$0$ to $4$ reduced SA from $3.36$ to $2.64$, reduced assembly validity by
roughly $8$ points, and increased DegradeMaster from $0.469$ to $0.536$. We
retained this setting because synthetic accessibility is relevant to
experimental follow-up.

We excluded the Morgan fingerprint term. Across the four weights, assembly
validity fell monotonically from $64.3\%$ to $50.7\%$, SA increased from
$3.00$ to $3.24$, and DegradeMaster decreased from $0.506$ to $0.447$.
Removal gave the best value for all three metrics. The individual
DegradeMaster cells overlap within one standard deviation, so we interpreted
the result as a monotone trend across four weights and three metrics. This
conclusion applies to the EGFR-VHL sweep. On BRD9-VHL and CDK6-CRBN the same
four settings are flat ($86.0$-$86.2\%$ assembly validity, DM
$0.530$-$0.535$). We attribute the effect to a pull toward training-set
chemotypes, which conflicts with the novelty results in
Table~\ref{tab:diversity_memorization}; across all settings we observed $0\%$
exact memorization of training molecules.

We also ablated hard constraints and soft guidance
(Table~\ref{tab:soft_hard_ablation}). Soft guidance alone reached
$0.3 \pm 0.5\%$ assembly validity, hard constraints alone reached
$11.0 \pm 3.7\%$, and their combination reached $85.0 \pm 3.6\%$. The
combined result supports the joint use of feasibility constraints and
directional guidance.

\begin{table}[!ht]
\centering
\caption{Soft guidance and hard constraint ablation on EGFR-VHL. Assembly
validity (\%), mean $\pm$ std over three seeds, $100$ linkers per seed. The
full configuration reaches $82.3$-$85.0\%$ across all three benchmark systems
(Table~\ref{tab:validity_degradmaster}).}
\label{tab:soft_hard_ablation}
\small
\begin{tabular}{lc}
\toprule
\textbf{Configuration} & \textbf{Assembly validity (\%)} \\
\midrule
Soft guidance only     & 0.3 $\pm$ 0.5 \\
Hard constraints only  & 11.0 $\pm$ 3.7 \\
\textbf{Both}          & \textbf{85.0 $\pm$ 3.6} \\
\bottomrule
\end{tabular}
\end{table}

\subsection{Computational Requirements}
\label{app:compute_PROTACs}

Sampling experiments were conducted on a single NVIDIA A6000 GPU
with 48 GB of VRAM. We generated 100 PROTACs per target in one parallel
batch. The reported sampling times cover each complete batch.

Under a matched compute budget, CupOFMoCA generated $0.247$
assembly-valid PROTACs per second, compared with $0.181$ for DiffPROTACs,
$0.072$ for LinkerNet, and $0.000$ for DiffLinker. We generated each batch
of $100$ linkers on one NVIDIA RTX A6000 with an AMD EPYC 7662 CPU and
eight CPU threads. CupOFMoCA completed the batch in $356$\,s, compared
with $442$, $636$, and $432$\,s for DiffPROTACs, LinkerNet, and
DiffLinker, respectively.

CPU-side constraint evaluation limited throughput. Each guided step
evaluated the classifier and soft-scorer stack (RDKit sanitization, SMARTS
pattern matching, SA score, MW/logP/RotBonds windows, connectivity check) on
every candidate type; GPU forward passes consumed under $20\%$ of per-step
wall-clock time. Vectorizing the CPU-bound checks reduced the original
$8$-$11$ minute implementation to $356$ seconds with byte-identical outputs
and unchanged validity and quality metrics.

%%%%%%%%%%%%%%%%%%%%%%%%%%%%%%%%%%%%%%%%%%%%%%%%%%%%%%%%%%%%

\section{Impact Statement}

This work introduces a generative framework for designing molecular conjugates such as peptide-drug conjugates (PDCs) and PROTACs, which are widely used in targeted therapeutics. By improving the ability to generate chemically valid and functionally consistent conjugates, this approach may accelerate early-stage drug discovery and reduce experimental screening costs. Potential benefits include more efficient design of targeted therapies for cancer and other diseases. However, as with all generative methods in molecular design, there is a risk of unintended misuse, such as the generation of harmful or toxic compounds. In this work, we restrict generation through chemical validity constraints and domain-specific objectives, and do not release any experimentally validated compounds. Future work should consider incorporating additional safety filters and oversight when deploying such models in practical settings.

\clearpage
\section*{NeurIPS Paper Checklist}

\begin{enumerate}

\item {\bf Claims}
    \item[] Question: Do the main claims made in the abstract and introduction accurately reflect the paper's contributions and scope?
    \item[] Answer: \answerYes{} % Replace by \answerYes{}, \answerNo{}, or \answerNA{}.
    \item[] Justification: The abstract and Introduction (Sec.~\ref{sec:problem}) state the paper's three contributions: (i) a coupled discrete flow framework for molecular conjugate assembly that enforces cross-component compatibility during generation, (ii) an objective-guided sampling scheme under hard validity constraints, and (iii) a unified instantiation across PDC and PROTAC modalities. Each claim is supported in the paper: the coupled flow formulation in Sec.~\ref{sec:problem} (Sec.~\ref{subsec:data_space}--\ref{subsec:flow_maps}); the validity-gated exponential-tilting sampler in Sec.~\ref{subsec:generation}, Appendix~\ref{sec:algorithm}, and Algorithm~\ref{alg:ExpTiltSimple}; PDC results across three targets in Sec.~\ref{sec:results} (Tables~\ref{tab:top20-aa},~\ref{tab:top20-smiles80}); and PROTAC results in Sec.~\ref{subsec:results_protac} (Tables~\ref{tab:validity_degradmaster},~\ref{tab:diversity_memorization}).
    \item[] Guidelines:
    \begin{itemize}
        \item The answer \answerNA{} means that the abstract and introduction do not include the claims made in the paper.
        \item The abstract and/or introduction should clearly state the claims made, including the contributions made in the paper and important assumptions and limitations. A \answerNo{} or \answerNA{} answer to this question will not be perceived well by the reviewers.
        \item The claims made should match theoretical and experimental results, and reflect how much the results can be expected to generalize to other settings.
        \item It is fine to include aspirational goals as motivation as long as it is clear that these goals are not attained by the paper.
    \end{itemize}

\item {\bf Limitations}
    \item[] Question: Does the paper discuss the limitations of the work performed by the authors?
    \item[] Answer: \answerYes{} % Replace by \answerYes{}, \answerNo{}, or \answerNA{}.
    \item[] Justification: Limitations are discussed in the Discussion section. The molecular-dynamics analysis addresses linker conformational dynamics and ternary-geometry compatibility for four generated EGFR--VHL candidates, but does not capture full ternary-complex formation or cellular context, and DegradeMaster is an in-domain post-hoc predictor rather than an experimental readout; experimental validation will be required to assess stability, target engagement, and functional efficacy of generated conjugates. PROTAC baselines are evaluated as single batches while CupOFMoCA is run across three seeds, so cross-method differences carry uneven uncertainty. The Discussion also flags the current assumption of fixed linker--payload chemistry as a limitation to be relaxed in future work, and identifies extension to additional conjugation modalities and pharmacokinetic objectives as open directions. Computational requirements per sample are reported in Appendix~\ref{app:compute}.
    \item[] Guidelines:
    \begin{itemize}
        \item The answer \answerNA{} means that the paper has no limitation while the answer \answerNo{} means that the paper has limitations, but those are not discussed in the paper.
        \item The authors are encouraged to create a separate "Limitations" section in their paper.
        \item The paper should point out any strong assumptions and how robust the results are to violations of these assumptions (e.g., independence assumptions, noiseless settings, model well-specification, asymptotic approximations only holding locally). The authors should reflect on how these assumptions might be violated in practice and what the implications would be.
        \item The authors should reflect on the scope of the claims made, e.g., if the approach was only tested on a few datasets or with a few runs. In general, empirical results often depend on implicit assumptions, which should be articulated.
        \item The authors should reflect on the factors that influence the performance of the approach. For example, a facial recognition algorithm may perform poorly when image resolution is low or images are taken in low lighting. Or a speech-to-text system might not be used reliably to provide closed captions for online lectures because it fails to handle technical jargon.
        \item The authors should discuss the computational efficiency of the proposed algorithms and how they scale with dataset size.
        \item If applicable, the authors should discuss possible limitations of their approach to address problems of privacy and fairness.
        \item While the authors might fear that complete honesty about limitations might be used by reviewers as grounds for rejection, a worse outcome might be that reviewers discover limitations that aren't acknowledged in the paper. The authors should use their best judgment and recognize that individual actions in favor of transparency play an important role in developing norms that preserve the integrity of the community. Reviewers will be specifically instructed to not penalize honesty concerning limitations.
    \end{itemize}

\item {\bf Theory assumptions and proofs}
    \item[] Question: For each theoretical result, does the paper provide the full set of assumptions and a complete (and correct) proof?
    \item[] Answer: \answerYes{} % Replace by \answerYes{}, \answerNo{}, or \answerNA{}.
    \item[] Justification: All theoretical results are stated formally with their assumptions and accompanied by complete proofs in Appendix~\ref{sec:proofs}. Specifically, Theorem~\ref{thm:exp_tilt} (exponential tilting solves KL-regularized objective improvement under hard validity constraints) is stated with explicit assumptions on the base kernel $K(\cdot\mid z)$, the score $S$, and the validity indicator $\mathbf{1}_{\mathrm{valid}}$, and is proved via Lagrangian optimization with a strict-concavity argument that establishes uniqueness. Corollary~\ref{cor:guided_limits} characterizes the $\beta\to 0$ and $\beta\to\infty$ limits of the guided kernel and is proved from log-sum-exp normalization properties.
    \item[] Guidelines:
    \begin{itemize}
        \item The answer \answerNA{} means that the paper does not include theoretical results.
        \item All the theorems, formulas, and proofs in the paper should be numbered and cross-referenced.
        \item All assumptions should be clearly stated or referenced in the statement of any theorems.
        \item The proofs can either appear in the main paper or the supplemental material, but if they appear in the supplemental material, the authors are encouraged to provide a short proof sketch to provide intuition.
        \item Inversely, any informal proof provided in the core of the paper should be complemented by formal proofs provided in appendix or supplemental material.
        \item Theorems and Lemmas that the proof relies upon should be properly referenced.
    \end{itemize}

\item {\bf Experimental result reproducibility}
    \item[] Question: Does the paper fully disclose all the information needed to reproduce the main experimental results of the paper to the extent that it affects the main claims and/or conclusions of the paper (regardless of whether the code and data are provided or not)?
    \item[] Answer: \answerYes{} % Replace by \answerYes{}, \answerNo{}, or \answerNA{}.
    \item[] Justification: The paper provides full details required for reproduction. Base-model architectures, optimizers, schedules, and training corpora for AA-PepReDi, UAA-PepReDi, and SMILESReDi are described in Appendix~\ref{subsec:pepredi_base} (with the SMILESReDi configuration in Table~\ref{tab:smilesredi_config}); the binding affinity oracle is documented with architecture (Table~\ref{tab:binding_affinity_arch}) and training configuration (Table~\ref{tab:binding_affinity_training}). Sampling is fully specified by Algorithm~\ref{alg:ExpTiltSimple} together with Phase~1 and Phase~2 hyperparameters in Tables~\ref{tab:phase1_config} and~\ref{tab:phase2_config}. Hard validity constraints used during generation are formalized in Sec.~\ref{subsec:objectives} for PROTACs and instantiated per conjugation chemistry in Appendix~\ref{sec:conjugation_validity_constraints} (Tables~\ref{tab:aoadau-validation},~\ref{tab:sscpt-validation},~\ref{tab:dspasn38-validation}). Benchmark targets, linker--payload pairings, and candidate counts ($100$ generations per seed across three seeds for CupOFMoCA and single batches of $N=100$ for baselines on PROTACs; top-20 over 100 generations for PDCs) are specified in Sec.~\ref{sec:results} and Sec.~\ref{subsec:results_protac}. An anonymous code repository is provided as part of the submission.
    \item[] Guidelines:
    \begin{itemize}
        \item The answer \answerNA{} means that the paper does not include experiments.
        \item If the paper includes experiments, a \answerNo{} answer to this question will not be perceived well by the reviewers: Making the paper reproducible is important, regardless of whether the code and data are provided or not.
        \item If the contribution is a dataset and/or model, the authors should describe the steps taken to make their results reproducible or verifiable.
        \item Depending on the contribution, reproducibility can be accomplished in various ways. For example, if the contribution is a novel architecture, describing the architecture fully might suffice, or if the contribution is a specific model and empirical evaluation, it may be necessary to either make it possible for others to replicate the model with the same dataset, or provide access to the model. In general. releasing code and data is often one good way to accomplish this, but reproducibility can also be provided via detailed instructions for how to replicate the results, access to a hosted model (e.g., in the case of a large language model), releasing of a model checkpoint, or other means that are appropriate to the research performed.
        \item While NeurIPS does not require releasing code, the conference does require all submissions to provide some reasonable avenue for reproducibility, which may depend on the nature of the contribution. For example
        \begin{enumerate}
            \item If the contribution is primarily a new algorithm, the paper should make it clear how to reproduce that algorithm.
            \item If the contribution is primarily a new model architecture, the paper should describe the architecture clearly and fully.
            \item If the contribution is a new model (e.g., a large language model), then there should either be a way to access this model for reproducing the results or a way to reproduce the model (e.g., with an open-source dataset or instructions for how to construct the dataset).
            \item We recognize that reproducibility may be tricky in some cases, in which case authors are welcome to describe the particular way they provide for reproducibility. In the case of closed-source models, it may be that access to the model is limited in some way (e.g., to registered users), but it should be possible for other researchers to have some path to reproducing or verifying the results.
        \end{enumerate}
    \end{itemize}

\item {\bf Open access to data and code}
    \item[] Question: Does the paper provide open access to the data and code, with sufficient instructions to faithfully reproduce the main experimental results, as described in supplemental material?
    \item[] Answer: \answerYes{} % Replace by \answerYes{}, \answerNo{}, or \answerNA{}.
    \item[] Justification: An anonymous code repository is provided alongside the submission, containing the CupOFMoCA framework, the per-conjugation builders and \texttt{PDCAnalyzer} for PDCs, the PROTAC validity classifier, the separate PROTAC-likeness gate, the three soft scorers $S_1$--$S_3$ defined in Sec.~\ref{subsec:objectives}, and the two-phase sampling pipeline of Algorithm~\ref{alg:ExpTiltSimple}. Experiments use publicly available resources, including PDCdb~\citep{sun2025pdcdb}, PROTAC-DB, the EGFR--VHL/BRD9--VHL/CDK6--CRBN PROTAC systems, PepNN~\citep{abdin2022pepnn}, BioLip2~\citep{zhang2024biolip2}, and PPIRef~\citep{bushuiev2023learning}, all cited in the corresponding sections. Pre-trained components (PepReDi/SMILESReDi base flows from~\citet{chen2025areurediannealedrectifiedupdates}, the ESM-2-650M tokenizer/encoder~\citep{lin2023evolutionary}, the binding-affinity oracle, and DegradeMaster~\citep{liu2025degrademaster}) and baselines (LinkerNet~\citep{guan2023linkernet}, DiffPROTACs~\citep{li2024diffprotacs}, DiffLinker~\citep{igashov2024difflinker}, PepTune~\citep{tang2025peptunenovogenerationtherapeutic}, AReUReDi~\citep{chen2025areurediannealedrectifiedupdates}) are publicly available and cited.
    \item[] Guidelines:
    \begin{itemize}
        \item The answer \answerNA{} means that paper does not include experiments requiring code.
        \item Please see the NeurIPS code and data submission guidelines (\url{https://neurips.cc/public/guides/CodeSubmissionPolicy}) for more details.
        \item While we encourage the release of code and data, we understand that this might not be possible, so \answerNo{} is an acceptable answer. Papers cannot be rejected simply for not including code, unless this is central to the contribution (e.g., for a new open-source benchmark).
        \item The instructions should contain the exact command and environment needed to run to reproduce the results. See the NeurIPS code and data submission guidelines (\url{https://neurips.cc/public/guides/CodeSubmissionPolicy}) for more details.
        \item The authors should provide instructions on data access and preparation, including how to access the raw data, preprocessed data, intermediate data, and generated data, etc.
        \item The authors should provide scripts to reproduce all experimental results for the new proposed method and baselines. If only a subset of experiments are reproducible, they should state which ones are omitted from the script and why.
        \item At submission time, to preserve anonymity, the authors should release anonymized versions (if applicable).
        \item Providing as much information as possible in supplemental material (appended to the paper) is recommended, but including URLs to data and code is permitted.
    \end{itemize}

\item {\bf Experimental setting/details}
    \item[] Question: Does the paper specify all the training and test details (e.g., data splits, hyperparameters, how they were chosen, type of optimizer) necessary to understand the results?
    \item[] Answer: \answerYes{} % Replace by \answerYes{}, \answerNo{}, or \answerNA{}.
    \item[] Justification: Full training and test details are reported across the appendices. Base-model training settings (Adam with $\mathrm{lr}=10^{-4}$, weight decay $10^{-5}$, $2\times 10^{-5}$ for rectified runs, $10$-epoch warmup, cosine decay, 16-step inference, and $80{:}10{:}10$ train/validation/test splits for the AA/UAA corpora) are given in Appendix~\ref{subsec:pepredi_base}; SMILESReDi configuration is in Table~\ref{tab:smilesredi_config}; the binding affinity oracle uses an Optuna-tuned configuration described in Tables~\ref{tab:binding_affinity_arch} and~\ref{tab:binding_affinity_training}. Sampling-time settings for PDCs (Phase 1: $N=16$ flow steps at temperature $1.0$ with \texttt{require\_valid\_x0}; Phase 2: $K=128$ guided steps, $\beta=50.0$, top-$K=200$, random position selection, acceptance criterion enabled) are reported in Tables~\ref{tab:phase1_config} and~\ref{tab:phase2_config}. PROTAC sampling uses $1500$ total steps ($100$ unguided warmup, $1400$ guided), $\beta=5.0$, top-$K=50$, random position selection, and monotonic acceptance disabled, because requiring non-decreasing scores reduced assembly validity (Table~\ref{tab:sampling_config}; Table~\ref{tab:acceptance_ablation}). PROTAC scorers, gating ranges, and the weights $(w_1, w_2, w_3)=(4.0, 2.0, 4.0)$ selected by the sweep in Appendix~\ref{sec:weight_ablation} are documented in Sec.~\ref{subsec:objectives}.
    \item[] Guidelines:
    \begin{itemize}
        \item The answer \answerNA{} means that the paper does not include experiments.
        \item The experimental setting should be presented in the core of the paper to a level of detail that is necessary to appreciate the results and make sense of them.
        \item The full details can be provided either with the code, in appendix, or as supplemental material.
    \end{itemize}

\item {\bf Experiment statistical significance}
    \item[] Question: Does the paper report error bars suitably and correctly defined or other appropriate information about the statistical significance of the experiments?
    \item[] Answer: \answerYes{} % Replace by \answerYes{}, \answerNo{}, or \answerNA{}.
    \item[] Justification: Quantitative results report mean and standard deviation across multiple generated samples per condition. PDC tables (Tables~\ref{tab:top20-aa},~\ref{tab:top20-smiles80},~\ref{tab:top20-smiles100}) report top-20 binding affinity as mean (std) over 100 generation runs per method per target. Full-valid-set tables (Tables~\ref{tab:full-aa},~\ref{tab:full-smiles80},~\ref{tab:full-smiles100}, and per-target tables~\ref{tab:full-aa-erbb2}--\ref{tab:full-smiles100-nrp1}) extend this to all valid conjugates. PROTAC results (Table~\ref{tab:validity_degradmaster}) report validity as mean $\pm$ std across three seeds of $100$ generations for CupOFMoCA and single batches of $N=100$ for baselines, with DegradeMaster as mean (std) over assembly-valid generations together with threshold-crossing fractions; Table~\ref{tab:diversity_memorization} reports diversity and training-set similarity over the same pooled sets. Figure~\ref{fig:validity-binding} additionally visualizes binding-affinity error bars ($\pm 1$ std) for the joint validity--binding tradeoff, and Figure~\ref{fig:property-distributions} reports physicochemical property distributions as box plots.
    \item[] Guidelines:
    \begin{itemize}
        \item The answer \answerNA{} means that the paper does not include experiments.
        \item The authors should answer \answerYes{} if the results are accompanied by error bars, confidence intervals, or statistical significance tests, at least for the experiments that support the main claims of the paper.
        \item The factors of variability that the error bars are capturing should be clearly stated (for example, train/test split, initialization, random drawing of some parameter, or overall run with given experimental conditions).
        \item The method for calculating the error bars should be explained (closed form formula, call to a library function, bootstrap, etc.)
        \item The assumptions made should be given (e.g., Normally distributed errors).
        \item It should be clear whether the error bar is the standard deviation or the standard error of the mean.
        \item It is OK to report 1-sigma error bars, but one should state it. The authors should preferably report a 2-sigma error bar than state that they have a 96\% CI, if the hypothesis of Normality of errors is not verified.
        \item For asymmetric distributions, the authors should be careful not to show in tables or figures symmetric error bars that would yield results that are out of range (e.g., negative error rates).
        \item If error bars are reported in tables or plots, the authors should explain in the text how they were calculated and reference the corresponding figures or tables in the text.
    \end{itemize}

\item {\bf Experiments compute resources}
    \item[] Question: For each experiment, does the paper provide sufficient information on the computer resources (type of compute workers, memory, time of execution) needed to reproduce the experiments?
    \item[] Answer: \answerYes{} % Replace by \answerYes{}, \answerNo{}, or \answerNA{}.
    \item[] Justification: Compute resources are reported in Appendix~\ref{app:compute} and the model-training appendices. AA-PepReDi training (Appendix~\ref{subsec:pepredi_base}) uses a single NVIDIA GPU with $128$~GB of GPU memory under SLURM. SMILESReDi training (Table~\ref{tab:smilesredi_config}) uses a $4\times$NVIDIA RTX A6000 Ada system with $48$~GB of VRAM per GPU. CupOFMoCA sampling (Appendix~\ref{app:compute}) is performed on a single NVIDIA A6000 GPU with $48$~GB of VRAM, with Phase~1 (16 flow steps) completing in under one minute and Phase~2 (128 guided optimization steps) taking $5$--$10$ minutes per sample depending on sequence length and candidate size. Matched-hardware PROTAC throughput is reported in Appendix~\ref{app:compute_PROTACs}.
    \item[] Guidelines:
    \begin{itemize}
        \item The answer \answerNA{} means that the paper does not include experiments.
        \item The paper should indicate the type of compute workers CPU or GPU, internal cluster, or cloud provider, including relevant memory and storage.
        \item The paper should provide the amount of compute required for each of the individual experimental runs as well as estimate the total compute.
        \item The paper should disclose whether the full research project required more compute than the experiments reported in the paper (e.g., preliminary or failed experiments that didn't make it into the paper).
    \end{itemize}

\item {\bf Code of ethics}
    \item[] Question: Does the research conducted in the paper conform, in every respect, with the NeurIPS Code of Ethics \url{https://neurips.cc/public/EthicsGuidelines}?
    \item[] Answer: \answerYes{} % Replace by \answerYes{}, \answerNo{}, or \answerNA{}.
    \item[] Justification: The research conforms with the NeurIPS Code of Ethics. The work does not involve human subjects or sensitive personal data, all datasets and pre-trained models used are publicly available and properly cited, and the submission preserves anonymity through the anonymous code repository. The Impact Statement explicitly addresses dual-use considerations for accelerated therapeutic conjugate design.
    \item[] Guidelines:
    \begin{itemize}
        \item The answer \answerNA{} means that the authors have not reviewed the NeurIPS Code of Ethics.
        \item If the authors answer \answerNo{}, they should explain the special circumstances that require a deviation from the Code of Ethics.
        \item The authors should make sure to preserve anonymity (e.g., if there is a special consideration due to laws or regulations in their jurisdiction).
    \end{itemize}

\item {\bf Broader impacts}
    \item[] Question: Does the paper discuss both potential positive societal impacts and negative societal impacts of the work performed?
    \item[] Answer: \answerYes{} % Replace by \answerYes{}, \answerNo{}, or \answerNA{}.
    \item[] Justification: The paper includes a dedicated Impact Statement in the Appendix that discusses both the positive societal impacts of CupOFMoCA, including its ability to generate chemically valid molecular conjugates that preserve predicted activity at the assembled-system level and to extend across PDC and PROTAC modalities for therapeutic discovery, and the negative considerations, including that improved \textit{in silico} validity and predicted activity do not guarantee \textit{in vivo} safety, target engagement, or efficacy, and that accelerated conjugate design can raise safety and dual-use considerations. The Impact Statement explicitly notes that CupOFMoCA must be coupled with careful experimental validation, safety screening, and responsible deployment.
    \item[] Guidelines:
    \begin{itemize}
        \item The answer \answerNA{} means that there is no societal impact of the work performed.
        \item If the authors answer \answerNA{} or \answerNo{}, they should explain why their work has no societal impact or why the paper does not address societal impact.
        \item Examples of negative societal impacts include potential malicious or unintended uses (e.g., disinformation, generating fake profiles, surveillance), fairness considerations (e.g., deployment of technologies that could make decisions that unfairly impact specific groups), privacy considerations, and security considerations.
        \item The conference expects that many papers will be foundational research and not tied to particular applications, let alone deployments. However, if there is a direct path to any negative applications, the authors should point it out. For example, it is legitimate to point out that an improvement in the quality of generative models could be used to generate Deepfakes for disinformation. On the other hand, it is not needed to point out that a generic algorithm for optimizing neural networks could enable people to train models that generate Deepfakes faster.
        \item The authors should consider possible harms that could arise when the technology is being used as intended and functioning correctly, harms that could arise when the technology is being used as intended but gives incorrect results, and harms following from (intentional or unintentional) misuse of the technology.
        \item If there are negative societal impacts, the authors could also discuss possible mitigation strategies (e.g., gated release of models, providing defenses in addition to attacks, mechanisms for monitoring misuse, mechanisms to monitor how a system learns from feedback over time, improving the efficiency and accessibility of ML).
    \end{itemize}

\item {\bf Safeguards}
    \item[] Question: Does the paper describe safeguards that have been put in place for responsible release of data or models that have a high risk for misuse (e.g., pre-trained language models, image generators, or scraped datasets)?
    \item[] Answer: \answerNA{} % Replace by \answerYes{}, \answerNo{}, or \answerNA{}.
    \item[] Justification: The paper does not release new foundation models, datasets, or general-purpose generators that pose a high risk for misuse. Released artifacts consist of the CupOFMoCA sampling framework, conjugation builders, and validity analyzers, together with domain-specific models derived from existing public resources. The PDC part uses the publicly available base models (PepReDi/SMILESReDi from~\citet{chen2025areurediannealedrectifiedupdates}) and publicly available conjugate database PDCdb. The PROTAC generator is fine-tuned from publicly released GuacaMol checkpoint of DeFoG  \citep{qin2025defog} using public database PROTAC-DB. The Impact Statement notes that downstream therapeutic use requires careful experimental validation and safety screening before deployment.
    \item[] Guidelines:
    \begin{itemize}
        \item The answer \answerNA{} means that the paper poses no such risks.
        \item Released models that have a high risk for misuse or dual-use should be released with necessary safeguards to allow for controlled use of the model, for example by requiring that users adhere to usage guidelines or restrictions to access the model or implementing safety filters.
        \item Datasets that have been scraped from the Internet could pose safety risks. The authors should describe how they avoided releasing unsafe images.
        \item We recognize that providing effective safeguards is challenging, and many papers do not require this, but we encourage authors to take this into account and make a best faith effort.
    \end{itemize}

\item {\bf Licenses for existing assets}
    \item[] Question: Are the creators or original owners of assets (e.g., code, data, models), used in the paper, properly credited and are the license and terms of use explicitly mentioned and properly respected?
    \item[] Answer: \answerYes{} % Replace by \answerYes{}, \answerNo{}, or \answerNA{}.
    \item[] Justification: All existing assets used in the paper are properly credited via citations to the original works, including PDCdb~\citep{sun2025pdcdb}, PROTAC-DB, PepNN~\citep{abdin2022pepnn}, BioLip2~\citep{zhang2024biolip2}, PPIRef~\citep{bushuiev2023learning}, ESM-2-650M~\citep{lin2023evolutionary}, the PepReDi and SMILESReDi base flows and PeptideCLM-23M tokenizer~\citep{chen2025areurediannealedrectifiedupdates}, the p2smi toolkit~\citep{feller2025p2smi}, RDKit~\citep{landrum2025rdkit}, Optuna~\citep{akiba2019optunanextgenerationhyperparameteroptimization}, DegradeMaster~\citep{liu2025degrademaster}, and the LinkerNet~\citep{guan2023linkernet}, DiffPROTACs~\citep{li2024diffprotacs}, DiffLinker~\citep{igashov2024difflinker}, PepTune~\citep{tang2025peptunenovogenerationtherapeutic}, and AReUReDi~\citep{chen2025areurediannealedrectifiedupdates} baselines.
    \item[] Guidelines:
    \begin{itemize}
        \item The answer \answerNA{} means that the paper does not use existing assets.
        \item The authors should cite the original paper that produced the code package or dataset.
        \item The authors should state which version of the asset is used and, if possible, include a URL.
        \item The name of the license (e.g., CC-BY 4.0) should be included for each asset.
        \item For scraped data from a particular source (e.g., website), the copyright and terms of service of that source should be provided.
        \item If assets are released, the license, copyright information, and terms of use in the package should be provided. For popular datasets, paperswithcode.com/datasets has curated licenses for some datasets. Their licensing guide can help determine the license of a dataset.
        \item For existing datasets that are re-packaged, both the original license and the license of the derived asset (if it has changed) should be provided.
        \item If this information is not available online, the authors are encouraged to reach out to the asset's creators.
    \end{itemize}

\item {\bf New assets}
    \item[] Question: Are new assets introduced in the paper well documented and is the documentation provided alongside the assets?
    \item[] Answer: \answerYes{} % Replace by \answerYes{}, \answerNo{}, or \answerNA{}.
    \item[] Justification: The new assets introduced are the CupOFMoCA framework and the associated conjugation builders, validity analyzers, and PROTAC scorers. The framework is documented in Sec.~\ref{sec:problem}, Sec.~\ref{subsec:objectives}, and Algorithm~\ref{alg:ExpTiltSimple}; the per-conjugation builders and \texttt{PDCAnalyzer} for Aoa--Dau, SSCpt, and DSPA--SN-38 are documented in Appendix~\ref{sec:conjugation_validity_constraints}, including SMARTS patterns and valency requirements (Tables~\ref{tab:aoadau-validation},~\ref{tab:sscpt-validation},~\ref{tab:dspasn38-validation}); the PROTAC validity classifier, the separate PROTAC-likeness gate, and the three soft scorers $S_1$--$S_3$ are documented in Sec.~\ref{subsec:objectives}; and the sampling pipeline is documented in Appendix~\ref{sec:algorithm} and Tables~\ref{tab:phase1_config} and~\ref{tab:phase2_config}. The accompanying anonymous code repository contains corresponding implementations.
    \item[] Guidelines:
    \begin{itemize}
        \item The answer \answerNA{} means that the paper does not release new assets.
        \item Researchers should communicate the details of the dataset/code/model as part of their submissions via structured templates. This includes details about training, license, limitations, etc.
        \item The paper should discuss whether and how consent was obtained from people whose asset is used.
        \item At submission time, remember to anonymize your assets (if applicable). You can either create an anonymized URL or include an anonymized zip file.
    \end{itemize}

\item {\bf Crowdsourcing and research with human subjects}
    \item[] Question: For crowdsourcing experiments and research with human subjects, does the paper include the full text of instructions given to participants and screenshots, if applicable, as well as details about compensation (if any)?
    \item[] Answer: \answerNA{} % Replace by \answerYes{}, \answerNo{}, or \answerNA{}.
    \item[] Justification: The paper does not involve crowdsourcing or research with human subjects. All experiments are computational and use publicly available molecular and peptide data resources.
    \item[] Guidelines:
    \begin{itemize}
        \item The answer \answerNA{} means that the paper does not involve crowdsourcing nor research with human subjects.
        \item Including this information in the supplemental material is fine, but if the main contribution of the paper involves human subjects, then as much detail as possible should be included in the main paper.
        \item According to the NeurIPS Code of Ethics, workers involved in data collection, curation, or other labor should be paid at least the minimum wage in the country of the data collector.
    \end{itemize}

\item {\bf Institutional review board (IRB) approvals or equivalent for research with human subjects}
    \item[] Question: Does the paper describe potential risks incurred by study participants, whether such risks were disclosed to the subjects, and whether Institutional Review Board (IRB) approvals (or an equivalent approval/review based on the requirements of your country or institution) were obtained?
    \item[] Answer: \answerNA{} % Replace by \answerYes{}, \answerNo{}, or \answerNA{}.
    \item[] Justification: The paper does not involve crowdsourcing or research with human subjects, so IRB approval is not applicable.
    \item[] Guidelines:
    \begin{itemize}
        \item The answer \answerNA{} means that the paper does not involve crowdsourcing nor research with human subjects.
        \item Depending on the country in which research is conducted, IRB approval (or equivalent) may be required for any human subjects research. If you obtained IRB approval, you should clearly state this in the paper.
        \item We recognize that the procedures for this may vary significantly between institutions and locations, and we expect authors to adhere to the NeurIPS Code of Ethics and the guidelines for their institution.
        \item For initial submissions, do not include any information that would break anonymity (if applicable), such as the institution conducting the review.
    \end{itemize}

\item {\bf Declaration of LLM usage}
    \item[] Question: Does the paper describe the usage of LLMs if it is an important, original, or non-standard component of the core methods in this research? Note that if the LLM is used only for writing, editing, or formatting purposes and does \emph{not} impact the core methodology, scientific rigor, or originality of the research, declaration is not required.
    \item[] Answer: \answerNA{} % Replace by \answerYes{}, \answerNo{}, or \answerNA{}.
    \item[] Justification: LLMs are not used as an important, original, or non-standard component of the core methodology. The framework concerns coupled discrete flows and validity-gated exponential-tilting sampling for molecular conjugate assembly, with Diffusion Transformer backbones for the PepReDi and SMILESReDi base flows. The pre-trained ESM-2-650M sequence encoder is used only as a tokenizer and embedding source for the binding-affinity oracle, as cited in Appendix~\ref{subsec:pepredi_base}.
    \item[] Guidelines:
    \begin{itemize}
        \item The answer \answerNA{} means that the core method development in this research does not involve LLMs as any important, original, or non-standard components.
        \item Please refer to our LLM policy in the NeurIPS handbook for what should or should not be described.
    \end{itemize}

\end{enumerate}

\end{document}